\documentclass[longauth]{aa}  
\usepackage{graphicx}
\usepackage{subfigure}
\usepackage[varg]{txfonts}
\usepackage{natbib}
\usepackage{amssymb,amsmath}
\bibpunct{(}{)}{;}{a}{}{,} 

\usepackage{multirow}
\usepackage{bigstrut}
\usepackage[small,bf,singlelinecheck=off]{caption}
\usepackage{pdflscape} 
\usepackage{longtable} 
\usepackage{capt-of}

\begin{document}
   \title{Star Formation in the Local Universe from the CALIFA sample}

   \subtitle{I. Calibrating the SFR using IFS data}

   \author{C. Catal\'an-Torrecilla\inst{1} \and A. Gil de Paz\inst{1} \and A. Castillo-Morales\inst{1} \and J. Iglesias-P\'aramo\inst{2,3} \and S.F. S\'anchez\inst{4} \and R. C. Kennicutt\inst{5} \and P.G. P\'erez-Gonz\'alez\inst{1}  \and  R. A. Marino\inst{6} \and C.J. Walcher\inst{7} \and B. Husemann\inst{8} \and R. Garc\'ia-Benito\inst{2} \and D. Mast\inst{9} \and R. M. Gonz\'alez Delgado\inst{2} \and J.C. Mu\~{n}oz-Mateos\inst{10} \and J. Bland-Hawthorn\inst{11} \and D. J. Bomans\inst{12} \and A. del Olmo\inst{2} \and L. Galbany\inst{13,14} \and J.M. Gomes\inst{15} \and C. Kehrig\inst{2} \and \'A. R. L\'opez-S\'anchez\inst{16,17} \and  M. A. Mendoza\inst{2} \and A. Monreal-Ibero\inst{18} \and M. P\'erez-Torres\inst{2,19} \and P. S\'anchez-Bl\'azquez\inst{20} \and J. M. Vilchez\inst{2} \and the CALIFA collaboration}
\institute{Departamento de Astrof\'{\i}sica y CC. de la Atm\'{o}sfera, Universidad Complutense de Madrid, E-28040, Madrid, Spain\\
\email{ccatalan@ucm.es}
\and Instituto de Astrof\'{\i}sica de Andaluc\'{\i}a-CSIC, Glorieta de la Astronom\'{\i}a, 18008 Granada, Spain
\and Estaci\'{o}n Experimental de Zonas \'Aridas (CSIC), Ctra. de Sacramento s/n, La Ca\~{n}ada, Almer\'{\i}a, Spain
\and Instituto de Astronom\'{\i}a,Universidad Nacional Auton\'{o}ma de M\'{e}xico, A.P. 70-264, 04510, M\'{e}xico,D.F.
\and Institute of Astronomy, University of Cambridge, Madingley Road, Cambridge, CB3 0HA, United Kingdom
\and CEI Campus Moncloa, UCM-UPM, Departamento de Astrof\'{\i}sica y CC. de la Atm\'{o}sfera, Universidad Complutense de Madrid, E-28040, Madrid, Spain
\and Leibniz-Institut f\"ur Astrophysik Potsdam (AIP), An der Sternwarte 16, D-14482 Potsdam, Germany
\and European Southern Observatory, Karl-Schwarzschild-Str. 2, 85748 Garching b. M\"unchen, Germany
\and Instituto de Cosmologia, Relatividade e Astrof\'{\i}sica - ICRA, Centro Brasileiro de Pesquisas F\'{\i}sicas, Rua Dr.Xavier Sigaud 150, CEP 22290-180, Rio de Janeiro, RJ, Brazil
\and European Southern Observatory, Casilla 19001, Santiago 19, Chile
\and Sydney Institute for Astronomy, School of Physics A28, University of Sydney, NSW2006, Australia
\and Astronomical Institute of the Ruhr-University Bochum Universitaetsstr, 150, 44801 Bochum, Germany       
\and Millennium Institute of Astrophysics, Universidad de Chile, Casilla 36-D, Santiago, Chile
\and Departamento de Astronom\'{\i}a, Universidad de Chile, Casilla 36-D, Santiago, Chile
\and Instituto de Astrof\'{\i}sica e Ci\^encias do Espa\c{c}o, Universidade do Porto, CAUP, Rua das Estrelas, PT4150-762 Porto, Portugal
\and Australian Astronomical Observatory, PO Box 915, North Ryde, NSW 1670, Australia
\and Department of Physics and Astronomy, Macquarie University, NSW 2109, Australia
\and GEPI, Observatoire de Paris, CNRS UMR8111, Universit\'e Paris Diderot, Place Jules Janssen, 92190 Meudon, France
\and Centro de Estudios de la F\'{\i}sica del Cosmos de Arag\'on (CEFCA), 44001 Teruel, Spain
\and Departamento de F\'{\i}sica Te\'{o}rica, Universidad Aut\'{o}noma de Madrid, Cantoblanco, E28049, Spain
}
   \date{}


  \abstract
     {The Star Formation Rate (SFR) is one of the main parameters used to analyze the evolution of galaxies through time. The need for recovering the light reprocessed by dust commonly requires the use of low spatial resolution far-infrared data. Recombination-line luminosities provide an alternative, although uncertain dust-extinction corrections based on narrow-band imaging or long-slit spectroscopy have traditionally posed a limit to their applicability. Integral Field Spectroscopy (IFS) is clearly the way to overcome such limitation.}
   {We obtain integrated H$\alpha$, ultraviolet (UV) and infrared (IR)-based SFR measurements for 272 galaxies from the CALIFA survey at 0.005\,<\,z\,<\,0.03 using single-band and hybrid tracers. We aim to determine whether the extinction-corrected H$\alpha$ luminosities provide a good measure of the SFR and to shed light on the origin of the discrepancies between tracers. 
   Updated calibrations referred to H$\alpha$ are provided. The well-defined selection criteria and the large statistics allow us to carry out this analysis globally and split by properties, including stellar mass and morphological type.}
   {We derive integrated extinction-corrected H$\alpha$ fluxes from CALIFA, UV surface and asymptotic photometry from GALEX and integrated WISE 22$\mu$m and IRAS fluxes.}
   {We find that the extinction-corrected H$\alpha$ luminosity agrees with the hybrid updated SFR estimators based on either UV or H$\alpha$ plus IR luminosity over the full range of SFRs (0.03-20 M$_{\odot}$\,yr$^{-1}$). The coefficient that weights the amount of energy produced by newly-born stars that is reprocessed by dust on the hybrid tracers, a$_{IR}$, shows a large dispersion. However, it does not became increasingly small at high attenuations, as expected if significant highly-obscured H$\alpha$ emission would be missed, i.e. after a Balmer decrement-based attenuation correction is applied. Lenticulars, early-type spirals and type-2 AGN host galaxies show smaller coefficients due to the contribution of optical photons and AGN to dust heating.}
   {In the Local Universe the H$\alpha$ luminosity derived from IFS observations can be used to measure SFR, at least in statistically-significant, optically-selected galaxy samples, once stellar continuum absorption and dust attenuation effects are accounted for. The analysis of the SFR calibrations by galaxies properties could be potentially used by other works to study the impact of different selection criteria in the SFR values derived and to disentangle selection effects from other physically motivated differences, such as environmental or evolutionary effects.}


   \maketitle
%

\section{Introduction}
\label{Introduction}

The measurement of the star formation rate (SFR) is crucial for understanding the birth and evolution of the galaxies \citep{Kennicutt_98} as it provides information on the amount of gas in galaxies and the efficiency in the formation of stars inside them, which depends strongly on the conditions of the interstellar medium in which they are formed \citep[][and references therein]{Kennicutt_Evans_2012}. The SFR is, together with galaxy mass, one of the most important parameters that define galaxies and their evolution across cosmic times \citep{Somerville_Dave_2014, Madau_Dickinson_2014}. Several authors have tried to quantify the rate of on-going star formation and its evolution with redshift \citep[e.g.,][]{Madau_1996, Lilly_1996, Perez_Gonzalez_2008, Bouwens_2007,Bouwens_2011,Bouwens_2014} using different tracers. These works have shown that the SFR density has declined by roughly a factor of six from z$=$2 to present day \citep{Hopkins_Beacom_2006}.

Until now, the study of the evolution of the SFR has focussed on the analysis of the integrated SFR in galaxies, with little attention being paid to where in galaxies (nuclei, bulges, disks) SFR takes places and how the SFR in each of these components evolves separately with redshift. It is remarkable that the use of NIR integral field spectroscopy on 8-10\,m class telescopes is now allowing us to measure the SFR in these different components in distant galaxies, up to z=1-3 \citep [e.g.,][]{Genzel_2008,forster_schreiber_2009,forster_schreiber_2011a,forster_schreiber_2011b,Nelson_2012,Nelson_2013,Wuyts_2013,Lang_2014} while the local benchmark for these and possible future studies is still missing except for a few studies rather limited in number and completeness \citep [e.g.,][]{pablo_m81,Kennicutt_2007,Leroy_2008,Bigiel_2008,Blanc_2009,Schruba_2011,Leroy_2012}. In this regard, a correct determination of the calibrators we use to calculate the spatially-resolved SFR is essential in order to compare how the star-formation of these different spatial components behave at different wavelength ranges and/or redshifts.

Although SFR calibrators have existed for almost 30 years, the last decade has been particularly fruitful thanks to the multi-wavelength surveys of nearby and distant galaxies. The development of the Integral Field Spectroscopy (IFS) technique has allowed us to combine the advantages of both imaging and spectroscopy at optical and near-infrared (NIR) wavelengths. In this paper we make use of a large and well-characterized sample of nearby galaxies from the Calar Alto Legacy Integral Field Area (CALIFA) survey \citep{Sanchez_2012} that spans the entire color-magnitude diagram to address this fundamental issue. CALIFA allows us to properly determine the H$\alpha$ and H$\beta$ fluxes using IFS spectroscopic data. This is particularly important in the case of galaxies with low equivalent widths in emission, specially in H$\beta$, like many of the objects in the CALIFA sample and in the Local Universe in general \citep{Gallego_1995,Brinchmann_2004}, where narrow-band imaging is not feasible. Furthermore, using these data we can separate the H$\alpha$ and [NII] flux while in narrow band imaging is only feasible if a [NII]/H$\alpha$ ratio is assumed. In this paper, the use of the uniqueness IFS data will allow us to obtain precise Balmer-decrement measurements to compute H$\alpha$ extinction-corrected luminosities. Various studies have shown the importance of computing the extinction using IFS data in nearby regions where the line ratios obtained from the integrated spectra are dominated by regions of lower surface brightness rather than by the brighter ones \citep{pellegrini_2010,relano_2010,monreal_ibero_2011}. Other advantage related with the IFS data is that we can cover the whole galaxy avoiding problems associated with the limited spatial coverage of long-slit spectroscopy. From these Balmer-corrected H$\alpha$ luminosities, we compute their corresponding SFRs which will be use as a fiducial measure of the current SFR. However, it is critical to first determine that at least, in a statistically sense, no significant fraction of the SFR is being missed when using the extinction-corrected H$\alpha$ luminosity as SFR estimator. This requires of a combined analysis of this estimator with other SFR estimators, including the continuum ultraviolet (UV) emission, recombination lines of hydrogen and other atomic species together with other estimators less affected by dust attenuation such as total infrared (TIR) luminosity, monochromatic infrared (IR) emission or radio emission. The combination of different SFR estimators is also needed to evaluate the potential differences between the current-day SFR given by H$\alpha$ and that given by tracers sensitive to intermediate-aged stellar population \citep{Kennicutt_Evans_2012,Calzetti_2012}. Ordered from less to more sensitive 22\,$\mu$m, FUV, NUV, TIR. Whether 22\,$\mu$m should precede FUV in this list is still controversial although some results indicate that should be the case \citep{pablo_m81,alonso_herrero_2006_paalpha,Calzetti_2007,Calzetti_2010,Kennicutt_2009}.

The SFR indicators considered in this paper come in two types: single-band and hybrid recipes \cite[see][for a recent compilation]{Kennicutt_Evans_2012,Calzetti_2012}. In the case of the recipes based on a single photometric band we have used the extinction-corrected UV (with a extinction correction based on the UV slope; \cite{treyer_2007,Cortese_2008,JuanCarlos_2009a}; a more precise dust-extinction correction is implicit to the use of UV+IR hybrid tracer), the extinction-corrected H$\alpha$ and the observed mid-infrared (MIR) or TIR luminosities. The hybrid ones combine luminosities measured directly (observed UV or H$\alpha$) with that of the light re-emitted by dust after being heated by young massive stars (in our case the MIR or TIR luminosities), assuming an approximate energy-balance approach \cite[see][for more details]{gordon_2000,inoue_2001,hirashita_2003,jorge_2006,Calzetti_2007, Kennicutt_2007,Kennicutt_2009,Hao_2011}

In this work, we derive integrated extinction-corrected H$\alpha$-based SFRs from the analysis of CALIFA IFS data and compare them with measurements from other SFR tracers. We provide new single-band and hybrid updated SFRs tracers (with and without type-2 AGN being considered) using our integrated extinction-corrected H$\alpha$ SFR as a reference, thanks to the quality of our attenuation correction via Balmer decrement. We pay special attention on the hybrids ones, providing for the first time, a set of hybrid calibrations for different morphological types and stellar masses. We also analyze the dependence with the color (SDSS $g-r$), axial ratio and ionized-gas attenuation. This analysis is the starting point for a series of papers in which we will study how the SFR in the Local Universe is distributed across galaxy components (bulge, disks, nuclei) and bidimensionally. Ultimately, we are interested in knowing how the local SFR density is spatially distributed over galaxies and how these results would compare to similar future studies at high redshift. 

This paper is organized as follows: in Section~\ref{The Sample} we describe the reference sample used in this article, in Section~\ref{Data and Analysis} we describe the data and the analysis apply to the data, in Section~\ref{Results} we discuss our results, and finally, in Section \ref{Conclusions} we summarize the main conclusions. Throughout this paper we use a cosmology defined by H$_{0}$ $=$ 70 km\,s$^{-1}$\,Mpc$^{-1}$, $\Omega$$_{\Lambda}$ $=$ 0.7 and a flat Universe.

\section{The Sample}
\label{The Sample}

\begin{figure*}
\centering
    \includegraphics[width=60mm]{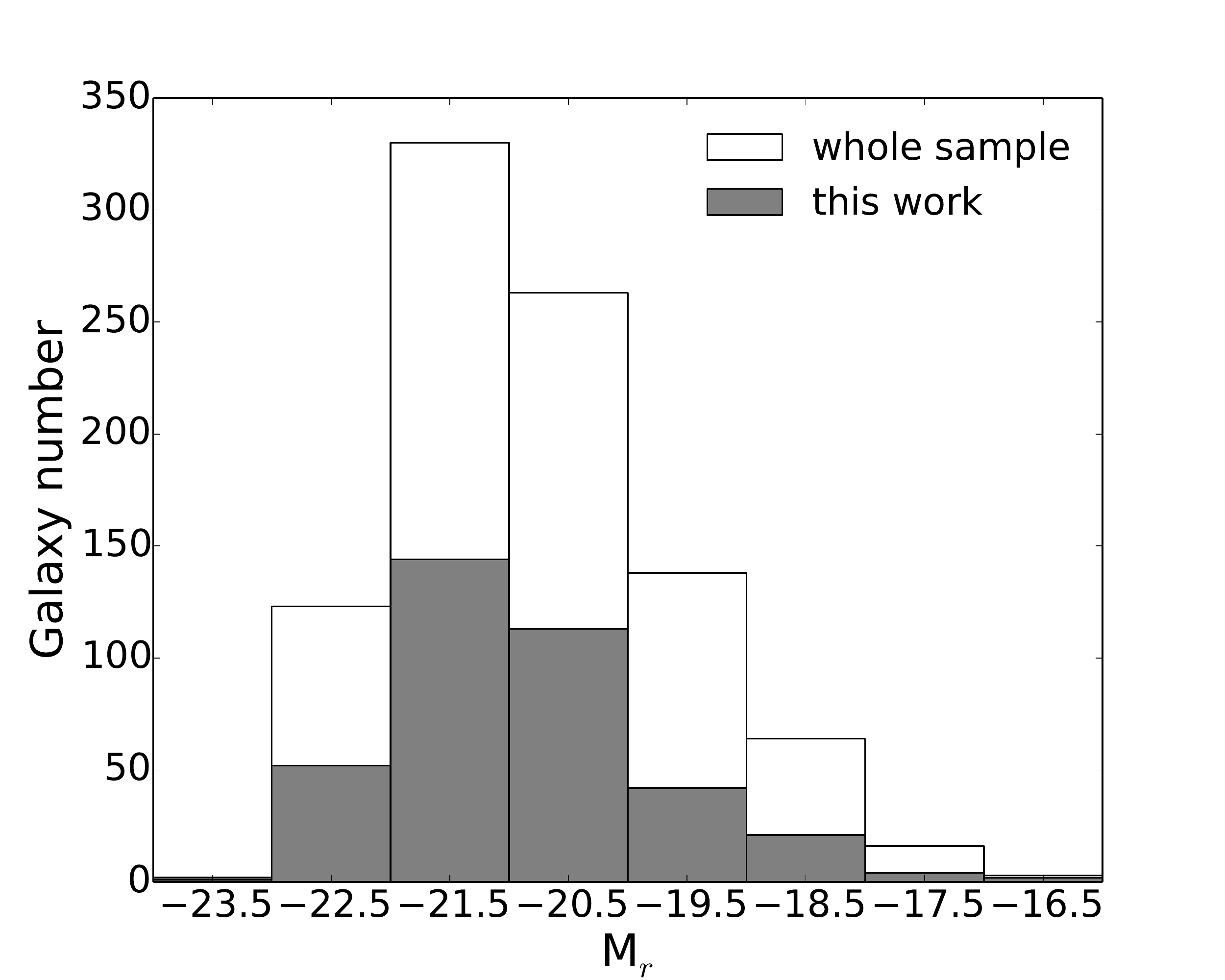} 
    \includegraphics[width=60mm]{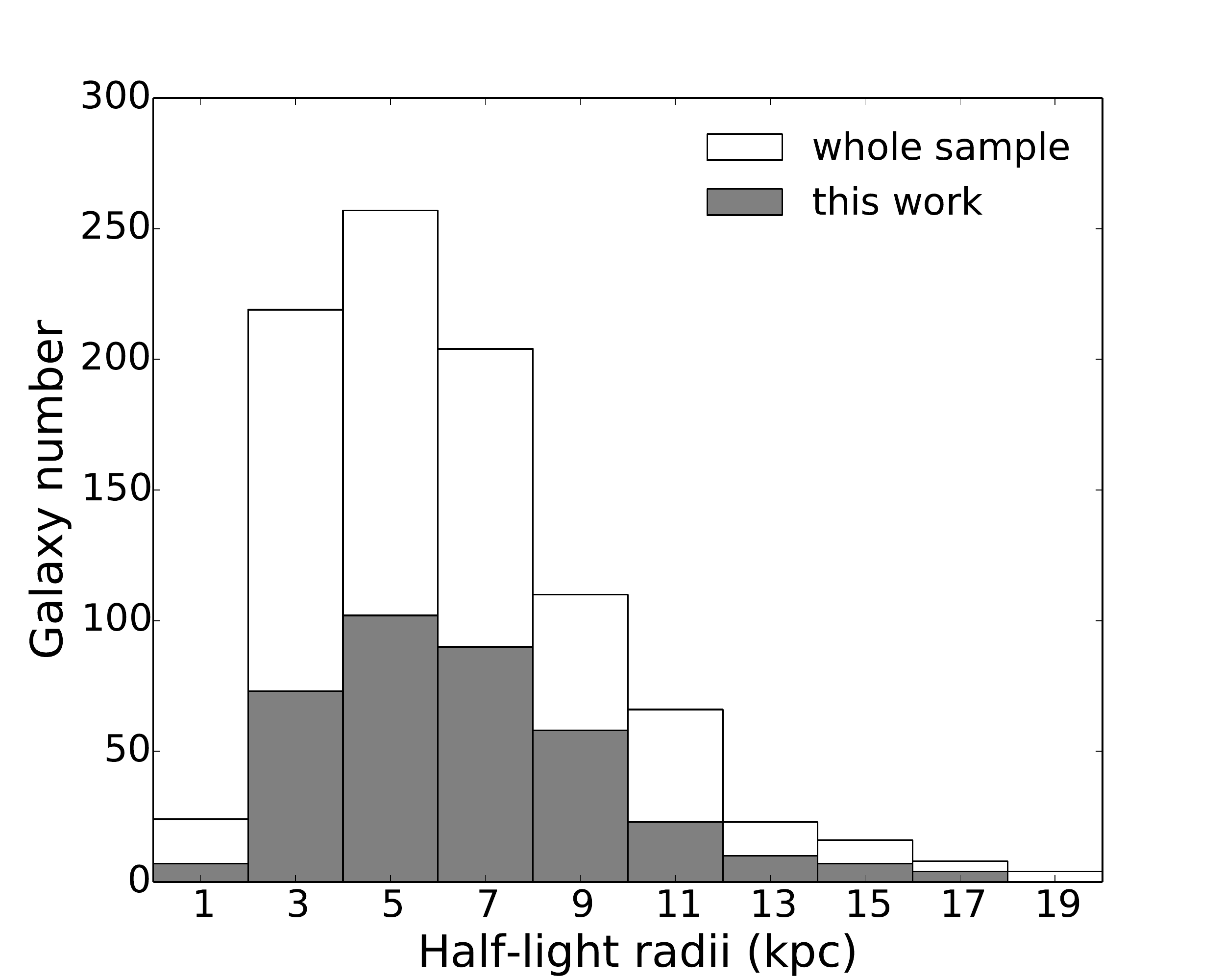} 
    \includegraphics[width=60mm]{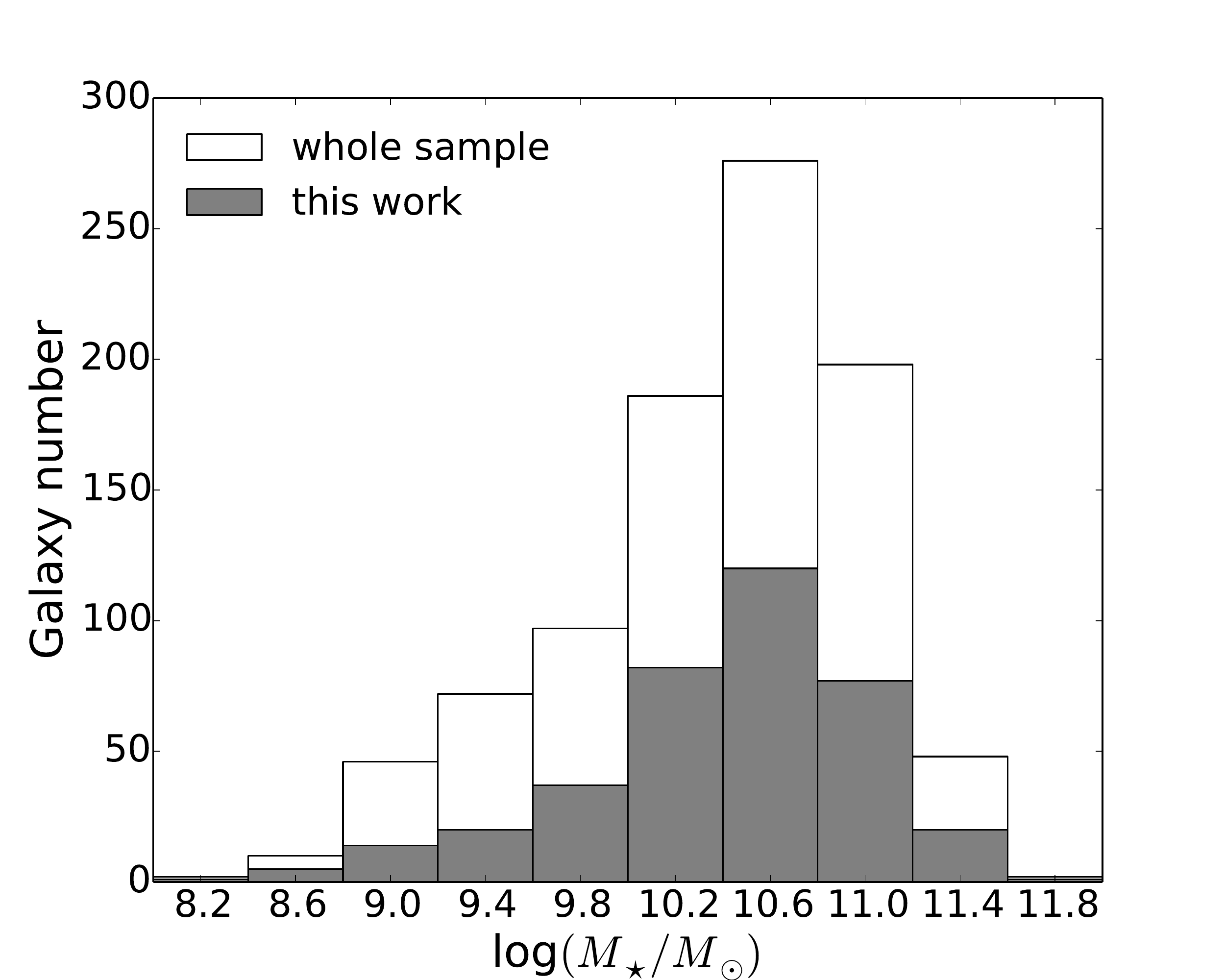} 
  \caption{From left to right. Distributions of the {\it r}-band absolute magnitude, half light radius and log(M$_{\star}$/M$_{\odot}$). The white histograms correspond to the complete CALIFA sample (939 galaxies), and the grey-filled areas correspond to our sample (380 galaxies). A visual inspection of these histograms together with the perform of the K-S test probability show that our sample is representative in terms of galaxy properties of the entire CALIFA sample.} 
\label{KS_test}
\end{figure*}

The galaxies studied in this work are part of the CALIFA survey \citep{Sanchez_2012}. The CALIFA mother sample includes 939 galaxies of all types. A total of $\sim$600 galaxies will be observed as part of CALIFA using the Potsdam Multi-Aperture Spectrophotometer \citep[PMAS,][]{roth_2005} in the PPak mode \citep{Kelz_2006} mounted at the Calar Alto 3.5 m telescope. The survey is described in detail in the presentation article \citep{Sanchez_2012}. As a summary, the CALIFA mother sample \citep{Walcher_2014} includes all galaxies in the DR7 SDSS photometric catalog \citep{Abazajian_2009} with declinations above 7$^{\circ}$, spectroscopic redshifts (from the SDSS spectroscopic catalog or elsewhere) in the range 0.005$<$$z$$<$0.03 and SDSS $r$-band diameters in the range 45''$<$D25$<$80'', where D25 refers to the isophote major axis at 25 magnitudes per square arcsecond in the SDSS $r$-band. The observations cover the optical wavelength range 3700-7000\,\AA, including the most relevant optical emission lines, such as the [OII]$\lambda\lambda$3726,3729\,\AA\, doublet, H$\alpha$ or the [NII]$\lambda\lambda$6549,6583\,\AA\, and [SII]$\lambda\lambda$6717,6731\,\AA\, doublets. The mother sample is representative of the general galaxy population with the following limits: $-$19.0 and $-$23.1 in {\it r}-band absolute magnitude, 1.7 and 11.5 kpc in half-light radius, 9.7 and 11.4 in log(M$_{\star}$/M$_{\odot}$) \citep{Walcher_2014}. 

This paper makes use of all 380 CALIFA galaxies that have been observed and processed up to Oct 27$^{th}$ 2013, including all those released as part of the Data Release 1 (DR1) \citep[see][]{dr1} and Data Release 2 (DR2) \citep[see][]{dr2}. We refer to this as our reference sample even though some objects do not show detectable line emissions and will not be used to derive the H$\alpha$-based SFR measurements. As this is a random sub-selection of the mother sample based only on visibility along the observing period, this should be representative in terms of galaxy properties of the entire CALIFA mother sample. To prove this statement we compare the whole mother sample (white areas in Figure~\ref{KS_test}) and the galaxies involved in this study (grey-filled areas). We use a K-S test to check whether the two data samples come from the same distribution. The K$-$S test probability is computed using the limits where the mother sample is representative of the general galaxy population as mentioned before. The values of the probabilities found by the K$-$S test are 40.25$\%$ in half light radius, 70.95$\%$ in {\it r}-band absolute magnitude and, finally, 75.55$\%$ in log(M$_{\star}$/M$_{\odot}$). From these values and from the visual inspection in Figure \ref{KS_test}, we conclude that the subsample we are using is representative of the mother sample, except for a marginal deficiency of intermediate luminosity objects in the range M$_r$=($-$20.5,$-$21.2) that might explain the low K$-$S values but that certainly does not bias the results against these systems.

\section{Data and Analysis}
\label{Data and Analysis}

\subsection{CALIFA Integral-Field Spectroscopy}
\label{CALIFA Integral-Field Spectroscopy}

\subsubsection{CALIFA Survey}
\label{CALIFA Survey}

The CALIFA spectra cover the range 3650-7500\,\AA\ in two overlapping setups, one in the red (3745-7500\,\AA) at a spectral resolution of R$\sim$850 (V500 setup) and one in the blue (3650-4840\,\AA) at R$\sim$1650 (V1200 setup), where the resolutions quoted are those at the overlapping wavelength range ($\lambda$$\sim$4500\,\AA). For the purpose of deriving extinction-corrected H$\alpha$ luminosities we make use of the V500 setup as we are interested in having both H$\beta$\,$\lambda$4861\,\AA\ and H$\alpha$\,$\lambda$6563\,\AA\ emission lines in the same observing range. The spectral resolution (FWHM$\sim$6\,\AA) is sufficient to deblend the H$\alpha$ emission line from the nearest [NII]$\lambda\lambda$6548,6584\,\AA\, doublet lines. We are using the v1.3c data products which yield the measured flux densities corrected for Galactic extinction. The data reduction is explained in detail in \cite{Sanchez_2012} and \cite{dr1}. 

\subsubsection{Aperture spectrophotometry}
\label{Aperture spectrophotometry}

For each galaxy for which the CALIFA V500 observations reached the full depth planned (3$\times$900 seconds exposures in a three-point dithered scheme), we generate an integrated spectrum within the largest common aperture possible between the CALIFA and the other complementary data (UV, IR). This aperture has an elliptical shape with a major axis radius of 36\,arcsec and the corresponding ellipticity of the galaxy, as given by the minor-to-major axis ratio listed in NED\footnote{https://ned.ipac.caltech.edu} for each object. The previous values and the position angle (PA) are measured at the 25.0 mag/arcsec$^{2}$ isophote at B-band provided by the RC3 catalogue \citep{rc3_1991}. When this information is not available we use the SDSS {\it g} or {\it r}-band isophotal photometry. As the extracted aperture is significantly larger ($\sim$4000$\times$) than the CALIFA pixel size (1\,arcsec$^2$), effects associated with the treatment of fractional pixels are negligible.

\subsubsection{Continuum subtraction and line-flux measurements}
\label{Continuum subtraction and line-flux measurements}

In order to minimize systematics associated to the stellar continuum subtraction at low-S/N regimes, we decide to first spatially integrate the datacube within these apertures. This is a particularly interesting use of the IFS data that allows both covering the whole galaxy and having a high-S/N in the integrated spectrum. Then, we carry out the necessary corrections to derive total extinction-corrected H$\alpha$ luminosities. The use of the H$\alpha$/H$\beta$ ratio derived from the integrated spectra is justified instead of correcting for extinction spaxel to spaxel and then coadding the flux in order to minimize systematics when adding up signal from very noisy individual spaxels, as is shown below. Thus, while \citet{Sanchez_2011} and \citet{Marino_2012} show that the dust attenuation from individual spaxels is a little larger than the one derived from the integrated spectra (1.24/1.04 and 1.19/1.03, respectively), \citet{Castillo_morales_2011} obtain rather similar values in each case. Nevertheless, the interesting point here is how the luminosity-weighted attenuation compares when using individual spaxels with the one from the integrated spectra. This question is more relevant as we are analyzing attenuation-corrected H$\alpha$ luminosities rather than attenuations themselves. For this matter, we select the galaxy NGC 5668 in \citet{Marino_2012} as it is a nearby spiral galaxy similar to the ones used in this work. We find that the difference between computing the average luminosity-weighted attenuation from individual spaxels and that derived from the integrated spectrum is less than 1$\%$. From this result, we conclude that we can safely use the H$\alpha$/H$\beta$ ratio derived from the integrated spectra to correct the H$\alpha$ flux in each galaxy. Besides, this way of obtaining the extinction-corrected H$\alpha$ luminosity would actually mimic what one could measure in more distant systems for which this work is intended to provide a local benchmark.

The first of those corrections to be applied to our data is to carefully remove the stellar continuum underlying the H$\beta$ and H$\alpha$ lines. This is done by means of adjusting a linear combination of two single stellar population (SSP) evolutionary synthesis models of \citet{Vazdekis_2010} based on the MILES stellar library \citep{Patricia_2006} to the spectrum obtained for each aperture. Two set of models with a Kroupa IMF \citep{Kroupa_2001} are combined. One set contains models (considered as a young stellar population) with ages of 0.10, 0.50 and 0.79 Gyr. A second set (considered as an old stellar population) involves ages of 2.00, 6.31 and 14.13 Gyr. For each age we considered five different metallicities with [M/H] values equal to 0.00, 0.20, $-$0.40, $-$0.71 and $-$1.31 dex offset from the solar value.

Different wavelength ranges corresponding to the emission lines from the ionized gas and sky-lines are masked and not included in the fit. The basic steps applied in this method are the following: (1) shift the SSP templates to match the systemic velocity of the integrated spectrum, (2) convolve each stellar population model with a Gaussian profile so the absorption features could be broadened to match those of the integrated spectrum, (3) redden the spectrum using a k($\lambda$) $=$ R$_{V}$ ($\lambda$/5500\,\AA)$^{-0.7}$ power law, where R$_{V}$ $=$ 5.9, as given by \citet{charlot_fall_2000} (4) Finally, the best linear combination of SSPs is determined by a $\chi$$^{2}$ minimization.

\begin{figure}
\centering
    \includegraphics[trim={0cm 1cm 0cm 0cm},width=\hsize]{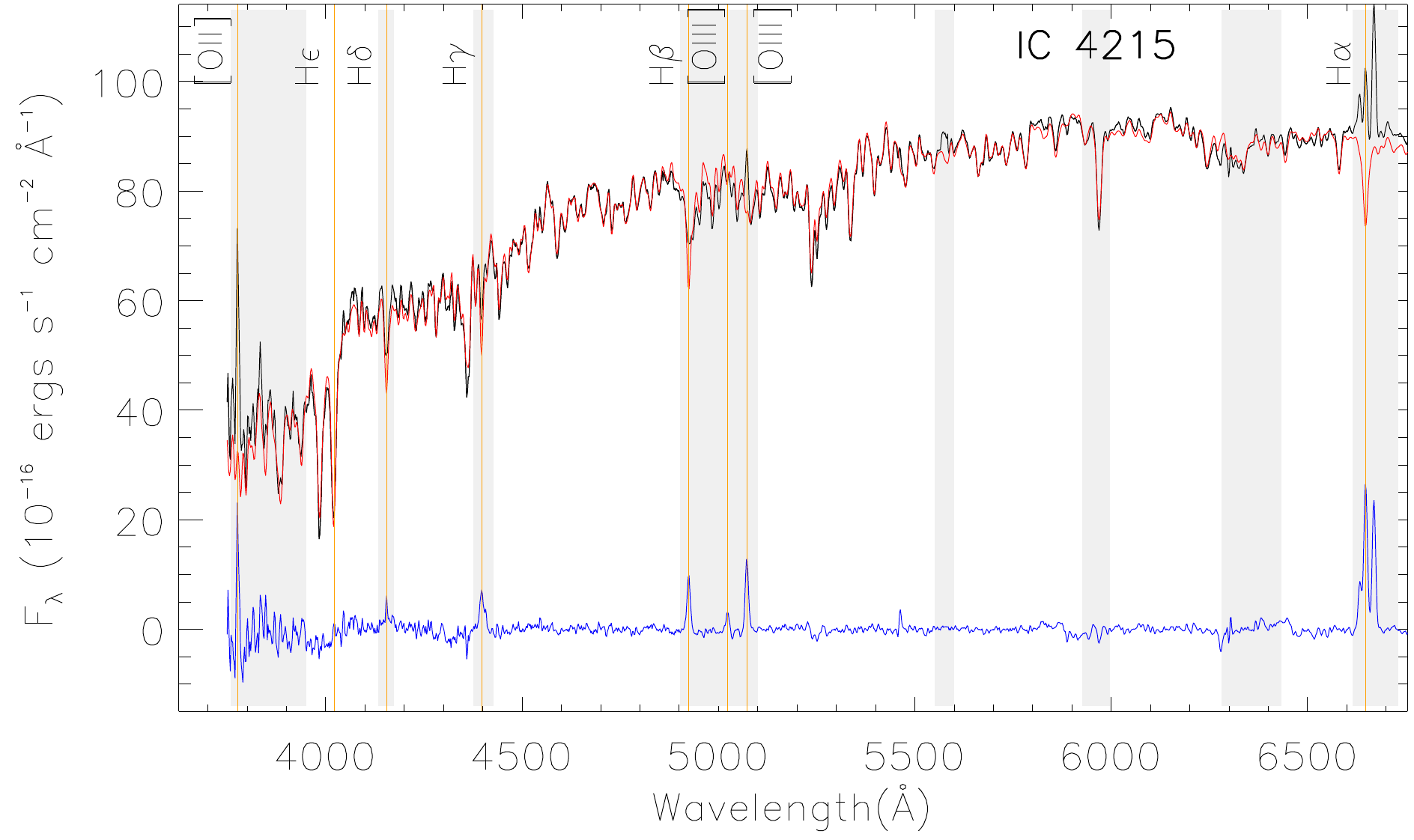} \\
    \vspace{1.0cm}        
    \includegraphics[trim={0cm 1cm 0cm 0cm},width=\hsize]{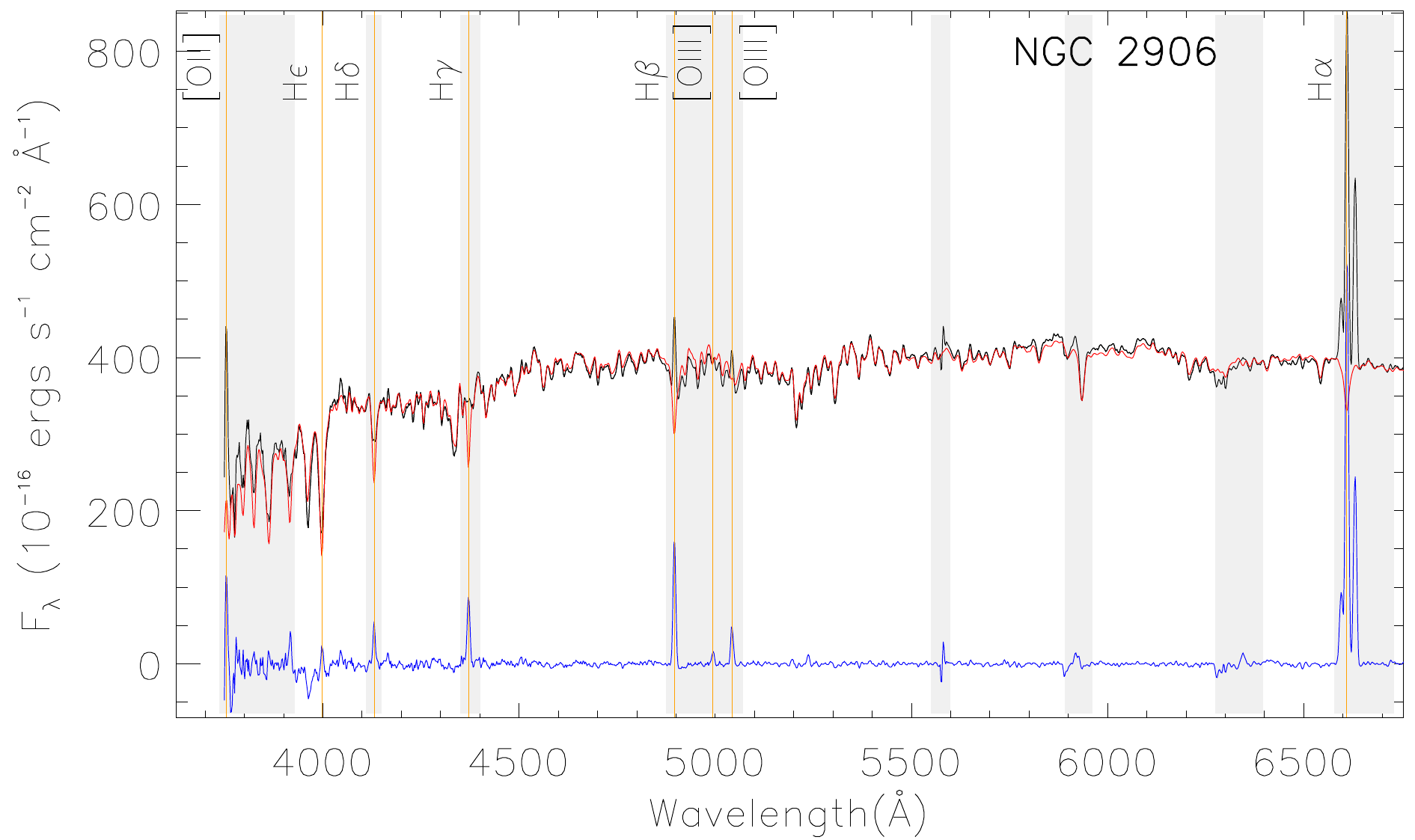} \\
    \vspace{1.0cm}
    \includegraphics[trim={0cm 0cm 0cm 0cm},width=\hsize]{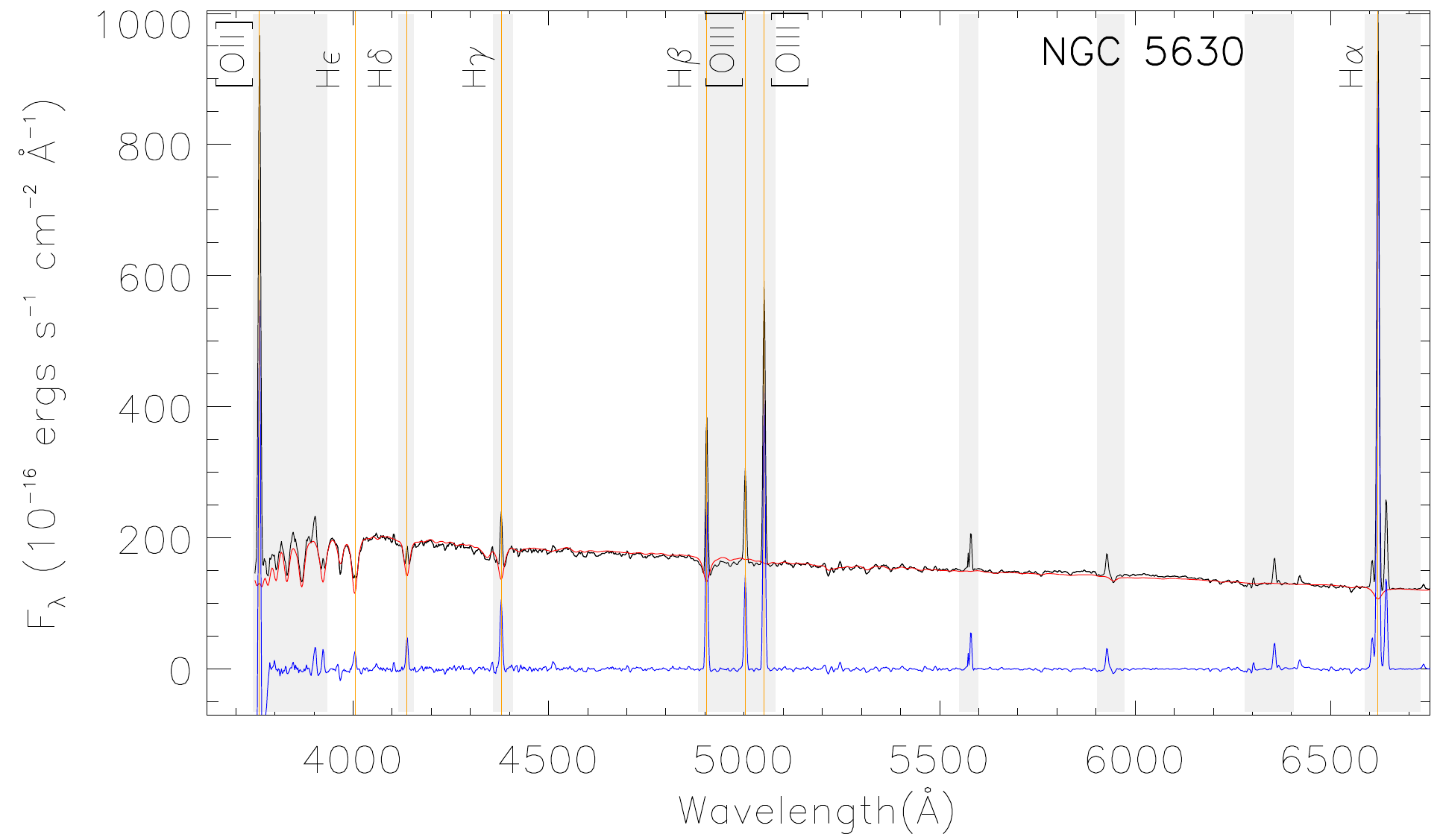} \\ 
\caption{Top panel: The original integrated spectrum for galaxy IC 4215 is shown in black, the best fit to the stellar population appears in red (using the 3745$-$7500\,\AA\, entire spectral range) and the pure emission line spectrum (after the subtraction of a residual continuum shape applying a smooth function) is shown in blue. This galaxy shows a small equivalent width in the H$\beta$ line. Center panel: Same as top panel for the galaxy NGC 2906. Note that in this case the spectrum shows more prominent H$\alpha$ and H$\beta$ emission lines. Bottom panel: In this case NGC 5630 shows a typical emission line dominated spectrum with very high EWs of H$\alpha$ and H$\beta$ emission lines. These spectra show the variety in levels of emission line strength in our galaxies. A proper subtraction of the underlying stellar population is required to obtain the estimation of the emission line fluxes.}
\label{espectros}
\end{figure}

Once we obtain the best underlying continuum of the stellar population, we subtract it from the original integrated spectrum to derive the pure emission line spectrum. The emission line fluxes are computed from this residual spectrum. As some residual continuum could still be present in some cases, we do not simply add all the flux in fixed windows in wavelength. Instead, we compute the H$\beta$ and H$\alpha$ emission line fluxes by fitting Gaussian functions plus a low order polynomial function. Figure~\ref{espectros} shows the original integrated spectrum for three galaxies with different levels of emission-line strength, IC 4215, NGC 2906 and NGC 5630 in black. The best fit to the spectrum of the underlying stellar population is shown in red and the emission-line spectrum produced by the ionized gas is shown in blue. Grey-coloured wavelength ranges correspond to the emission-lines and sky-lines masked out in the fitting procedure.

A proper estimation of the H$\beta$ emission line flux is crucial to obtain a reliable Balmer decrement and, from it, the correction for extinction of the H$\alpha$-based SFR. The method applied here is expected to be a robust procedure as long as a relatively wide wavelength coverage is available \citep[see][]{marmol_queralto_2011} and the models contain an extensive range of ages and metallicities. However, when the whole spectral range (3750$-$7000)\,\AA\ is used for the stellar continuum fitting we still detect systematic residuals around the H$\beta$ absorption line. The treatment of these spectral features is particularly critical. They could be real, due to the limitation of the models in reproducing simultaneously a broad wavelength range and the H$\beta$ region, or introduced during the data reduction. We have also checked that adding an intermediate age population in the linear combination of the SSPs does not change the overall results. For that reason, the stellar continuum fitting around the H$\beta$ line is done for other wavelength ranges using the method explained before. The new spectral ranges used are (3700$-$5500)\,\AA, (4100$-$5500)\,\AA\, and (4800$-$5500)\,\AA. Given that the residual continuum around H$\alpha$ and H$\gamma$ emission lines does not show systematic uncertainties, we determine the H$\beta$ flux by anchoring to H$\alpha$ and H$\gamma$ fluxes based on theoretical line ratios and extinction coefficients. In high S/N spectra, this H$\beta$ emission line flux estimation is compared with the values obtained when different spectral ranges for the stellar continuum fitting are used. Finally, we obtain that the H$\beta$ emission fluxes calculated using the spectral range (4800$-$5500)\,\AA\, are in best agreement with the theoretical ones.

We impose a minimum S/N for both H$\alpha$ and H$\beta$ emission lines fluxes in order to obtain a precise measurement of the extinction using the Balmer decrement. The S/N emission line estimation is done using a formal method calculating the ratio between the Gaussian amplitude at H$\beta$ and the root mean square in the near featureless continuum. A visual inspection of the continuum-subtracted spectra at H$\beta$ is performed for all the analyzed galaxies and a minimum S/N$>$ 5 is considered for H$\beta$ emission line detection. The number of galaxies with detected H$\beta$ emission is 272 over the initial 380 galaxies. This is the sample (listed in Table~\ref{table_fluxes} for reference) that will be used in the rest of the analysis. 

The spectrophotometric accuracy in CALIFA DR1 was checked using SDSS {\it g} and {\it r}-band photometry which are both entirely covered by the V500 setup. \cite{dr1} found a systematic offset of  $\Delta$(g $-$r) $=$ $-$0.06 mag (median) with a scatter of only 0.05 mag. This means that the spectrophotometric accuracy across most of the covered wavelength range is 6$\%$ for the CALIFA data. This value is included in our error estimation for the emission line fluxes. 

\subsubsection{Flux corrections and uncertainties}
\label{Flux corrections}

We also apply some corrections to our data such as aperture corrections and those associated with the spatial masking of field stars and background galaxies applied to the datacubes. We use aperture corrections for the galaxies whose line emission is expected to be more extended that the CALIFA field-of-view (FOV). Our main criterion is to select the band that would first trace the SFR, and second, that it would do it with the best spatial resolution possible. Also, we want to have them available for most galaxies in our sample. Here, we assume that the H$\alpha$ light distribution beyond the FOV is similar to that of the ultraviolet light in each galaxy individually. Besides, the UV band is the one with highest spatial resolution within the bands we are using in this work that is closely related to the SFR. Thus, for that purpose we use the GALEX NUV images (deeper and available for a few more objects than the FUV ones). We compute the difference between the NUV magnitudes obtained for the 36-arcsec-aperture and the asymptotic magnitudes. We fit the aperture correction data for the whole mother sample (those galaxies having NUV data) as a function of the galaxy size, given by the $isoA$ in the {\it r}-band from the SDSS. The correlation between aperture correction and isophotal diameter is the strongest of all those analyzed and it will be used for galaxies without NUV magnitudes. The resulting median correction is around 1.4. The observed H$\alpha$ luminosities already corrected for aperture effects are listed in Table~\ref{table_fluxes}.

The spatial masking is applied over the datacubes before performing the stellar continuum fitting. That means that the light from spaxels contaminated by field stars and background objects is not summed up at this stage. Then, we correct the emission-line fluxes for the flux coming from those missing spaxels. The mean value for the correction factor is 2.2\,$\%$. The corrections are applied only over 44.7\,$\%$ of the galaxies, the ones that have contaminating sources. These correction factors are obtained comparing the aperture fluxes between two sets of synthetic continuum-subtracted narrow-band images. One of them without the flux from the corresponding contaminated pixels and the other where the flux from those pixels is obtained by local interpolation.

Once H$\alpha$ and H$\beta$ emission line fluxes are computed, we correct the H$\alpha$ flux for dust-attenuation assuming that the relation between H$\alpha$ reddening and extinction follows the foreground dust screen approximation. Although this could be a possible source of systematic error in the analysis, some models have shown that when applied to normal star-forming galaxies it does not introduce significant systematic errors \citep{Jonsson_2010}. See a detail discussion about the use of attenuation corrections based on Balmer decrements with a Galactic extinction curve and a foreground screen dust geometry in sections 3.3 and 6.3 in \citet{Kennicutt_2009}. For the attenuation correction we use an intrinsic Balmer ratio of 2.86 for case B recombination \citep{Osterbrock_1989} at electron temperature T$_{e}$ $=$ 10,000 K and density n$_{e}$ $=$ 100 cm$^{-3}$ \citep{Hummer_Storey_1987} using the following expression \ref{attenuation_formula}, where K$_{H{\alpha}}$ $=$ 2.53 and K$_{H{\beta}}$ $=$ 3.61 are the extinction coefficients for the Galactic extinction curve from \cite{Cardelli_1989}.

\begin{equation}
A(H{\alpha}) = \frac{K_{H{\alpha}}}{-0.4 \times (K_{H{\alpha}}-K_{H{\beta}})} \times log_{10} \frac{F_{H{\alpha}}/F_{H{\beta}}}{2.86}
\label{attenuation_formula}
\end{equation}

As an example of how little this attenuation correction would vary among extinction curves and dust-to-stars geometries, we compare the ratio between the A(H$\alpha$) attenuations for the same A(H$\beta$) using the \cite{Cardelli_1989} (R$_{V}$ $=$ 3.1) law above and the \cite{Calzetti_2000} (R$_{V}$ $=$ 4.05) attenuation law. We obtain A(H$\alpha$)$_{Calzetti}$/A(H$\alpha$)$_{Cardelli}$ $=$ 1.03.

Note that the standard H$\alpha$/H$\beta$ ratio used in equation \ref{attenuation_formula} is only valid for the particular ionization conditions indicated above, but values below 2.86 are also physically possible in HII regions, depending on the electron density, effective temperature and therefore on the chemical abundance. This leads to a number of galaxies for which we assumed A(H$\alpha$) $=$ 0. The computed extinction values A(H$\alpha$) are listed in Table~\ref{table_fluxes}. The H$\alpha$ luminosity corrected by attenuation and by the effects mentioned along this section will be referred to hereafter as H$\alpha_{corr}$.

\begin{figure}
\centering
\includegraphics[trim=2cm 1.8cm 0cm 1.2cm, clip, width=100mm]{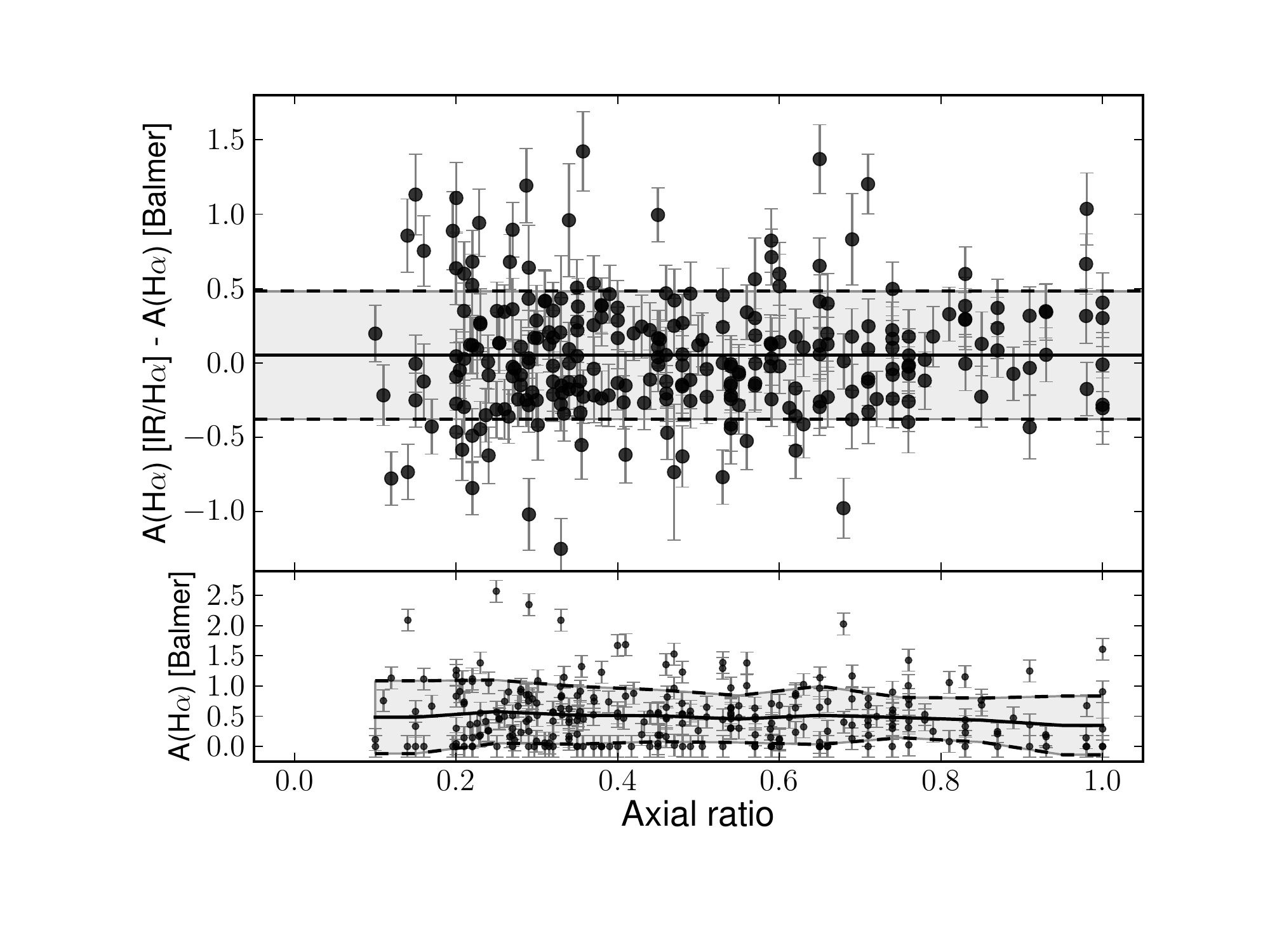}
\caption{Top panel: Difference between H$\alpha$ attenuations derived from the ratio of IR/H$\alpha$ and those obtained using the Balmer decrement as a function of galaxy axial ratio. Solid and dashed black lines correspond to the mean and 1$\sigma$ values, (0.05 $\pm$ 0.43) mag, after applying a rejection of 4$\sigma$. Due to the good agreement between both attenuations computed from different methods, we can safely assume that our Balmer decrement attenuations yield proper results. Besides, we do not find any systematic residuals against the axial ratio parameter associated with highly-inclined galaxies. Bottom panel: A(H$\alpha$) derived from the Balmer decrement as a function of the axial ratio (used as a proxy for inclination). Grey shadow corresponds to the 1$\sigma$ intervals around the mean value showed in black solid line.}
\label{residuals_attenuation_axial_ratio}
\end{figure}

We test if the foreground dust screen approximation has an effect on edge-on galaxies. For that purpose, we plot the difference between A(H$\alpha$) derived from the Balmer decrement and the A(H$\alpha$) values derived from the ratio of IR/H$\alpha$ as a function of galaxy axial ratio (see top panel in Figure~\ref{residuals_attenuation_axial_ratio}). The expression used to derive the attenuation from the ratio of IR/H$\alpha$ is A(H$\alpha$) $=$ 2.5 $\times$ log[1+a$_{IR}$$\times$L(IR)/L(H$\alpha$$_{obs}$)] \citep[see] [equation 2]{Kennicutt_2009}. L(IR), in this case, corresponds to L(22$\mu$m) available for a larger number of galaxies in our sample than L(TIR). The coefficient a$_{IR}$ is equal to 0.015$^{+0.018}_{-0.006}$ (being this coefficient the average value derived from our sample in Section \ref{Hybrid tracers}). The value given by \citet{Kennicutt_2009} is a$_{IR}$ $=$ 0.020 $\pm$ 0.001r $\pm$ 0.005s which is in good agreement with ours even taking into account that they are obtained from different samples. Finally, the difference between both A(H$\alpha$) estimations yields a mean and 1$\sigma$ values of (0.05 $\pm$ 0.43) mag after doing a rejection of 4$\sigma$. This value shows that both methods produce compatible results. As we do not see systematic residuals against the axial ratio parameter, we conclude that we do not find a different behaviour in the case of highly-inclined galaxies.

Finally, the uncertainty in the H$\alpha$ flux is estimated doing a random redistribution of the residuals obtained after fitting a Gaussian function to the pure emission line spectrum in the spectral range around H$\alpha$ emission. The new residual spectrum is added to the pure emission line spectrum and a new Gaussian fit is performed. This procedure is repeated 1000 times and the standard deviation of the computed H$\alpha$ fluxes is considered as the error in the H$\alpha$ flux. On the other hand, the comparison between the measured H$\beta$ line fluxes and those expected from the H$\alpha$/H$\gamma$ Balmer decrements for the same ionized-gas physical conditions gives us an estimation for the H$\beta$ flux uncertainty. A dispersion of $\sigma$ $=$ 7 $\%$ centered around unity is obtained across the whole sample. This method provides much larger uncertainties compared with the one using the redistribution of the residuals around the H$\beta$ emission line. The reported error includes the potential uncertainties in the modeling of the stellar continuum and it is taken as a conservative upper limit for the error in the H$\beta$ flux. This H$\beta$ flux uncertainty propagates to a much larger one in the corrected H$\alpha$ flux. A standard error propagation method is used to compute the uncertainties in other quantities such as extinction or luminosity.

\subsection{GALEX UV imaging}
\label{GALEX UV imaging}

For nearly two thirds of the galaxies in the CALIFA mother sample, we could collect UV observations available from the Galaxy Evolution Explorer (GALEX\footnote{http://galex.stsci.edu/GalexView/}) archive \citep[see][]{Martin_2005}. In most cases (655 out of the 663 objects with GALEX UV data) this includes both far-ultraviolet (FUV, effective wavelength $\lambda_{\rm{eff}}$$\sim$1516\,\AA) and near-ultraviolet (NUV, $\lambda_{eff}$$\sim$2267\,\AA) bands, 200 of them included in the sample of the 272 galaxies analyzed in this work.

The GALEX archive provides simultaneous co-aligned FUV and NUV images with a pixel scale of 1.5\,arcsec per pixel and a spatial resolution (FWHM) of 4 $-$ 5 arcsec. We have selected galaxies located within the central 0.5-degree radius of the 1.2-degrees circular GALEX FOV. We also imposed that the whole galaxy is included in the GALEX FOV. In order to calculate the integrated mean flux of the galaxy, foreground stars and other targets in the field were identified and removed by averaging the interpolation along rows and columns from the GALEX images. 

The typical background in the GALEX UV images is very low so the distribution of count rates in each image typically follows a non-Gaussian distribution. Because of this peculiarity, we estimate the background using the mean instead of the median or the mode used at high background levels, such as ground-based optical or NIR imaging \citep[see][]{Armando_2007}. Surface and aperture photometry was then carried out for each galaxy, using the IRAF task ELLIPSE as described in \cite{Armando_2007}, within elliptical isophotes with fixed ellipticity and position angle (the same ones used for the extraction of the spectra from the CALIFA datacubes). In addition to the 36-arcsec aperture mentioned above we also extracted UV photometry in other concentric elliptical apertures until the error in the surface photometry reached 0.8\,mag (including both background-subtraction and photon noise). From each set of concentric elliptical apertures we finally obtained asymptotic magnitudes for the whole sample \citep{Armando_2007}. These asymptotic magnitudes are the ones applied previously in Section \ref{Flux corrections} to obtain the aperture corrections.

As the UV luminosity suffers from severe attenuation by dust this has to be corrected in order to properly estimate the SFR. The most commonly accepted method to estimate the dust attenuation at UV wavelengths is to use the ratio between the IR (22-25\,$\mu$m MIR, FIR or TIR) and the UV flux (also known as infrared excess or simply IRX). This is equivalent to the use of hybrid SFR estimators that include information from these two wavelengths, which is the approach used later in this work. The IR/UV ratio is almost independent of the dust properties and the relative distribution of dust and stars \citep{Buat_2005}. However, it depends on the age of the dust-heating populations \citep[see][]{Cortese_2008}. 
In the context of this section, we analyzed only the case when no IR data is available. Should that be the case, a relation between the FUV$-$NUV color and the infrared excess could be used instead \citep[see][equation 2]{JuanCarlos_2009a}. For the sake of simplicity, and given the intrinsic large dispersion of the IRX-$\beta$ relationship (see Section \ref{Hybrid tracers} for the IRX-$\beta$ relationship in our sample), we make use of the following linear relation between A(FUV) and FUV$-$NUV and the corresponding $\pm$1$\sigma$ prediction intervals: 

\begin{equation}
A(FUV) = 0.556 + 2.292 \times (FUV - NUV)
\label{attenuation_uv_equation}
\end{equation}

This linear empirical relation is based on the analysis of UV and infrared surface photometry of the SINGS sample \citep{Kennicutt_2003} carried out by \citet{JuanCarlos_2009a}. These authors use FUV, NUV and TIR luminosity profiles with the same spatial resolution to compute both, FUV$-$NUV colors and A(FUV) attenuations via the L(TIR)/L(FUV) ratio using the expression given by \citet{Buat_2005}. This is similar to the IRX-$\beta$ relationship first studied by \citet{Meurer_1995} and calibrated for starburst galaxies. However, \citet{JuanCarlos_2009a} use star forming galaxies that have lower values of the extinction for a given FUV-NUV color. We emphasize that such relations (based on the UV color alone) should be used only as a rough estimate of the UV light attenuation with some (limited) statistical meaning but of very little use in a case-by-case basis. More recently, \citet{Hao_2011} provided a physical motivation for such linear relationship between UV color and attenuation and yielded a y-intercept of $-$0.084\,mag (that corresponds to a FUV-NUV color in the absence of dust of 0.022 $\pm$ 0.024 mag) and a slope of 3.83. Taking into account that the intrinsic FUV$-$NUV color for zero attenuation is different in both cases (due to the noisy relation between A(FUV) and FUV$-$NUV color), we decide to use Equation \ref{attenuation_uv_equation} (J. C. Mu\~noz-Mateos, priv$.$ comm$.$) as in this case we have prediction intervals as a function of the UV color.

We apply Equation \ref{attenuation_uv_equation} to galaxies that have a FUV-NUV color less or equal to 1 mag. Galaxies with colors FUV$-$NUV$>$1 could correspond to either red galaxies with old stellar populations or galaxies with large amounts of dust reddening. In our sample the mean value of the dust attenuation in the FUV is 1.73 magnitudes and vary from 0.81 to 2.80 magnitudes, as it is found by other authors \citep[e.g.,][]{Buat_2005,Burgarella_2005}. The FUV$-$NUV colors, L(FUV) and L(NUV) in ergs s$^{-1}$ for 200 galaxies over the 272 galaxies analyzed in this work are listed in Table~\ref{table_fluxes}.

\subsection{WISE MIR imaging}
\label{WISE mid-infrared imaging}

\begin{figure}
   \centering
   \includegraphics[trim=0.1cm 0.0cm 1.0cm 1.0cm, clip, width=90mm]{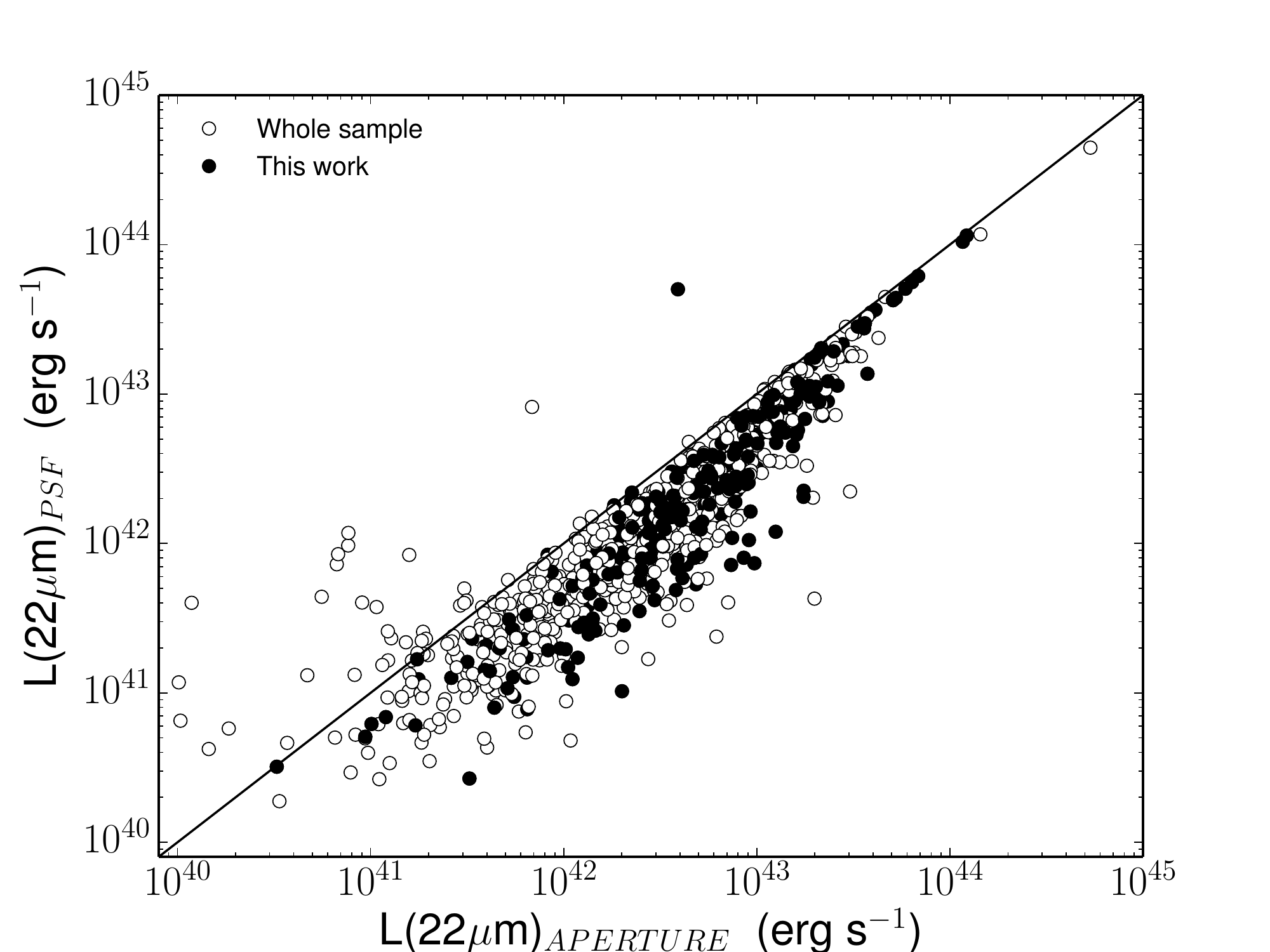}
      \caption{Comparison between the WISE 22\,$\mu$m PSF and aperture photometry for the entire CALIFA mother sample (white points) and for the galaxies used in this work (black points). Solid line corresponds to the 1:1 line and it is given for reference. Although the WISE Source Catalog is ideal for point sources and the resolution of the WISE 22\,$\mu$m band is wide enough (FWHM $\sim$ 11 arcsec) it seems not to be suitable for the CALIFA sample.}
         \label{psf_vs_aperture}
   \end{figure}

The Wide-field Infrared Survey Explorer \citep[WISE,][]{wright_2010} surveyed the entire sky at MIR wavelengths 3.4, 4.6, 12 and 22 $\mu$m (W1 through W4 bands) with 5 $\sigma$ point-source sensitivities of $\sim$ 0.08, 0.11, 0.8, and 4 mJy, respectively. The WISE All-sky Data Release is available through the Infrared Science Archive (IRSA\footnote{http://irsa.ipac.caltech.edu}). It includes imaging (Image Atlas) and PSF-photometry source catalogs (Source Catalog) for all four WISE bands for the entire CALIFA mother sample.
For the purpose of this work we make use of the WISE 22\,$\mu$m data (W4-band) in order to have information on the amount of (mainly UV) photons being processed through dust absorption and re-emitted. The WISE Source Catalog is optimized for point sources and in spite of the resolution of the 22\,$\mu$m data (FWHM $\sim$ 11 arcsec) and the size of the CALIFA galaxies (limited in diameter to $\sim$ 1 arcmin), this photometry catalog might not be appropriate for our sample (see Figure~\ref{psf_vs_aperture}). Therefore, we decide to perform aperture photometry using the Image Atlas in order to calculate the integrated 22\,$\mu$m fluxes and magnitudes. We obtain aperture photometry in circular apertures that enclosed the entire flux from the source. A circular annulus around this aperture is used to compute the sky. We derive L(22\,$\mu$m) in ergs s$^{-1}$ for 265 objects out of the 272 galaxies with detected H$\beta$ emission included in this work (see Table~\ref{table_fluxes}). It has been pointed out by several authors \citep{wright_2010,Jarrett_2013,Brown_2014} that star forming galaxies measured with the WISE 22\,$\mu$m filter are systematically brighter by $\sim$ 10 $\%$ than what is inferred from Spitzer IRS and 24\,$\mu$m data. This factor has been applied in our 22\,$\mu$m luminosities along the article. Values of L(22\,$\mu$m) in Table~\ref{table_fluxes} should be multiplied by this correction factor.

\subsection{TIR fluxes: WISE, IRAS and AKARI}
\label{Total infrared fluxes: WISE, IRAS and AKARI}

Although the longest WISE band already provides relevant information on the attenuation of the UV light associated to regions of star formation, a significant fraction of the energy re-radiated in the IR by dust emerges at longer wavelengths. In order to account for possible differences in the dust temperature or grain-size distribution that could hamper the use of WISE 22\,$\mu$m alone, we have also collected IRAS photometry for the entire CALIFA mother sample. 

The Infrared Astronomical Satellite \citep[IRAS,][]{neugebauer_1984} surveyed 96$\%$ of the sky in four wavelength bands at 12, 25, 60 and 100\,$\mu$m. The main data source for this paper is the IRAS Faint Source Catalog v2.0 (FSC) \citep{Moshir_1990} with a detection limit about one magnitude deeper than the Point Source Catalog (PSC) \citep{Beichman_1988}, reaching a depth of $\sim$0.2 Jy at 12, 25 and 60\,$\mu$m and greater than 1.0 Jy at 100\,$\mu$m. The FSC is at least 98.5\%\ reliable at 12 and 25\,$\mu$m and $\sim$94\%\ at 60\,$\mu$m .

We have performed a cross-match of the CALIFA mother sample with the IRAS Faint Source Catalog (closest IRAS source within 40 arcsec) finding 488 galaxies in common for the four IRAS bands. Within the IRAS FSC catalog a flux density measurement can be either high quality (FQUAL=3), moderate quality (FQUAL=2) or just an upper limit (FQUAL=1) \citep{suplemento_iras}. In this work we make use of only high and moderate quality measurements available for the CALIFA mother sample galaxies, that yielded 12, 25, 60 and 100\,$\mu$m detections for 200, 203, 486 and 443 sources, respectively. Note that poor spatial resolution of IRAS in any of these bands ensures that the flux measurements in the FSC are accurate for the CALIFA objects as long as the object is relatively isolated, but could have an impact on the TIR measurements of galaxies in pairs or close groups.

\begin{figure}
   \centering
   \includegraphics[trim=2cm 1.4cm 0.6cm 1.1cm, clip, width=95mm]{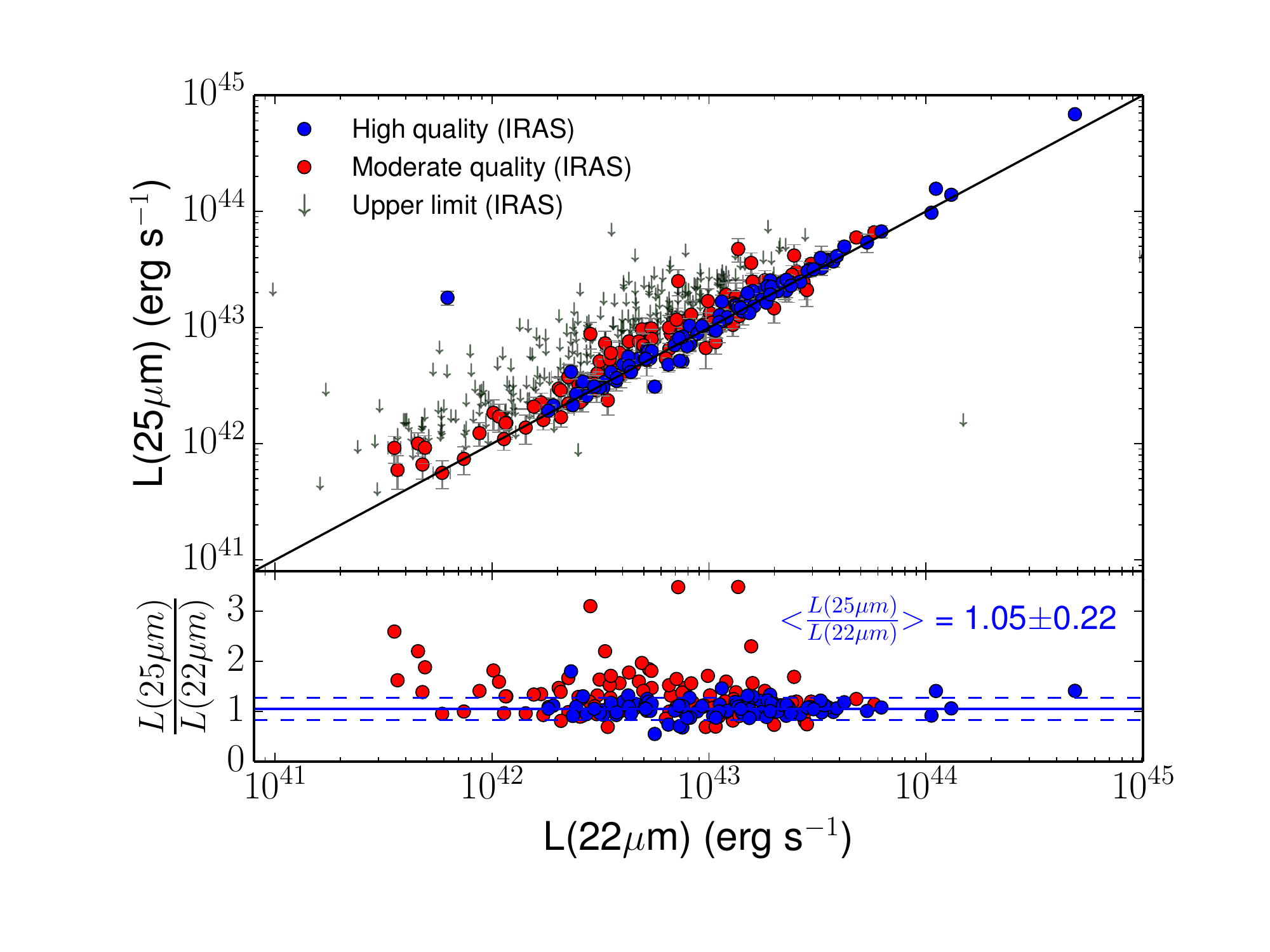}
      \caption{Comparison between 22\,$\mu$m WISE and 25\,$\mu$m IRAS luminosities for the CALIFA mother sample. Blue and red points correspond to high- and moderate-quality IRAS 25\,$\mu$m data, respectively. Arrows represent upper limits for the same IRAS band. The solid line shows a linear 1:1 relation for reference. The average ratio between 25\,$\mu$m and 22 $\mu$m luminosities is 1.05 $\pm$ 0.22 when using high quality 25\,$\mu$m IRAS measurements. This tight relation allows us to use them interchangeably using the previous conversion factor.}
         \label{iras_wise_all}
   \end{figure}

As noted above, the fraction of galaxies with 25\,$\mu$m measurements is significantly lower than that of galaxies with 60\,$\mu$m and 60+100\,$\mu$m measurements. This is due to the comparable detection limit of IRAS at 25 and 60\,$\mu$m but larger flux densities of nearby star-forming galaxies at these latter wavelengths. In order to recover a larger fraction of galaxies with total infrared flux densities (TIR, i.e$.$ 8-1000\,$\mu$m) we decide to combine the WISE 22\,$\mu$m photometry with that from IRAS to determine the galaxies TIR luminosity. The reliability of this procedure is demonstrated by the tight correlation between our WISE 22\,$\mu$m luminosities and those detected at 25\,$\mu$m with high and moderate quality flux by IRAS (blue and red points in Figure~\ref{iras_wise_all}). Therefore, we can confidently use our WISE 22\,$\mu$m photometry to increase the number of CALIFA galaxies with TIR measurements. In addition, we are going to use the WISE 22\,$\mu$m measurements instead of the IRAS 25\,$\mu$m in the corresponding IR SFR tracers as both are found compatible and there are significantly more measurements from WISE 22\,$\mu$m. Previous studies \citep{Kennicutt_2009,Calzetti_2010} show that the average ratio between 24\,$\mu$m and 25 $\mu$m luminosities is 0.98 $\pm$ 0.06. In our case, we find that the average ratio between 25\,$\mu$m and 22 $\mu$m luminosities is 1.05 $\pm$ 0.22 when using high quality 25\,$\mu$m IRAS measurements. Note that the galaxies used in our work are more distant and, therefore, the photometric errors tend to be larger.

As we are interested in estimating the TIR luminosity for our sample of galaxies, we add AKARI photometry measurements at 140 and 160 $\mu$m from the AKARI/FIS All-Sky Survey Bright Source Catalogue \citep{akari}. Using a cone search of 90 arcsec, we find 247 galaxies at 140\,$\mu$m and 70 galaxies at 160\,$\mu$m with high-quality data that implies confirmation of the source detection and a reliable flux estimation. By adding these measurements, we include information at wavelengths at the peak of the spectral energy distribution (SED) and beyond. 

We test the consistency of the IRAS and AKARI measurements when possible (IRAS 60\,$\mu$m vs. AKARI 65\,$\mu$m and IRAS 100\,$\mu$m vs. AKARI 90\,$\mu$m). In general, AKARI gives lower flux values than expected from the IRAS photometry measurements for those wavelengths in common. Besides, we also find that AKARI 140 and 160\,$\mu$m fluxes tend to fall (quite systematically) below the values predicted by our best-fitting infrared SEDs at these wavelengths when data at all IRAS bands is also available. For this reason, we use AKARI 140 and 160\,$\mu$m bands as lower-limits to help discriminate between different dust SEDs (which still provides information for the fits in cases where some IRAS fluxes could be missing).

Finally, by fitting a set of IR templates from \cite{Chary_Elbaz_2001}, \cite{Dale_Helou_2002} and \cite{Rieke_2009} to the WISE 22\,$\mu$m, IRAS and AKARI photometry we derive TIR fluxes for 547 (out of 939) galaxies in the CALIFA mother sample, 221 of them included in the sample of the 272 galaxies analyzed in this work. The fitting procedure used to derive TIR fluxes is described in detail in \cite{Perez_Gonzalez_2008}. As a brief explanation, the code carries out a template-fitting procedure using the rest-frame effective wavelengths (i.e., $\lambda$$_{eff}$/(1+z)). Then, it integrates the best-fit spectra in the wavelength range (8-1000)\,$\mu$m for a total of 100 Monte-Carlo simulated SEDs per galaxy. The average of the TIR individual MC-simulated SEDs luminosities obtained for each galaxy is listed in Table~1. We adopt these values as the best measure of the TIR luminosity.

\section{Results}
\label{Results}

Our aim is to determine the different SFR estimators (single-band and hybrids) for the CALIFA sample and, in particular, ensure that H$\alpha_{corr}$ can be safely used for future statistical spatially-resolved studies, at least in the Local Universe. 

There are three different ways to carry out this analysis: by comparing fluxes, luminosities (or SFRs) and surface brightnesses (or SFR surface densities). Ideally, we would also like to include in this comparison as many SFR tracers as possible. In the latter case, this implies having good spatial resolution to identify the area in the galaxy responsible for the current activity of star formation. Specifically, in the case of the IR measurements this is usually not possible as the WISE and IRAS data do not provide such a high spatial resolution. 
For this reason, the analysis of the SFR surface density can not be carried out in all tracers. Therefore, for most of this section we will rely on the comparison between integrated SFR measurements. The use of fluxes for this comparison is excluded as the correlations in such case would be mainly driven by the wide range in distances spanned by our sample  (0.005$<$$z$$<$0.03).

However, the linear correlation of the integrated SFR between different tracers could be partly due to galaxies with different total SFR but similar SFR surface density (scaled-up versions of low-SFR surface density objects), even more than the SFR itself. Using the tracers with higher spatial resolution, UV and H$\alpha$, (see Section ~\ref{Recipes for determining the SFR in galaxies} for a description of the SFR calibrators found in the literature), we compare in Section~\ref{SFR surface density} the predictions of the SFR surface density.

In the majority of cases, the recipes used to determine the integrated SFR found in the literature are based on samples with ill-defined selection criteria, where the bias towards or against low-metallicity, low-extinction galaxies or highly-extinct systems has not been accounted for. We compare their predictions for the CALIFA sample in Section ~\ref{Comparison of the different SFR tracers}. We describe the possible discrepancies among the different SFR tracers used until now in Section ~\ref{Origin of the discrepancies among SFR tracers}. 

Finally, in Section \ref{Updated SFR tracers for the SDSS-based diameter-limited CALIFA sample} we provide updated calibrations for the CALIFA sample assuming that (as proven across this section) the H$\alpha$ extinction-corrected SFR provides a reliable SFR estimator in the Local Universe. Thus, we anchor both single-band and hybrid tracers to the H$\alpha$ extinction-corrected SFR tracer. We also explore the origin of the difference between the SFR tracers used as a function of galaxy properties such as morphological type, stellar mass, SDSS {\it g}$-${\it r} color, axial ratio or attenuation.

As we are interested in separating star-forming galaxies from the galaxies hosting an AGN, the plots provided in the following sections show SF galaxies in blue and type-2 AGN host galaxies in orange. The same color-coding will be used in the rest of the paper. The information regarding the optical AGN classification can be found in \citet{Walcher_2014}. Briefly, the authors use the emission-line fluxes for all SDSS spectra of DR7. They create a classical [O III]$\lambda$5007/H$\beta$ vs. [N II]$\lambda$6583/H$\alpha$ diagram \citep{Baldwin_1981} to classify the objects and discriminate between different ionization sources at the center of CALIFA galaxies \citep[see][Figure 17]{Walcher_2014}. For the galaxies with no classification we extract the same 3"-diameter circular apertures in the nuclear part. Then, we follow the same criteria as described in \citet{Walcher_2014} to classify them into their corresponding activity type, either SF or type-2 AGN host galaxies.

\subsection{Recipes for determining the SFR in galaxies}
\label{Recipes for determining the SFR in galaxies}

The SFR indicators considered are of two types: single-band and hybrid. In the case of the recipes based on a single photometric band we use the extinction-corrected UV (from the UV slope), extinction-corrected H$\alpha$ (from the Balmer decrement) and the observed MIR or TIR luminosities. The hybrid tracers combine luminosities measured directly (observed UV or H$\alpha$) with that of the light emitted by dust after being heated by young massive stars \cite[see][for more details]{gordon_2000,inoue_2001,hirashita_2003,jorge_2006,Calzetti_2007,Kennicutt_2007,Kennicutt_2009,Hao_2011,Kennicutt_Evans_2012,Calzetti_2012,dominguez_sanchez_2014}.

The most widely-used recipes for SFR tracers are included in \citet{Calzetti_2012} and are listed here for convenience. These expressions are used to compute the SFR from different data, both for single-band and hybrid recipes, scaled to the same IMF \citep{Kroupa_2001}. The mass range varies from 0.1 to 100 M$_{\odot}$. The value of the timescale over which the star formation must remain constant depends on each tracer, being up to 100 Myr for the UV, MIR or TIR and having a lower value for the H$\alpha$ tracer, equal or larger than 6 Myr. The expressions listed below are for global scales as we are using integrated fluxes for the whole galaxy in each case. Also, recipes for determining the SFR at local scales could be found in the review of \citet{Calzetti_2012}. It is worth noting that for the case of Equations \ref{equation_sfr_22}, \ref{equation_fuv_22} and \ref{equation_ha_22}, we have rescaled the coefficients that multiply L(22\,$\mu$m) taking into account the L(25\,$\mu$m)/L(22\,$\mu$m) ratio obtained for our sample and the average ratio between L(24\,$\mu$m) and L(25\,$\mu$m) derived in previous studies (Kennicutt et al. 2009; Calzetti et al. 2010) as explained in Section \ref{Total infrared fluxes: WISE, IRAS and AKARI}. The non-linear behavior for galaxies with L(22\,$\mu$m) $>$ 5$\times$10$^{43}$ erg s$^{-1}$ present in the original recipe \citep[see][]{Calzetti_2012} is not included here as we only find four galaxies in that range. L(TIR) is the total infrared emission in the range 8$-$1000\,$\mu$m. 

First we list those based on single-band, where all the luminosities are in units of ergs s$^{-1}$ :

\begin{equation}
\label{equation_sfr_fuv}
\mathrm{SFR} \thickspace (M_{\odot}  \ yr^{-1}) = 4.6 \times 10^{-44} \times L(FUV_{corr})  
\end{equation}

\begin{equation}
\mathrm{SFR} \thickspace (M_{\odot} \ yr^{-1}) = 5.5 \times 10^{-42} \times L(H\alpha_{corr})  
\label{equation_sfr_ha}
\end{equation}

\begin{equation}
\mathrm{SFR} \thickspace (M_{\odot} \ yr^{-1}) = 2.8 \times 10^{-44} \times L(TIR)
\label{equation_sfr_tir}
\end{equation}

\begin{equation}
\mathrm{SFR} \thickspace (M_{\odot} \ yr^{-1}) = 2.10 \times 10^{-43} \times L(22\mu m)
\label{equation_sfr_22}
\end{equation}

The hybrid tracers are obtained assuming an approximate energy-balance approach. The expressions for the hybrid tracers are shown below where the luminosities are observed and are in units of ergs s$^{-1}$. The global coefficients, 4.6 $\times$ 10$^{-44}$ and 5.5 $\times$ 10$^{-42}$ [M$_{\odot}$yr$^{-1}$/ergs$^{-1}$], correspond to the calibration of the single-band or monochromatic indicators shown before, UV and H$\alpha$ respectively. On the other hand, the coefficients that multiply the IR luminosity, either L(22\,$\mu$m) or L(TIR), are dependent on this tracer and on the one used for the direct stellar light emission. We are going to calibrate empirically these coefficients in Section \ref{Hybrid tracers} to create dust-corrected SFRs:

\begin{equation}
\mathrm{SFR} \thickspace (M_{\odot} \ yr^{-1}) = 4.6 \times 10^{-44}\thickspace [L(FUV_{obs}) + 4.08 \times L(22\mu m)] 
\label{equation_fuv_22}
\end{equation}

\begin{equation}
\mathrm{SFR} \thickspace (M_{\odot} \ yr^{-1}) = 4.6 \times 10^{-44}\thickspace [L(FUV_{obs}) + 0.46 \times L(TIR)]
\label{equation_fuv_tir}
\end{equation}

\begin{equation}
\mathrm{SFR} \thickspace (M_{\odot} \ yr^{-1}) = 5.5 \times 10^{-42}\thickspace [L(H\alpha_{obs}) + 0.021 \times L(22\mu m)] 
\label{equation_ha_22}
\end{equation}

\begin{equation}
\mathrm{SFR} \ (M_{\odot} \ yr^{-1}) = 5.5 \times 10^{-42}\thickspace [L(H\alpha_{obs}) + 0.0024 \times L(TIR)] 
\label{equation_ha_tir}
\end{equation}

The original recipes for the hybrid tracers make use of the 25\,$\mu$m luminosity but we are interested in using our 22\,$\mu$m luminosities instead as we have a large number of these measurements. The reader is referred to Section \ref{Total infrared fluxes: WISE, IRAS and AKARI} where we justify the use of L(22) instead of L(25) after a 1.05 $\pm$ 0.22 conversion factor is applied. This factor is computed as the average ratio between 25\,$\mu$m and 22\,$\mu$m luminosities when using high quality 25\,$\mu$m IRAS measurements.

\subsection{SFR surface density}
\label{SFR surface density}

As mentioned before, the only tracers with enough spatial resolution to compute SFR surface densities across the CALIFA sample are the UV (FWHM $\sim$ 4.5 arcsec) and H$\alpha$ (FWHM $\sim$ 2.5 arcsec) measurements. We calculate the SFR surface density in both as the SFR per unit area measured in the largest elliptical apertures (semi-major axis = 36 arcsec) fitting the PPaK FOV with the ellipticity and PA of the corresponding galaxy. The H$\alpha$ data are corrected for extinction using the Balmer decrement measured within these apertures. In the case of the UV, we use the hybrid tracer (Equation \ref{equation_fuv_22}). This tracer combines UV-observed luminosities with 22\,$\mu$m luminosities (FWHM $\sim$ 11 arcsec). Because the negative dust extinction gradients found in star-forming galaxies virtually all the flux at 22\,$\mu$m  was found to come from inside these elliptical apertures \citep[see][]{JuanCarlos_2009a}. Nevertheless, in order to avoid systematic offsets we de-corrected for aperture the total 22\,$\mu$m fluxes using the same aperture correction as described in Section \ref{Flux corrections}. This means that now all the fluxes, 22\,$\mu$m, H$\alpha$ and FUV, are calculated for the same area.

Thus, the SFR surface density is computed using the following expression:

\begin{equation}
\Sigma_{SFR} = \frac{SFR}{\pi a^{2}  \left( \frac{d}{206265} \right)^{2}} 
\end{equation}

Where the expressions used for estimating the SFR values are Equations \ref{equation_sfr_ha} and \ref{equation_fuv_22} given in Section \ref{Recipes for determining the SFR in galaxies}. The parameter {\it a} corresponds to the semi-major axis set to 36 arcsec in all cases as described in Section \ref{Continuum subtraction and line-flux measurements} and {\it d} is the distance in Mpc to the galaxy calculated from its redshift (listed in Table~\ref{table_fluxes}).

\begin{figure}
\centering
\includegraphics[trim=0.8cm 1.4cm 0.6cm 1.2cm, clip, width=95mm]{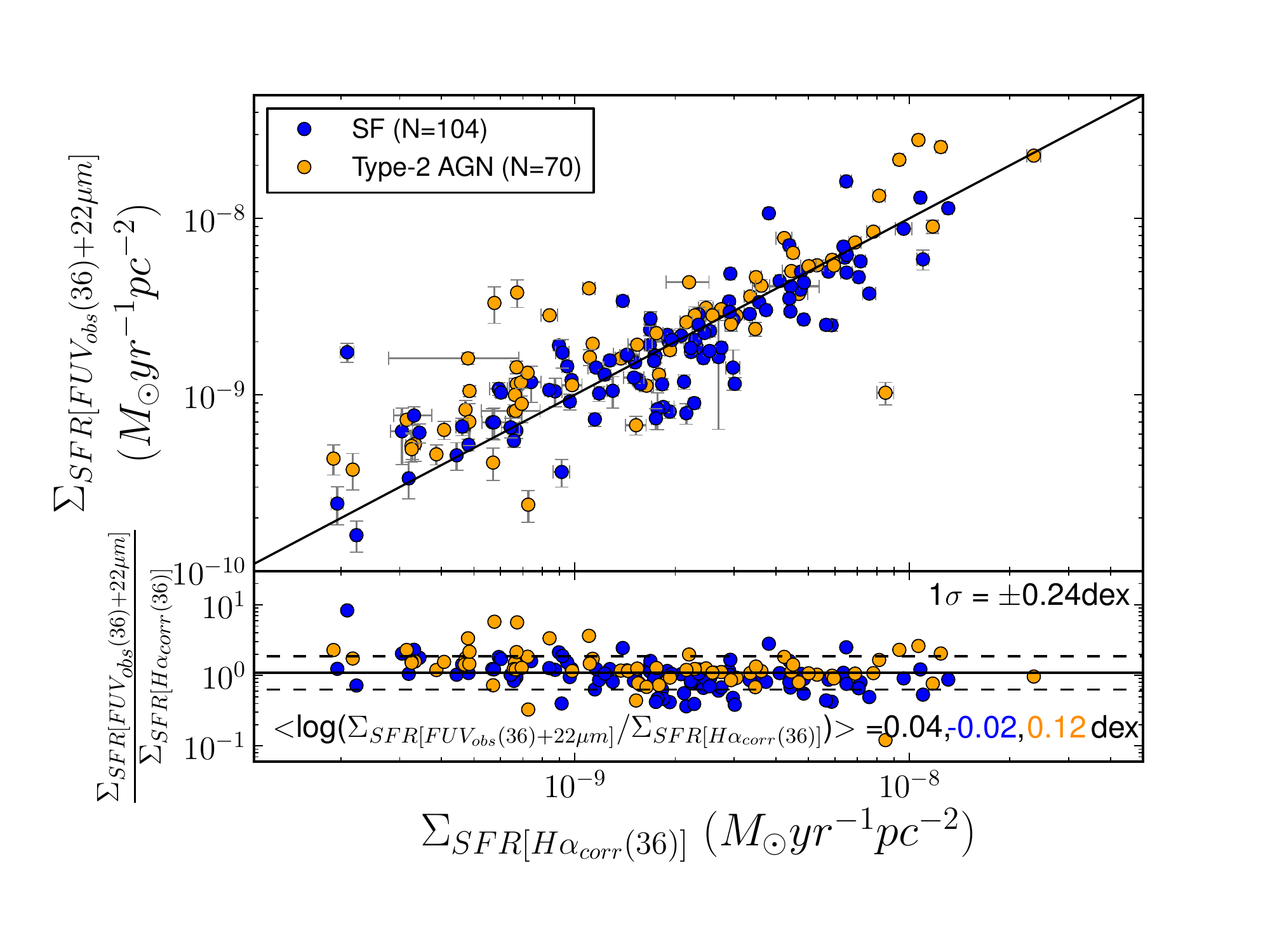}
\caption{Star formation rate surface density derived using a combination of observed-FUV and 22\,$\mu$m luminosities compared to Balmer attenuation-corrected H$\alpha$ star formation rate surface density. The values are obtained within an area of semi-major axis equal to 36 arcsec. The solid line corresponds to equal $\Sigma_{SFR}$ on both axis. Bottom part of this plot represents the residuals as a function of the Balmer-corrected H$\alpha$ star formation rate surface density. The mean value is shown with the solid line while dashed lines correspond to the 1$\sigma$ dispersion in dex around the mean value. Black, blue and orange numbers correspond to the mean values for the whole sample, SF and type-2 AGN host galaxies, respectively. The tight relation found for these two SFR density tracers shows that there are not systematic differences between them. Then, we can safely use the SFR instead of these measurements along this work.}
\label{sfr_density_hb}
\end{figure}

Figure \ref{sfr_density_hb} compares the hybrid star formation surface density using the observed FUV and 22\,$\mu$m fluxes with their corresponding H$\alpha$ attenuation-corrected star formation surface density. We have excluded elliptical and lenticular galaxies in this plot where part of the UV emission could come from HB stars responsible for the UV-upturn \citep{Brown_1997,Yi_1997}.  

We found a good linear correlation between both measurements in a wide range of values of $\sim$2 dex, specially, at $\Sigma$$_{SFR[H\alpha_{corr}(36)]}$$>$10$^{-9}$ M$_{\odot}$yr$^{-1}$pc$^{-2}$. The mean value of $<$log($\Sigma$$_{SFR[FUV_{obs}(36)+22\mu m]}$$/$$\Sigma$$_{SFR[H\alpha_{corr}(36)]}$)$>$ is 0.04 and the dispersion is $\pm$0.24 dex r.m.s. (see Figure \ref{sfr_density_hb}). There are a number of galaxies at low surface brightnesses which correspond to galaxies with null A(H$\alpha$) values. 

The consistency between the two star formation surface density values and the large range involve shows that there are no systematic differences between the two tracers when SFR surface densities are used or, at least, these are of the order of the object-to-object variation. Thus, we can use the SFR instead of SFR density surface from now on in this work, which will allow us to use all TIR measurements confidently.

\subsection{Comparison of the different SFR tracers}
\label{Comparison of the different SFR tracers}

As CALIFA provides an excellent H$\alpha$-integrated luminosity and a precise Balmer decrement we are going to study the SFR tracers found in the literature and provide updated calibrations (Section \ref{Updated SFR tracers for the SDSS-based diameter-limited CALIFA sample}).     

Once we have verified that the extinction-corrected H$\alpha$ SFR surface density behaves linearly with the hybrid SFR surface density (FUV$_{obs}$ + 22$\mu$m) within the errors (previous section), we can safely assume that any correlation between the integrated SFR is not primarily driven by scaling effects. 

Thus, in the rest of Section \ref{Results} we describe the results from the analysis of the galaxies' total SFR. We first analyze the behavior when using different SFR indicators independently, including the UV and IR-continuum luminosities and, of course, extinction-corrected emission-line H$\alpha$ luminosity. Then, we compare the results of the different tracers among themselves assuming that the ones combining directly observable luminosities (either UV or H$\alpha$) and those associated to dust re-emission (monochromatic or TIR) should be able to recover the entire energy budget from recently-formed massive stars.

\subsubsection{Single-band SFR tracers}
\label{Single-band SFR tracers}

\paragraph{Comparison between Mid-IR and extinction-corrected H$\alpha$} \hspace{0pt} \\

\begin{figure}
\centering
\includegraphics[trim=1.3cm 1.4cm 0.6cm 1.2cm, clip, width=95mm]{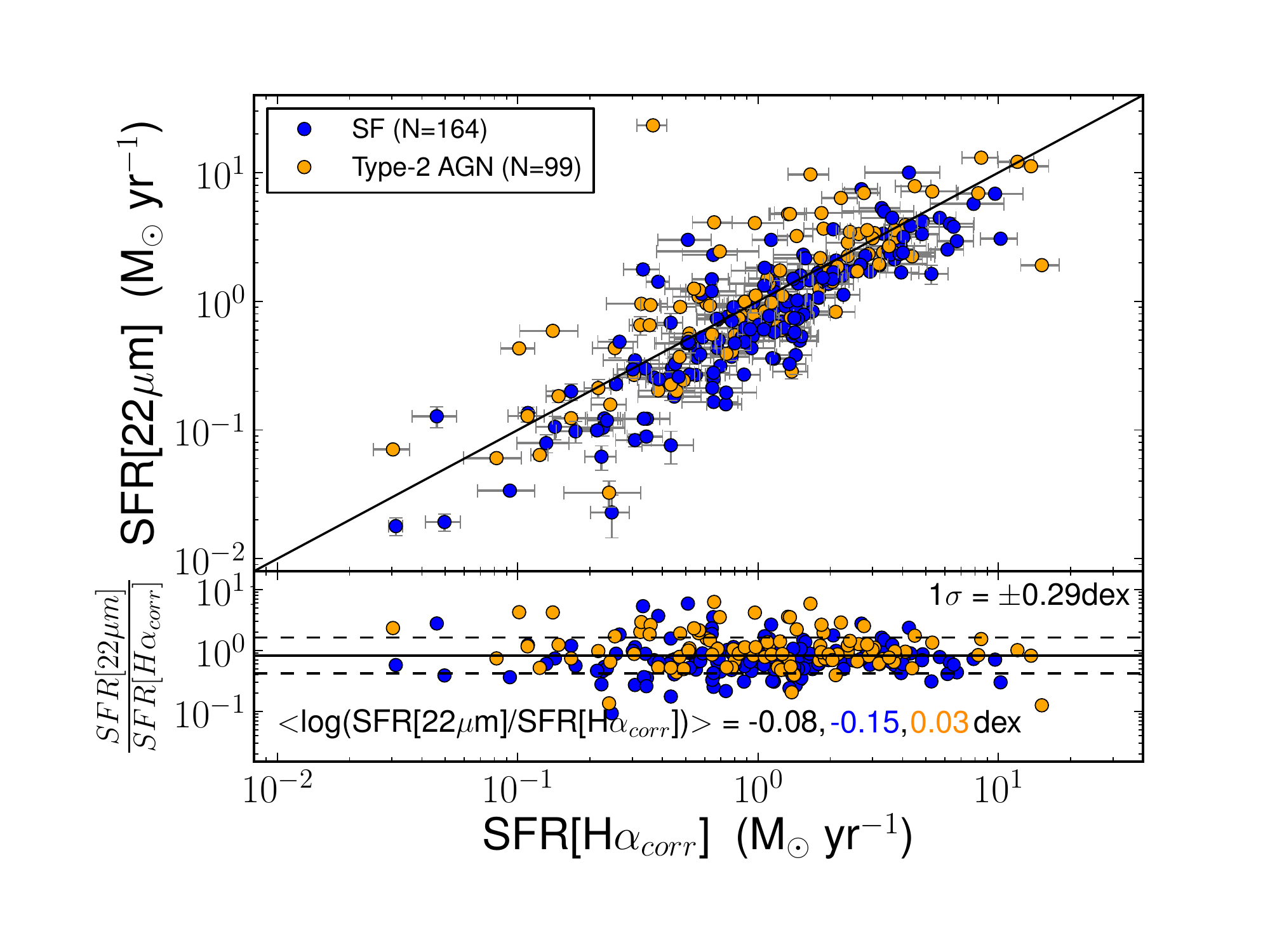}
\caption{Comparison between the MIR (22\,$\mu$m) and the Balmer-corrected H$\alpha$ SFR tracers (recipes from \citet{Calzetti_2012}; see Equations \ref{equation_sfr_22} $\&$ \ref{equation_sfr_ha}). Solid line corresponds to the 1:1 line. Orange points show type-2 AGN galaxies and blue points represent star-forming galaxies. The legend gives the number of objects available in both cases. Residuals appear in the bottom part of this figure as a function of the Balmer-corrected H$\alpha$ SFR tracer. Solid line shows the mean value of -0.08 when all the galaxies are included and the dashed lines are referred to the 1$\sigma$ dispersion ($\pm$0.29 dex) around it. For SF galaxies the mean values is -0.15 while for type-2 AGN host galaxies is 0.03.}
\label{sfr_22_plot}
\end{figure}

In this section we compare the SFRs using warm-dust sensitive 22\,$\mu$m WISE luminosities with Balmer attenuation-corrected H$\alpha$ SFRs (Equations \ref{equation_sfr_22} and \ref{equation_sfr_ha}, respectively) in Figure \ref{sfr_22_plot}. We find that at high luminosities 22\,$\mu$m reproduces the SFR measured with extinction-corrected H$\alpha$. Nevertheless, L(22\,$\mu$m) underestimates the SFR expected at low luminosities, where these galaxies are supposed to have very little dust and consequently weak L(22\,$\mu$m) emission. This could be the reason why the mean value of the residuals, expressed as $<$log(SFR[22\,$\mu$m]/SFR[H$\alpha_{corr}$])$>$ in the sub-panel of Figure \ref{sfr_22_plot}, is equal to -0.08. 

We do not have included the non-linear behavior for galaxies with L(22\,$\mu$m) $>$ 5$\times$10$^{43}$ erg s$^{-1}$ present in the original recipe \citep[see][]{Calzetti_2012}, as we only find four galaxies in that range. Three of them have similar values of the SFR(22\,$\mu$m), 11.22, 12.14 and 13.07, making this SFR range rather small to determine whether a non-linear fit would be more appropriate in this case.

\paragraph{Comparison between $\beta$-based extinction-corrected UV and extinction-corrected H$\alpha$} \hspace{0pt} \\

\begin{figure*}
\centering
\includegraphics[trim=1.0cm 1.4cm 0.2cm 0.3cm, clip, width=93mm]{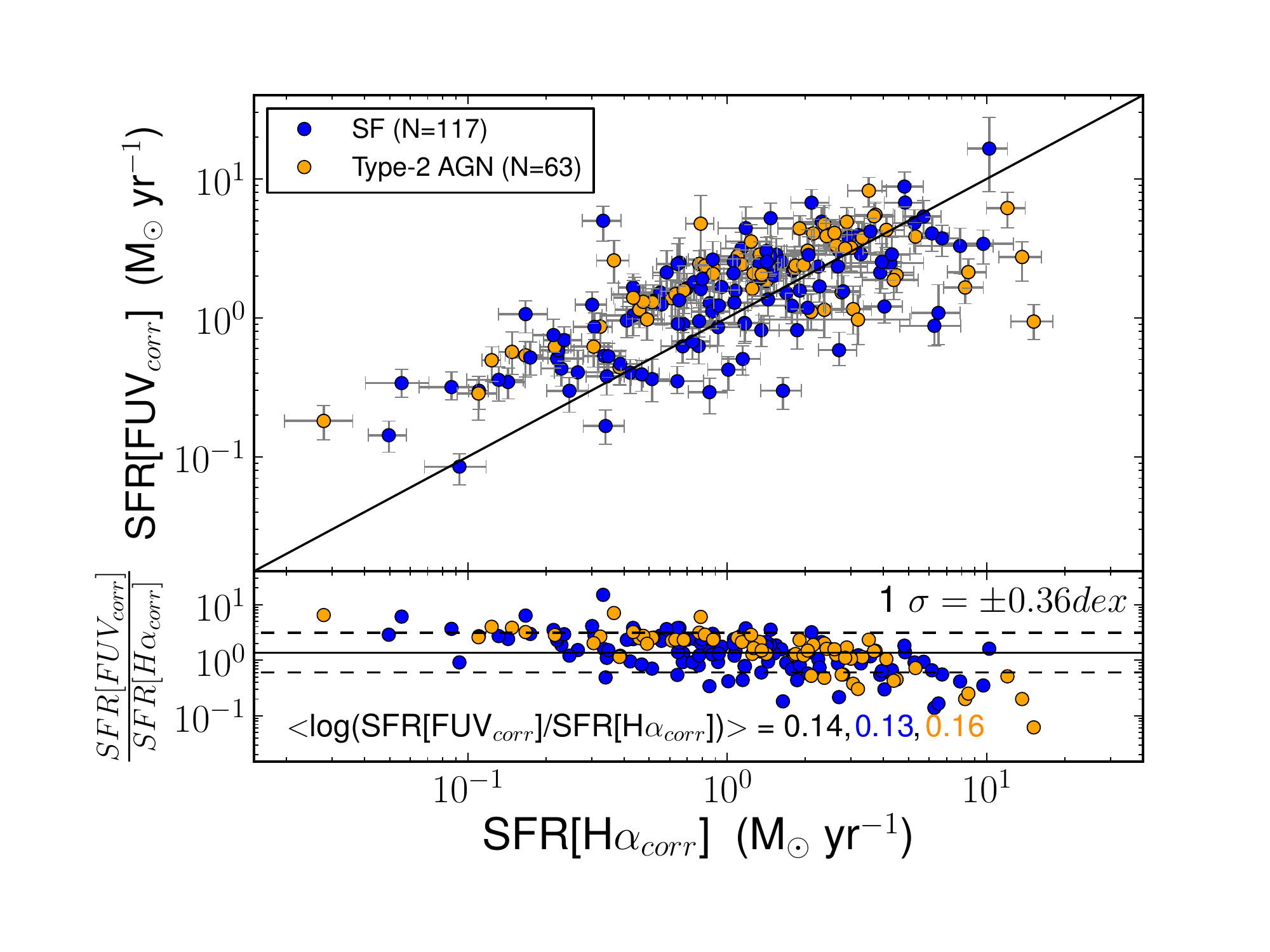} 
\includegraphics[trim=0.0cm 0.0cm 0.0cm 0.0cm, clip, width=90mm]{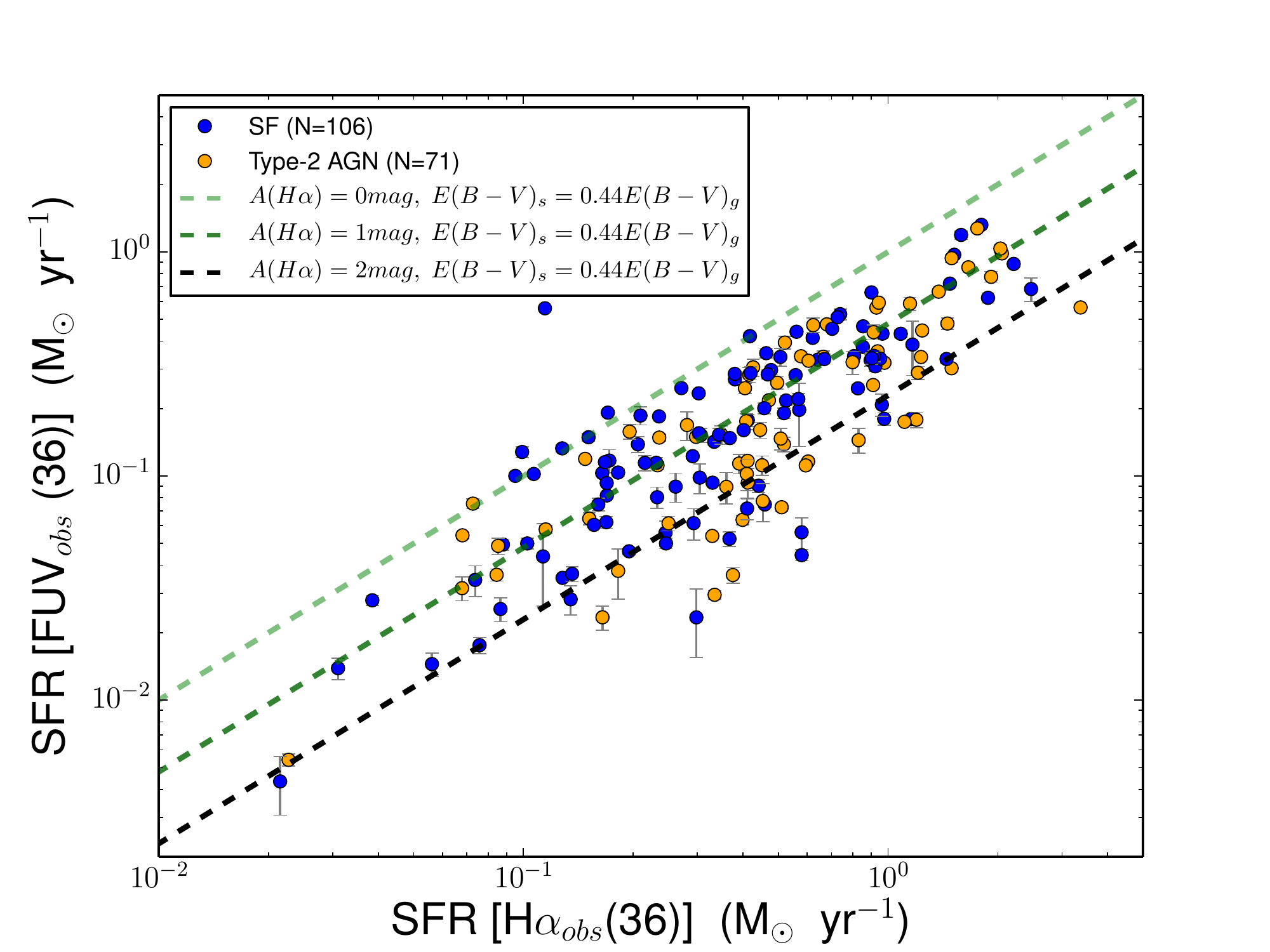} 
\includegraphics[trim=0.0cm 0.0cm 0.0cm 0.0cm, clip, width=90mm]{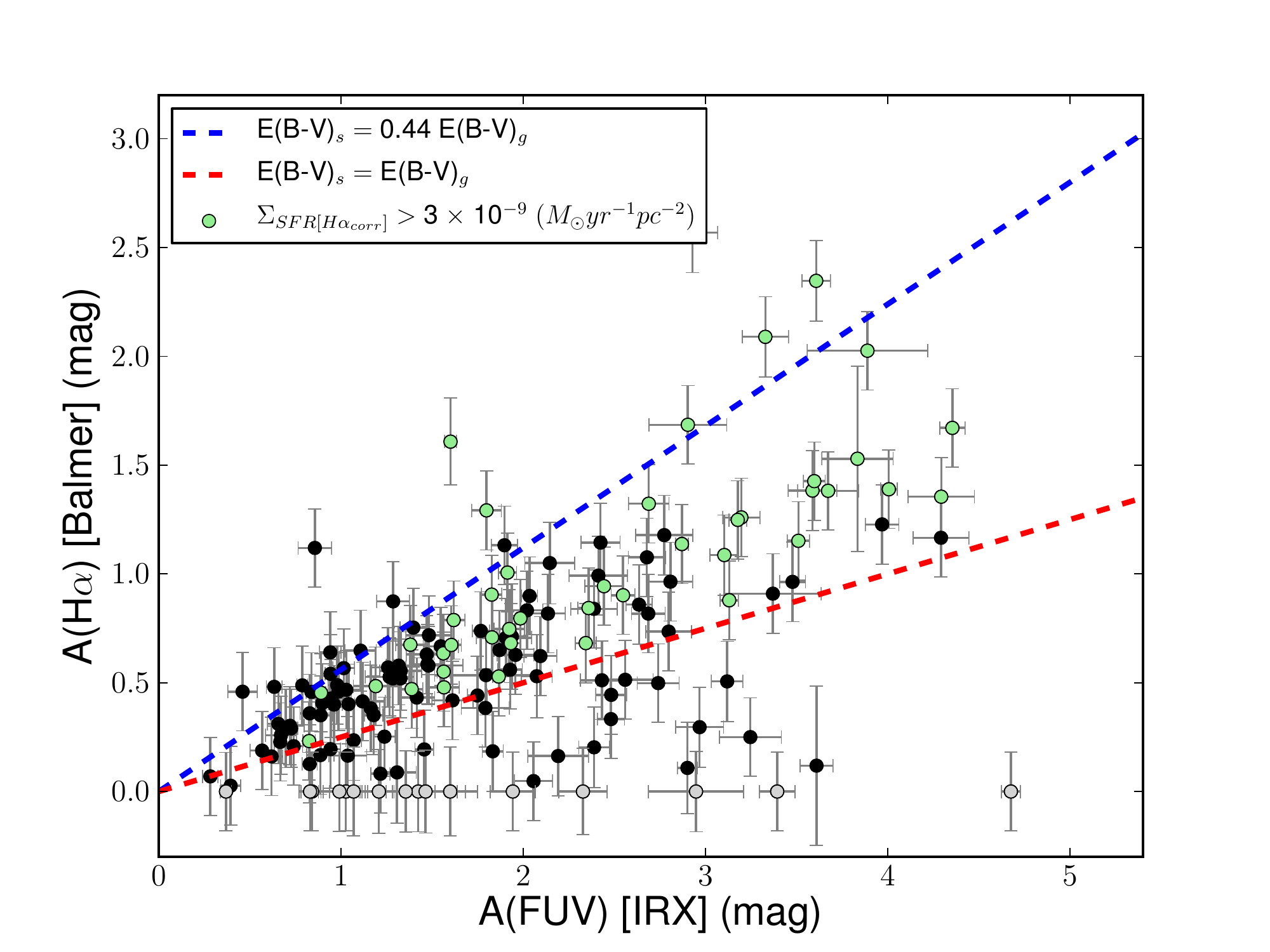} 
\includegraphics[trim=0.0cm 0.0cm 0.0cm 0.0cm, clip, width=90mm]{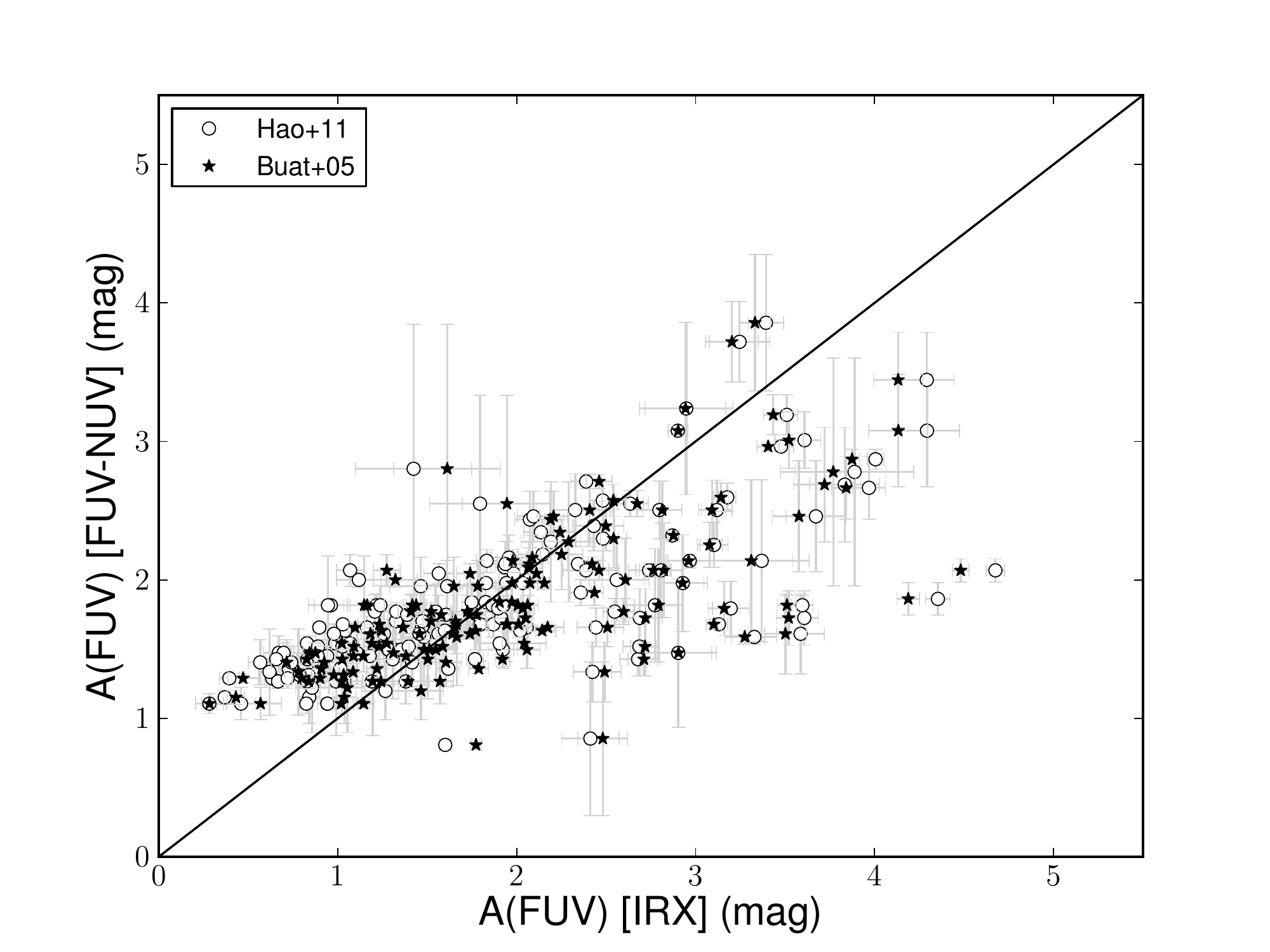} 
\caption{Top left panel: FUV-corrected SFR as a function of Balmer-corrected H$\alpha$ SFR, using Equations \ref{equation_sfr_fuv} $\&$ \ref{equation_sfr_ha}, respectively. Color-coding, solid and dashed lines have the same meaning as in Figure \ref{sfr_22_plot}. At high L(H$\alpha$)$_{corr}$ the SFR derived from the UV alone is underestimated. It might be that the extinction correction using the FUV$-$NUV color traces only the most superficial and less extinct part of the SFR. Consequently, the higher SFRs associated with higher values of the extinction are being underestimated. Top right panel: Relation between observed-FUV SFR and observed-H$\alpha$ SFR. Light-green, dark-green and black dashed lines corresponds to values of A(H$\alpha$) equal to 0, 1 and 2 magnitudes, respectively. All cases are based on the assumption that the relation between the color excess of the stars and the gas is E(B-V)$_{s}$$=$0.44E(B-V)$_{g}$. Bottom left panel: Comparison between A(H$\alpha$) from the Balmer decrement (Equation \ref{attenuation_formula}) and A(FUV) from IR/FUV flux ratio using the expression by \citet{Hao_2011}. Blue dashed line represents a relation between color excess of E(B-V)$_{s}$$=$0.44E(B-V)$_{g}$ while red dashed line assumes that the relation is E(B-V)$_{s}$$=$E(B-V)$_{g}$. Grey points show cases of A(H$\alpha$) equal to zero magnitudes while light-green points show galaxies with the highest values of the SFR surface density. This plot suggest that we might apply a higher value for the relation between the color excess of the gas and the stellar continuum that the one found by \citet{Calzetti_2000} for our sample, although galaxies with higher values of the SFR surface density are more similar to this previous relation. Bottom right panel: Comparison between A(FUV) derived using the FUV$-$NUV color (Equation \ref{attenuation_uv_equation}) and A(FUV) from IR/FUV flux ratio. A discrepancy between these two expressions is found for the lowest and highest values of the attenuation. Star-like symbols show the values when the expression used to compute the A(FUV) is the one from \cite{Buat_2005} while open circles show the values from \citet{Hao_2011}. The solid line shows the 1:1 line for reference.}
\label{sfr_fuv_new}
\end{figure*}

We analyze the FUV continuum and the H$\alpha$ emission-line luminosities as tracers of recent star formation (Equations \ref{equation_sfr_fuv} and \ref{equation_sfr_ha}) since both are linked to the presence and amount of massive (i.e$.$ young) stars (see top left panel in Figure \ref{sfr_fuv_new}). The non-ionizing UV emission is mainly photospheric direct emission from O and B stars formed over the past 10$-$200 Myr and the optical emission lines from ionized gas surrounding massive young stars with lifetimes of $\sim$ 3$-$10 Myr. We apply the attenuation relation given by Equation \ref{attenuation_uv_equation} mentioned in Section \ref{GALEX UV imaging} to correct the FUV luminosity.

We find a rather noisy relation of $\pm$0.36 dex around the mean value, $<$log(SFR[FUV$_{corr}$]$/$SFR[H$\alpha_{corr}]$)$>$ $=$ 0.14. This likely reflects the large uncertainties in the correction for dust attenuation at UV wavelengths using only UV data. They are associated with uncertainties in our knowledge of the slope of the attenuation curve in the UV and with the slope of the underlying stellar continuum. Besides, whether the reddening of the UV continuum can recover all dust-processed SFR is not free for systematics. Figure \ref{sfr_fuv_new} (top left panel) shows that at high L(H$\alpha$)$_{corr}$ (SFR[H$\alpha$$_{corr}$] $>$ 5 M$_{\odot}$ yr$^{-1}$) the SFR derived from the UV alone is underestimated. This fact could be explained by taking into account that higher SFRs are associated with higher values of the attenuation \citep{Kennicutt_1998_otro, Calzetti_2007}. It might be that the extinction correction using the FUV$-$NUV color traces only the most superficial and less extinct part of the SFR. Consequently, the higher SFRs associated with higher values of the extinction are being underestimated.

In order to establish whether other effects could be present, such as an intrinsic discrepancy between the light emitted in the ionizing and non-ionizing UV light from galaxy to galaxy, we also compare the SFR[H$\alpha_{obs}$] and the SFR[FUV$_{obs}$] in the top right panel of Figure \ref{sfr_fuv_new}. Although one might think that dust attenuation should erase any linear correlation between these quantities, the fact that one comes from emission from stars and the other from the ionized-gas should partly compensate for the difference in wavelength. In principle, this makes the two quantities not very different for the whole range of SFRs involved with a slope close to unity. We emphasize that this numerical agreement does not imply, of course, that there is physical reason for them to be equal in any galaxy.

As these luminosities are observed quantities we can estimate the expected extinction for these measurements to match. We assume that the color excess of the stellar continuum is related to the color excess of the gas by E(B-V)$_{s}$ $=$ 0.44 E(B-V)$_{g}$ \citep{Calzetti_1997b,Calzetti_2000}. For the color excess of the ionized gas we use a standard extinction curve such as the Galactic extinction curve proposed by \citet{Cardelli_1989} and R$_{V}$ $=$ 3.1. For the case of the color excess of the stellar continuum we use the attenuation law derived by \citet{Calzetti_2000} and R$_{V}$ $=$ 4.05 $\pm$ 0.80. Finally, we obtain the relation for the stars attenuation in FUV and that of the gas in H$\alpha$ as A(FUV)$_{s}$ $=$ 1.79 A(H$\alpha)_{g}$. This value is similar to the ones obtained by other authors using differente samples, such as the case of \citet{Hao_2011} that found A(FUV)$_{s}$ $=$ 1.82 A(H$\alpha)_{g}$. 

If we suppose that the SFR deduced from the FUV continuum and the SFR from H$\alpha$ emission line (Equations \ref{equation_sfr_fuv} $\&$ \ref{equation_sfr_ha} in Section \ref{Recipes for determining the SFR in galaxies}), both corrected by extinction, are equal, then:

\begin{equation}
\log(\mathrm{SFR}[FUV_{obs}]) = \log(\mathrm{SFR}[H\alpha_{obs}]) + 0.4 [A(H\alpha)_{g} - A(FUV)_{s}]
\end{equation}

Using the previous relation between the corresponding attenuations yields:

\begin{equation}
\log(\mathrm{SFR}[FUV_{obs}]) = \log(\mathrm{SFR}[H\alpha_{obs}]) - 0.32 A(H\alpha)_{g}
\end{equation}

Light-green dashed line in Figure \ref{sfr_fuv_new} (top right panel) corresponds to values of A(H$\alpha$) equal to 0 magnitudes. Nearly every galaxy falls below this line. As it is expected, having no attenuation correction applied to neither Halpha nor to FUV luminosities will imply lower values of the SFR(FUV) as it suffers from higher attenuation. Dark-green and black dashed lines in the same figure correspond to values of H$\alpha$ attenuation of 1 and 2 magnitudes, respectively. It seems like values of A(H$\alpha$) around 1 magnitude are in relatively good agreement with our data. However, the A(H$\alpha$) values in this work vary from 0 to 2.57 magnitudes (Table~\ref{table_fluxes}) with a mean value of 0.49 magnitudes which does not match the expected value. One possibility for this offset could be that the assumption E(B-V)$_{s}$ = 0.44 E(B-V)$_{g}$ is not obeyed for our galaxy sample. Alternatively, the corrected SFR could be different when the UV and H$\alpha$ tracers are used. As we will see in Section \ref{Hybrid SFR tracers}, the latter does not look to be the cause. On the contrary, as pointed out previously, the main aim with this comparison is showing that the real problem when comparing SFR[H$\alpha_{corr}$] and SFR[FUV$_{corr}$] is the difficult estimation of the A(FUV) and also the importance of the attenuation corrections. 

To explore the possibility that our sample might have a different assumption than E(B-V)$_{s}$ = 0.44 E(B-V)$_{g}$, we compare de A(H$\alpha$) from the Balmer decrement and the A(FUV)[IRX] in the bottom left panel in Figure \ref{sfr_fuv_new}. Blue dashed line shows the relation between these two quantities when the expression applied for the stellar continuum and the gas color excess is E(B-V)$_{s}$ $=$ 0.44 E(B-V)$_{g}$. Red dashed line is plotted assuming that the color excess from the stellar continuum and the gas are equal, E(B-V)$_{s}$ $=$ E(B-V)$_{g}$. This comparison suggests that we could apply a higher value than the once found by \citet{Calzetti_2000} for our sample and that the values of A(H$\alpha$) would not be as higher as those expected from the top right panel in Figure \ref{sfr_fuv_new}. Nevertheless, as we find many points below the blue line that could be due to a deviation from the screen foreground model used to compute the ionized gas extinction, we decide to explore this behavior using another parameter such as the SFR surface density. Light-green points show where the galaxies with higher values of the SFR surface density are located in this plot. Clearly, these galaxies are between both lines and they never appear below the red line. This result goes in the line that galaxies with higher values of the SFR surface density (starburst-like) have a relation between the color excess of the stellar continuum and the gas more similar to the one found by \citet{Calzetti_2000} than galaxies with lower values of the SFR surface density.

Finally, we compare A(FUV) derived using the UV-slope (FUV$-$NUV color) with those obtained using the IR/FUV flux ratio (IRX) in Figure \ref{sfr_fuv_new} (bottom right panel). For the IRX case we use the expression by \cite{Buat_2005} (star-like symbols) and the one in \cite{Hao_2011} (open circles). It is clear from this representation that A(FUV)[FUV$-$NUV] gives higher values than A(FUV)[IRX] for the lowest values of attenuations. On the other hand, A(FUV)[FUV$-$NUV] gives lower values than A(FUV)[IRX] when the highest values of attenuations are involved. Both expressions, \cite{Buat_2005} and \cite{Hao_2011}, yield similar results. As explained before, this plot suggests that using the FUV$-$NUV color to recover the dust-processed SFR is not the best option.

\paragraph{Comparison between TIR and extinction-corrected H$\alpha$} \hspace{0pt} \\

The main problem in using the SFR based only on TIR luminosity is that we are assuming that there is a negligible fraction of the light coming directly from the stars without being reprocessed by dust. Besides, even if there are no UV photons escaping directly (without being processed by dust) to the observer, the calibration of the SFR[TIR] assumes that the light reprocessed by dust comes from young stars, i.e. the ones linked to the current SF we want to trace. Nevertheless, optical photons from old stars contribute to the heating of the dust \cite[see][]{Johnson_2007} and thus, to the TIR luminosity. Indeed, based on constant star formation (CSF) models, \citet{Calzetti_2012} found a reduction in this constant of almost a factor of 2 from models with a CSF lasting for 100 Myr compared to those CSF models lasting for over 10 Gyr. Besides, according to \citet{Cortese_2008}, for star formation timescales (equivalently the age of the Universe at which the SFR peaks in their 'a la Sandage' SFH) larger than $\sim$6-7\,Gyr the UV radiation dominates the dust heating with a contribution of $>$75$\%$ to the total energy absorbed and then re-emitted in the infrared. On the other hand, the same authors derive that if $\tau$$<$\,5\,Gyr the UV light contributes less than 50$\%$ to the TIR emission.

\begin{figure}
\centering
\includegraphics[trim=1.3cm 1.4cm 0.6cm 1.2cm, clip, width=95mm]{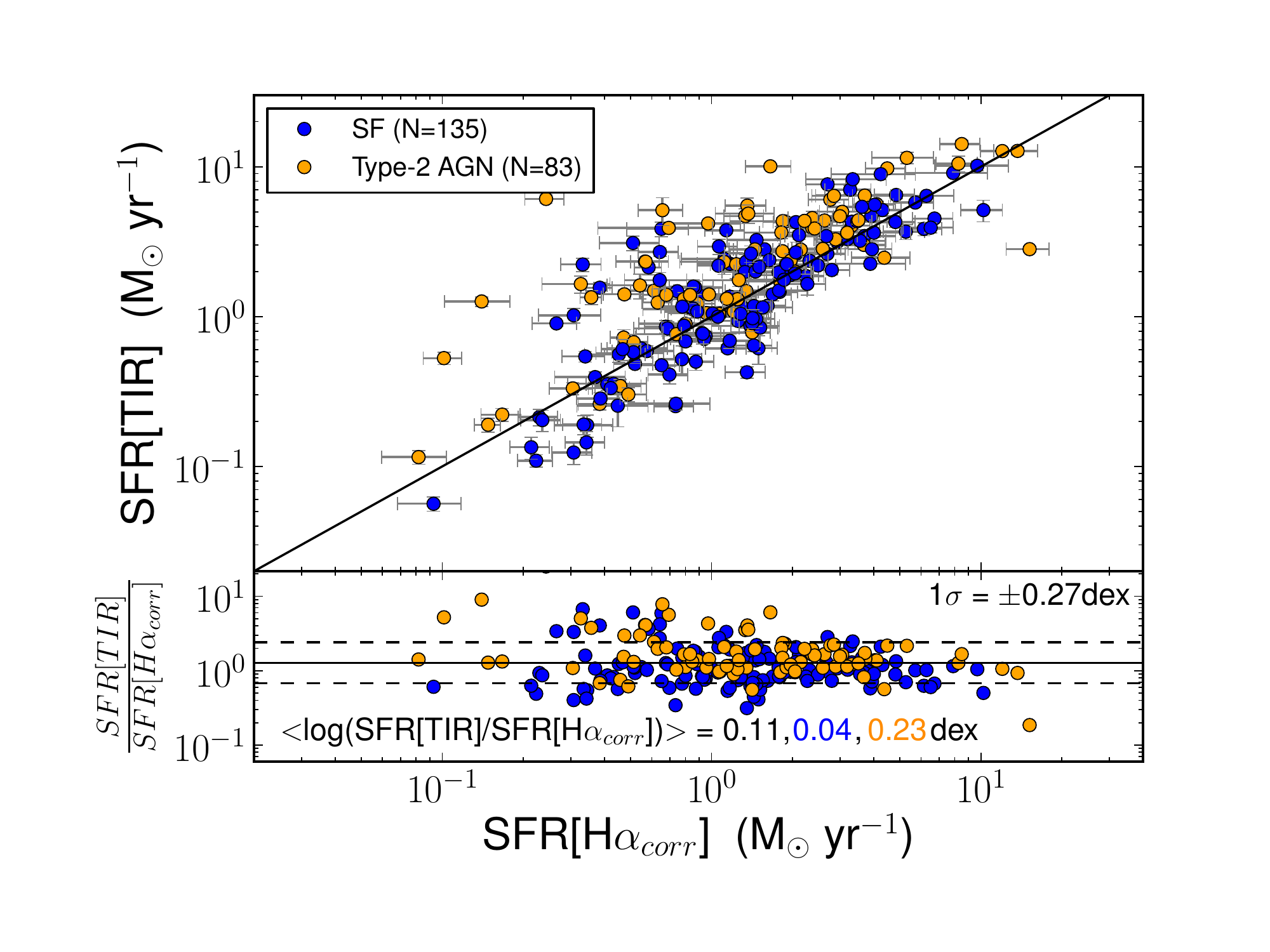}
\caption{Comparison between the SFR tracer using the TIR luminosity and the extinction-corrected H$\alpha$ SFR tracer, after applying Equations \ref{equation_sfr_tir} and \ref{equation_sfr_ha}, respectively. Color-coding, solid and dashed lines have the same meaning as in Figure \ref{sfr_22_plot}.}
\label{sfr_tir}
\end{figure}

The comparison between the SFR[TIR] and SFR[H$\alpha$$_{corr}$] (Figure \ref{sfr_tir}) shows that at low TIR luminosities the SFR[TIR] is being underestimated \citep[in the line of the results of][]{Rieke_2009}. We find that for values of the SFR[TIR] below 0.3 M$_{\odot}$ yr$^{-1}$ the average value of A$_{H\alpha}$ is 0.28 $\pm$ 0.04 mag. On the other hand, at high luminosities (SFR[TIR] $>$ 1 M$_{\odot}$ yr$^{-1}$) the TIR seems to provide SFRs somewhat higher than those obtained from H$\alpha$. In fact, a large number of galaxies appear in this regime making the mean value of the ratio between these tracers larger than zero in the residuals, $<$log(SFR[TIR]/SFR[H$\alpha$$_{corr}$])$>$ $=$ 0.11. This is either because the contribution of heating due to optical photons or nuclear activity becomes relevant at those luminosities and/or because a fraction of the H$\alpha$ recombination line luminosities are not recovered when correcting for dust attenuation using the Balmer decrement. The analysis of the hybrid calibrations (see section below) favors the former scenario.

\subsubsection{Hybrid SFR tracers}
\label{Hybrid SFR tracers}

A powerful way of determining the SFR is to combine a photometric band that is sensitive to the light directly emitted by young stars (i.e. observed UV or H$\alpha$ fluxes) with that reprocessed by dust, usually in the mid- or far-infrared (or, alternatively, the TIR emission). This is an alternative approach to correct the UV or H$\alpha$ fluxes for attenuation.

In both cases, the validity of these hybrid tracers is that the observed IR emission comes from light whose optical depth (or attenuation) is of the order of that in the UV or H$\alpha$, otherwise it would not be possible to write the total SFR as a sum of the two luminosities, observed and dust-processed \citep[see][]{Kennicutt_2009}. This assumption could not be valid if (1) the heating of the dust is dominated by optical photons particularly important at long IR wavelengths where the contribution of low-temperature dust emission is most relevant or (2) by UV photons more energetic than those observed directly (e.g. if the bluest observed band is in the NUV) or (3) in the case of a significant AGN contamination, where any of these bands could be actually tracing a UV radiation field that is not merely due to recently-formed massive stars. If there is a significant difference between the $\tau_{FUV}$ (or $\tau_{H\alpha}$) and the opacity of the photons that lead to the IR emission used in the corresponding tracer, a linear relation between the SFR and the two (emitted and dust-absorbed) luminosities should not be present. In the particular case of UV and H$\alpha$ the $\tau$ are similar to the one that comes from the dust component so the approach of using a linear relation between L(FUV$_{obs}$) or L(H$\alpha_{obs}$) combined with the L(IR) luminosity can be safely done. That implies that for the IR tracer both bands should suffer the same attenuation \citep[see a detail analysis in][]{Kennicutt_2009,Hao_2011}

\begin{figure}
\centering
\includegraphics[trim=0.5cm 1.4cm 0.6cm 1.2cm, clip, width=95mm]{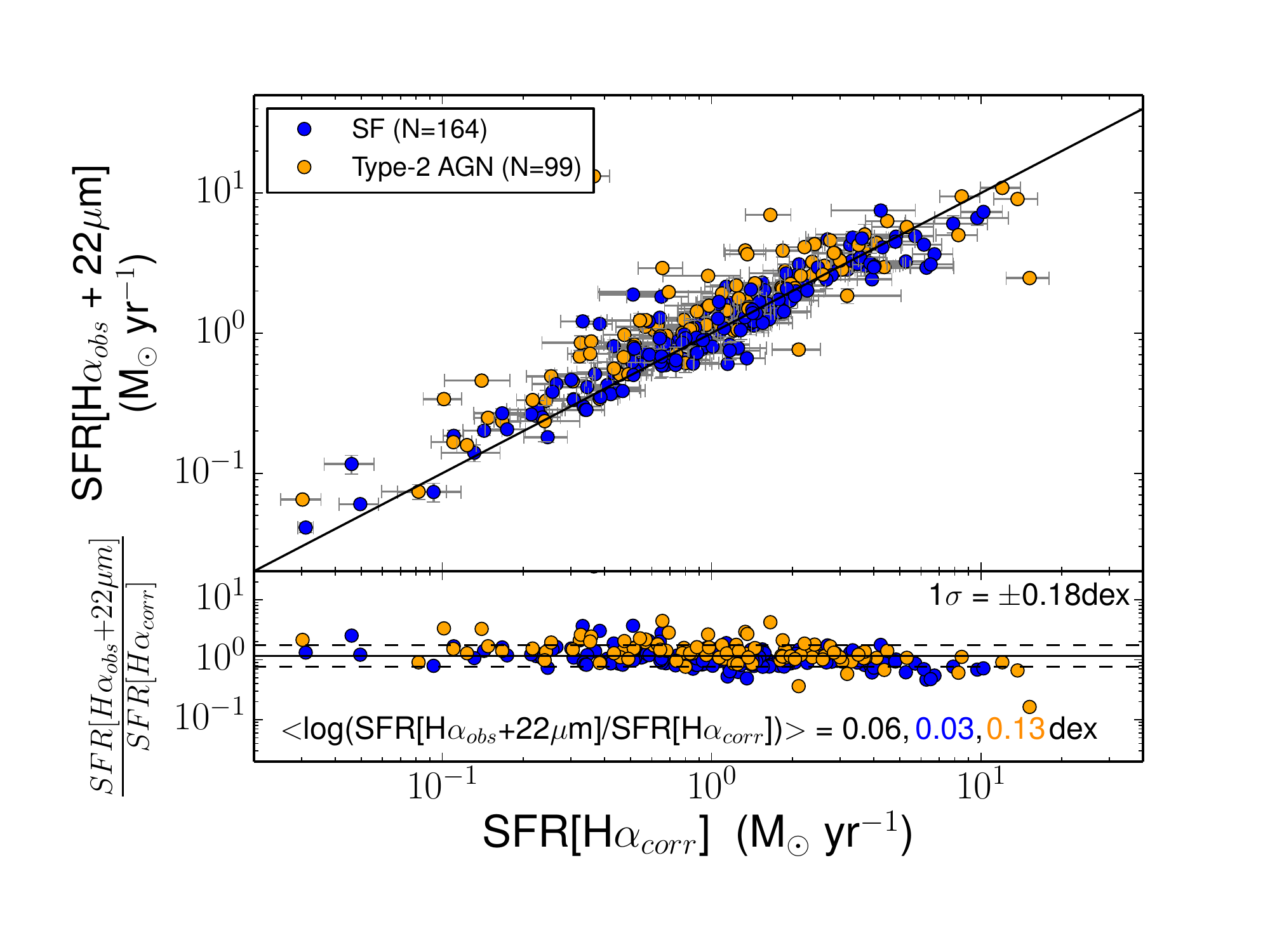} \\
\includegraphics[trim=0.5cm 1.4cm 0.6cm 0cm, clip, width=95mm]{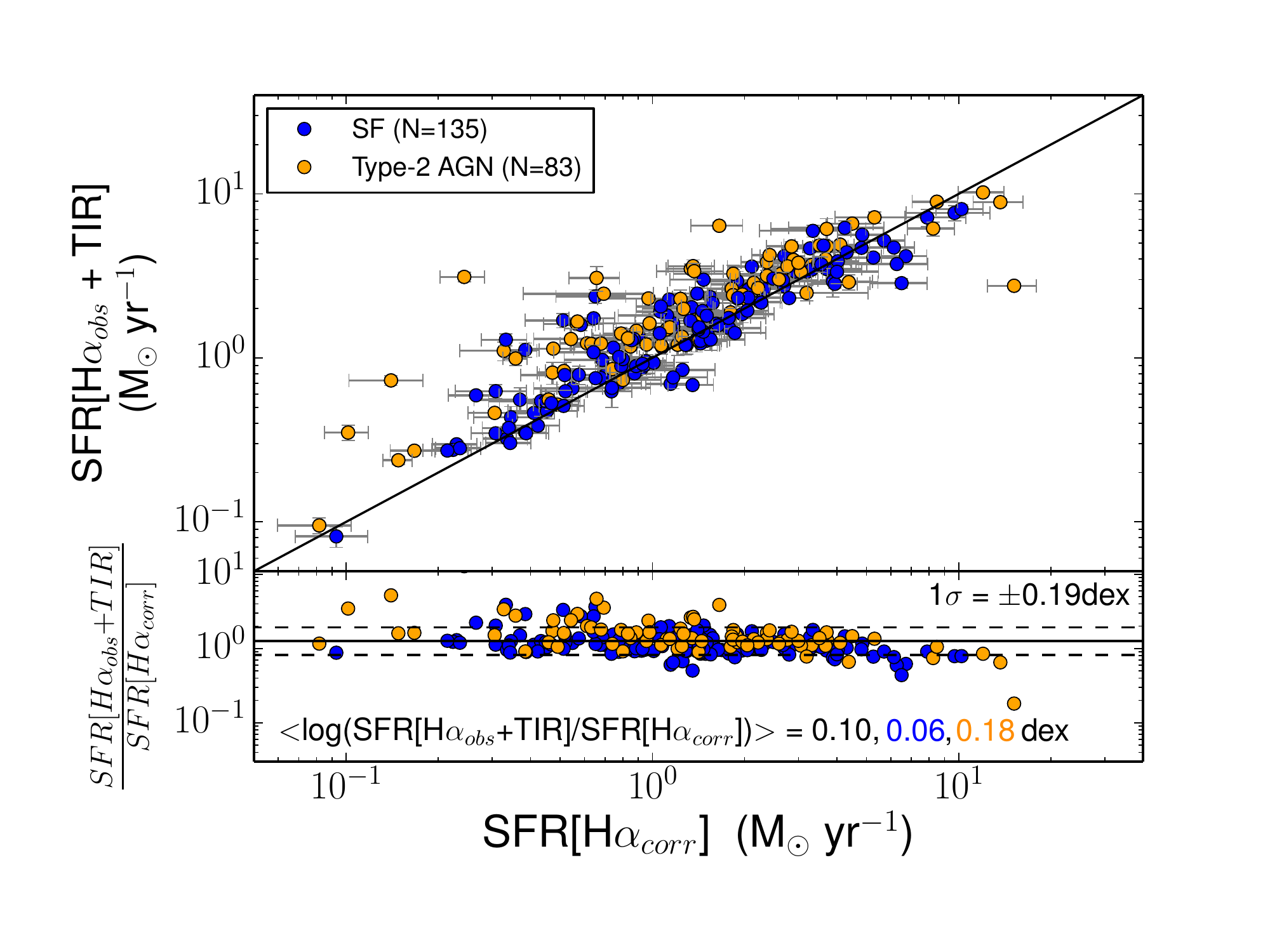} \\
\caption{Top panel: H$\alpha$$_{obs}$ $+$ 22\,$\mu$m hybrid tracer as a function of Balmer-corrected H$\alpha$ SFR, using Equations \ref{equation_ha_22} and \ref{equation_sfr_ha}, respectively. Color-coding and lines have the same meaning as in Figure \ref{sfr_22_plot}. Bottom part shows the residuals as a function of H$\alpha$-corrected SFR being the mean value 0.06, 0.03 and 0.13 for all the galaxies, SF galaxies and type-2 AGN host galaxies, respectively. Dashed lines represent the 1$\sigma$ dispersion in dex around the mean value. Bottom panel: Same as the top panel but showing the H$\alpha$$_{obs}$ $+$ TIR hybrid tracer as a function of Balmer-corrected H$\alpha$ SFR instead (Equations \ref{equation_ha_tir} and \ref{equation_sfr_ha}).}
\label{sfr_hybrid_ha}
\end{figure}

\begin{figure}
\centering
\includegraphics[trim=0.5cm 1.4cm 0.6cm 1.2cm, clip, width=95mm]{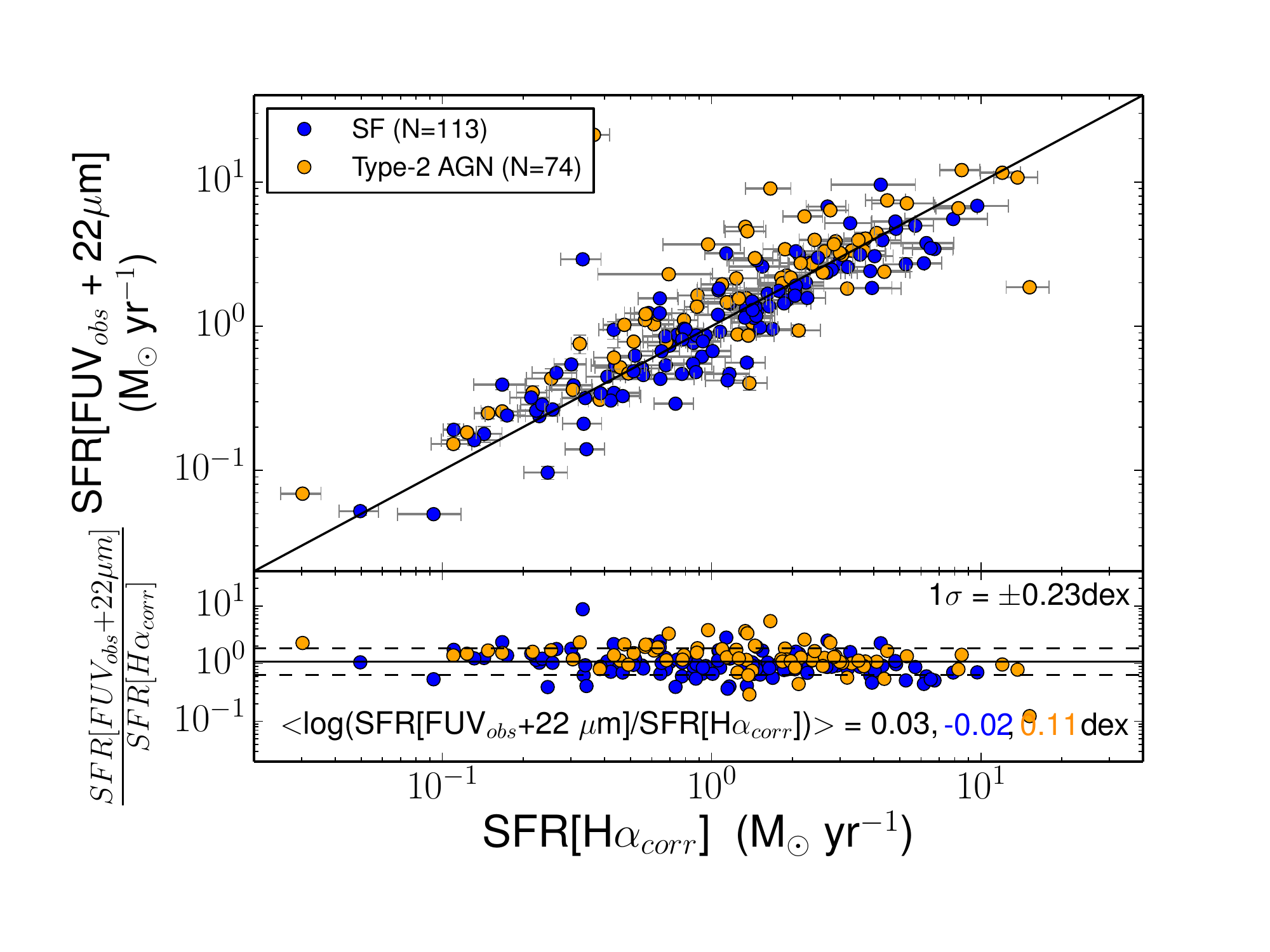} \\
\includegraphics[trim=0.5cm 1.4cm 0.6cm 0cm, clip, width=95mm]{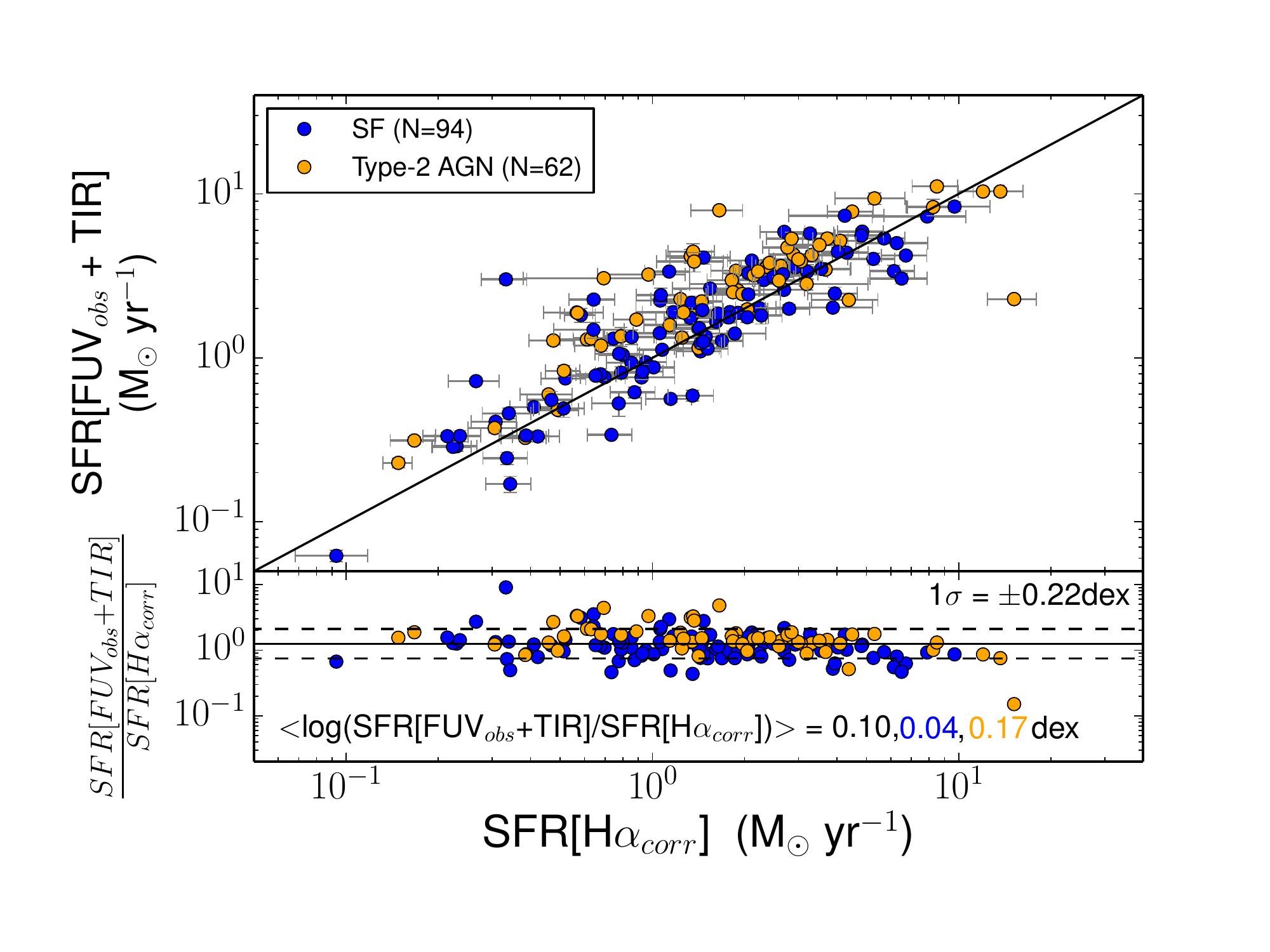} \\
\caption{Top panel: Comparison between FUV$_{obs}$ $+$ 22\,$\mu$m hybrid tracer and the Balmer-corrected H$\alpha$ SFR, using Equations \ref{equation_fuv_22} and \ref{equation_sfr_ha}, respectively. Color-coding, lines and residuals have the same meaning as in Figure \ref{sfr_hybrid_ha}. Bottom panel: Same as the top panel this time showing FUV$_{obs}$ $+$ TIR hybrid tracer as a function of Balmer-corrected H$\alpha$ SFR (Equations \ref{equation_fuv_tir} and \ref{equation_sfr_ha}). The hybrid tracers shown here and in Figure \ref{sfr_hybrid_ha} reduce the dispersion when compared with single-band tracers.}
\label{sfr_hybrid_fuv}
\end{figure}

Once we have explained the assumptions imposed to the use of the hybrid tracers, we compare them with our H$\alpha$ extinction-corrected SFR tracer. In the first place, we examine the behaviour using H$\alpha$ observed luminosity combined with 22\,$\mu$m and TIR luminosity. Figure \ref{sfr_hybrid_ha} shows that applying the method explained in \cite{Kennicutt_2009} we now obtain very similar results to theirs but using a larger sample and for the first time with IFS data. Secondly, we replace H$\alpha$ observed luminosity with FUV observed luminosity combined with the IR luminosities (Figure \ref{sfr_hybrid_fuv}). In both cases we find a very good correlation across 2.5\,dex in SFR but with an offset in the mean ratio of SFRs of 25 per cent. This offset goes in the sense that SFR derived from the hybrid H$\alpha$+IR and FUV+IR SFR tracers is larger than for the extinction-corrected H$\alpha$ one. As explained before, one possibility could be the presence of optical photons from old stellar populations heating the dust specially at long IR wavelengths or the effects of AGN. In particular, if we discriminate between star-forming and type-2 AGN galaxies when computing these ratios, type-2 AGN host galaxies yield larger offsets than the ones reported for SF galaxies. We refer the reader to Section \ref{Hybrid tracers} for an extensive discussion on this issue.

We conclude that when comparing the hybrid calibrators with Balmer decrement attenuation-corrected H$\alpha$ SFR tracer, we find tighter correlations than those obtained with single-band tracers (see the 1$\sigma$ dispersions around the mean values in Figures \ref{sfr_22_plot}, \ref{sfr_fuv_new}, \ref{sfr_tir}, \ref{sfr_hybrid_ha} and \ref{sfr_hybrid_fuv}.)

\paragraph{}

\subsection{Origin of the discrepancies among SFR tracers}
\label{Origin of the discrepancies among SFR tracers}

As we have seen in the previous section, there is a general good agreement between the SFR tracers considered, single-band and especially hybrids, compared to the attenuation-corrected H$\alpha$ SFR tracer. Nevertheless, we can appreciate some differences if we take a closer look at these relations. In the case of the single-band tracers the main problems appear when using FUV$_{corr}$ luminosities as the extinction correction is a big problem to deal with, specially for high SFR values. However, we can mitigate this effect using hybrid tracers combining the FUV$_{obs}$ luminosity with the IR luminosity, both 22\,$\mu$m and TIR. In the latter case we are assuming that we can recover all the light that it has been re-emitted by the dust. 
Similar cases appear when using the single-band tracers for 22\,$\mu$m or TIR luminosities, where we apparently lose some SF in galaxies with low values of the SFR. Again, when using hybrid tracers the agreement between calibrators improves.

One of the main reasons behind these discrepancies is the different selection criteria used in the process of determining the SFR calibrators in the literature. Now, we have the opportunity to re-calibrate these tracers for a diameter-limited sample of 380 galaxies. Moreover, we are able to use integral field spectroscopy data to assure a proper determination of the attenuation using the Balmer decrement avoiding the problems associated with narrow-band imaging. Thus, we are going to provide updated SFR tracers based on our state-of-the-art attenuation-corrected H$\alpha$ luminosities.

\subsection{Updated SFR tracers for the diameter-limited CALIFA sample}
\label{Updated SFR tracers for the SDSS-based diameter-limited CALIFA sample}

We now provide updated calibrations for the global current SFR in external galaxies by means of anchoring the different tracers (single-band and hybrid ones) to the SFR derived from the extinction-corrected H$\alpha$ luminosity measured in our sample of CALIFA galaxies. Seminal works in this context include \citet{Kennicutt_98}, \citet{Kennicutt_2009} and \citet{Hao_2011}. 

As we are interested in calibrating the SFR tracers we need to exclude galaxies that have type-1 AGN signatures to avoid contamination of sources that are not star-forming (only galaxies UGC 00987 and UGC 03973 are classified as type-1 AGN within our sample). As explained in Section \ref{Comparison of the different SFR tracers}, the information regarding the optical AGN classification can be found in \citet{Walcher_2014}. We provide separate calibrations for the sample when type-2 AGN galaxies are included and when they are not. The reason for this is that, despite numerous efforts \citep{alonso_herrero_2006, diaz_santos_2008, diaz_santos_2010, castro_2014}, the fraction of UV or line emission arising from circumnuclear star formation in type-2 AGN is still highly uncertain. We remind the reader that the nuclear emission in type-2 AGN includes the contribution of both a dusty torus \citep[external radius of a few parsecs, see][]{ramos_almeida_2009} and a circumnuclear region that could expand until 1kpc from the central region. We estimate the level of contamination of the emission from the AGN host galaxies to the total SFR. We find that the contribution of the attenuation-corrected H$\alpha$ luminosity in the nucleus (measured in a 3"-diameter aperture) over the total one for galaxies classified as type-2 AGN is 8.3$\%$ while for the purely SF galaxies this contribution is 5.1$\%$. Galaxies classified as type-2 AGN are shown in our plots as orange points.

We first provide updated calibrations in the case of the single-band tracers. We do not perform this analysis in the case of the SFR[FUV$_{corr}$] because, as we have explained before, the attenuation correction is highly uncertain and the SFR tracer proposed would not be reliable. On the other hand, the estimation of the hybrid tracers using FUV$_{obs}$ luminosity will be pursued.

\begin{figure*}
\centering
\includegraphics[trim=2.1cm 0.8cm 0cm 1cm, clip, width=90mm]{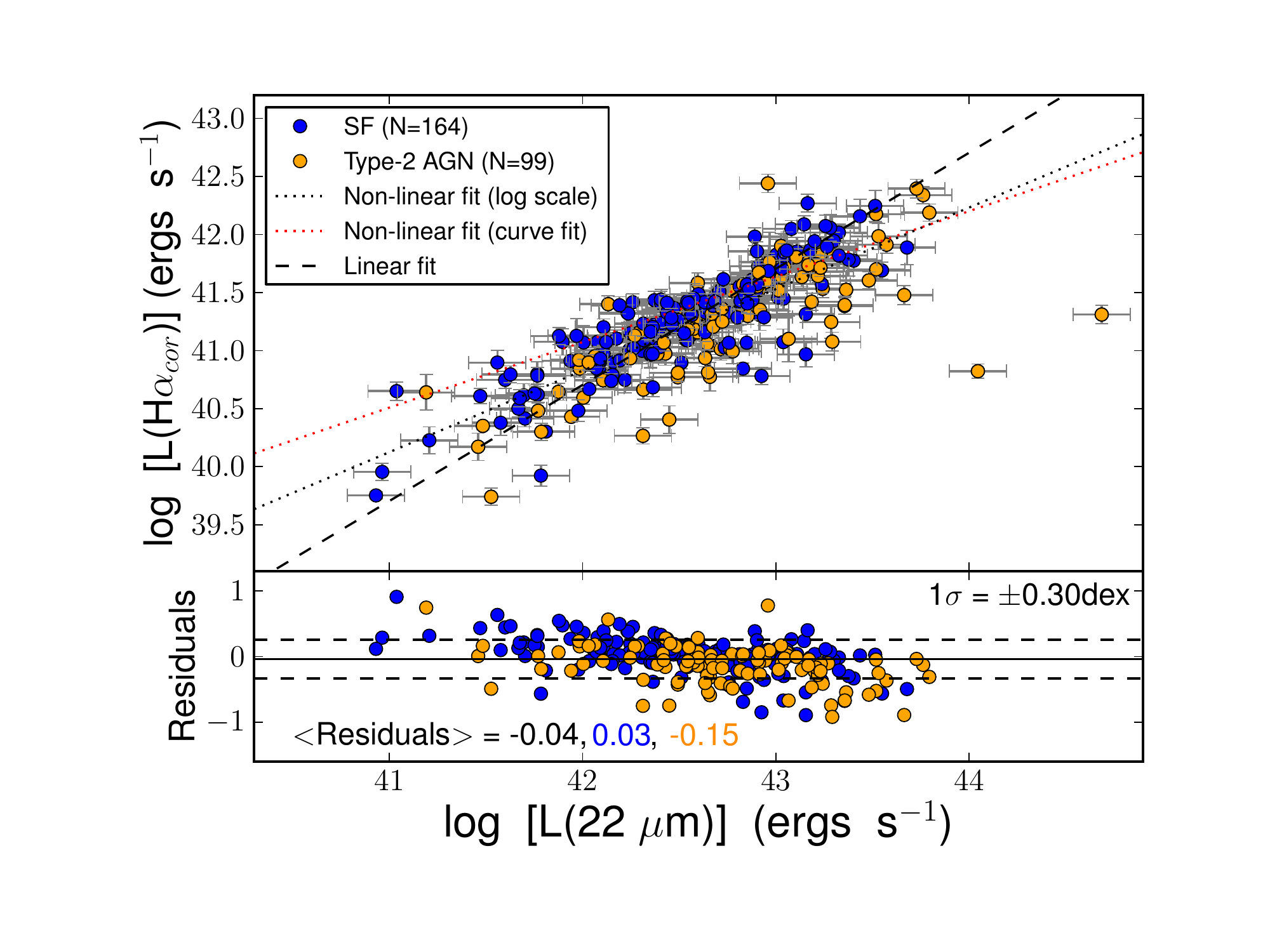} 
\includegraphics[trim=1.8cm 0.8cm 0.3cm 1cm, clip,width=90mm]{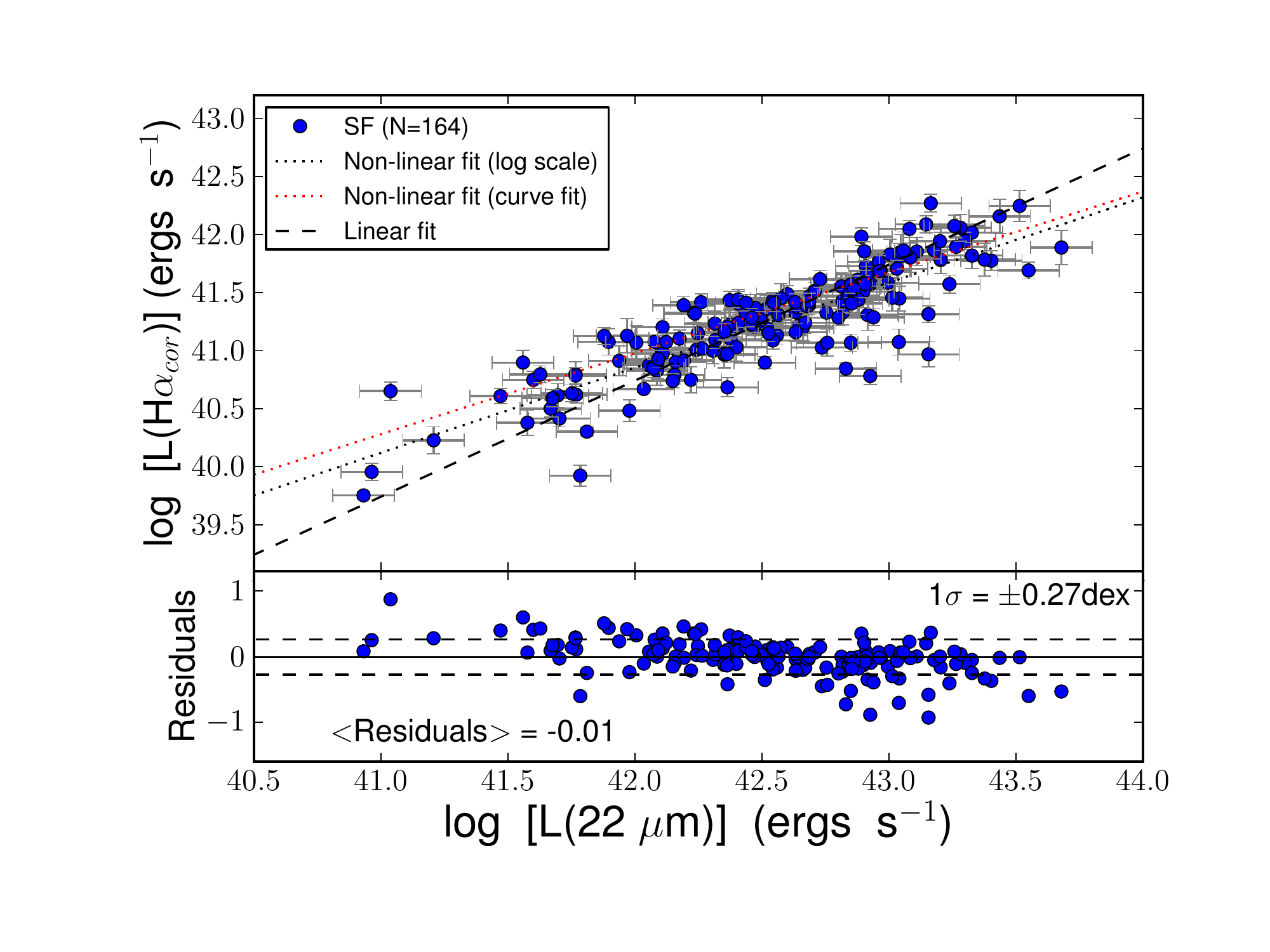} \\
\includegraphics[trim=2.1cm 1.2cm 0cm 0.5cm, clip,width=90mm]{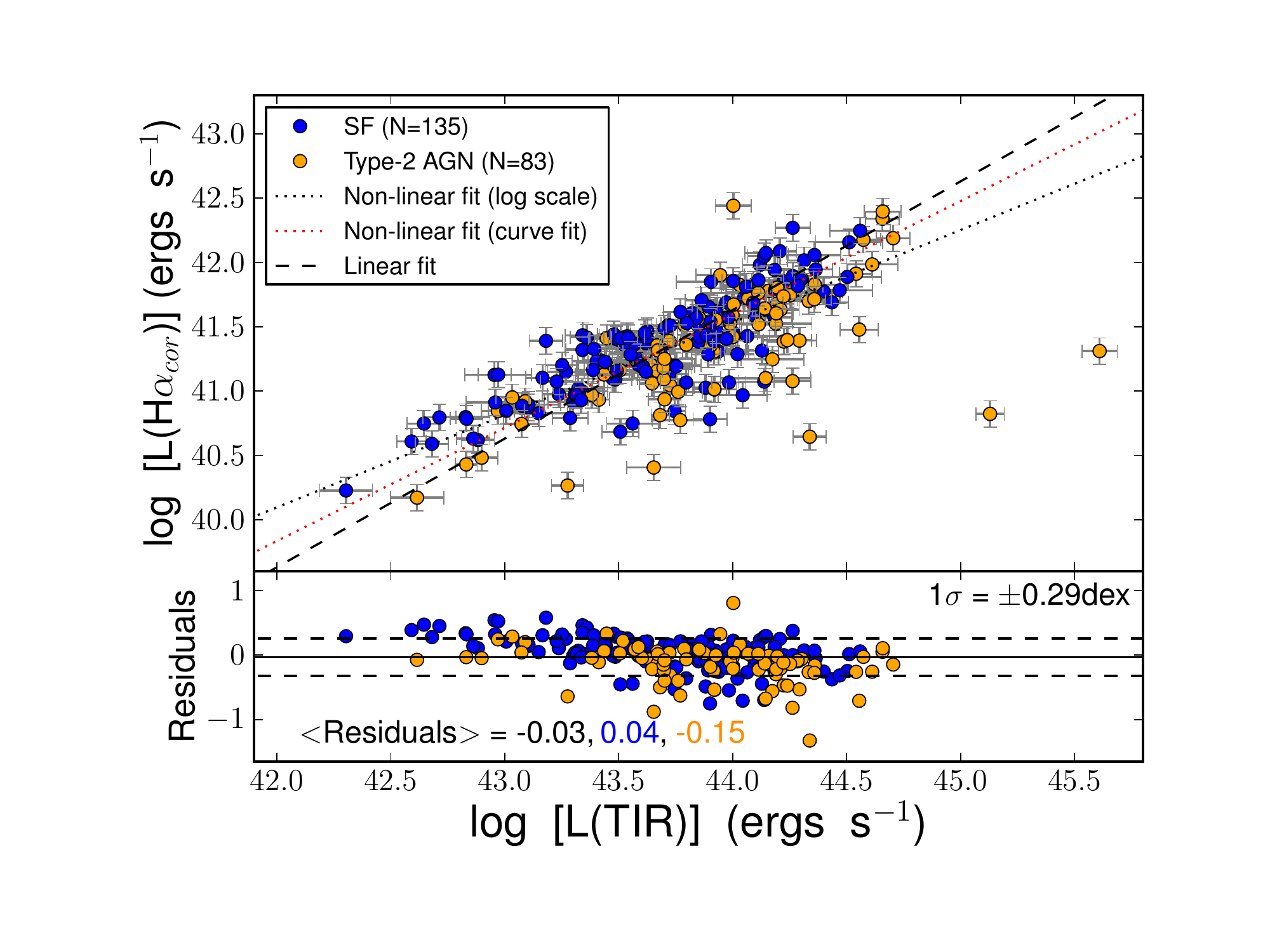}
\includegraphics[trim=1.8cm 1.2cm 0.3cm 0.5cm,clip, width=90mm]{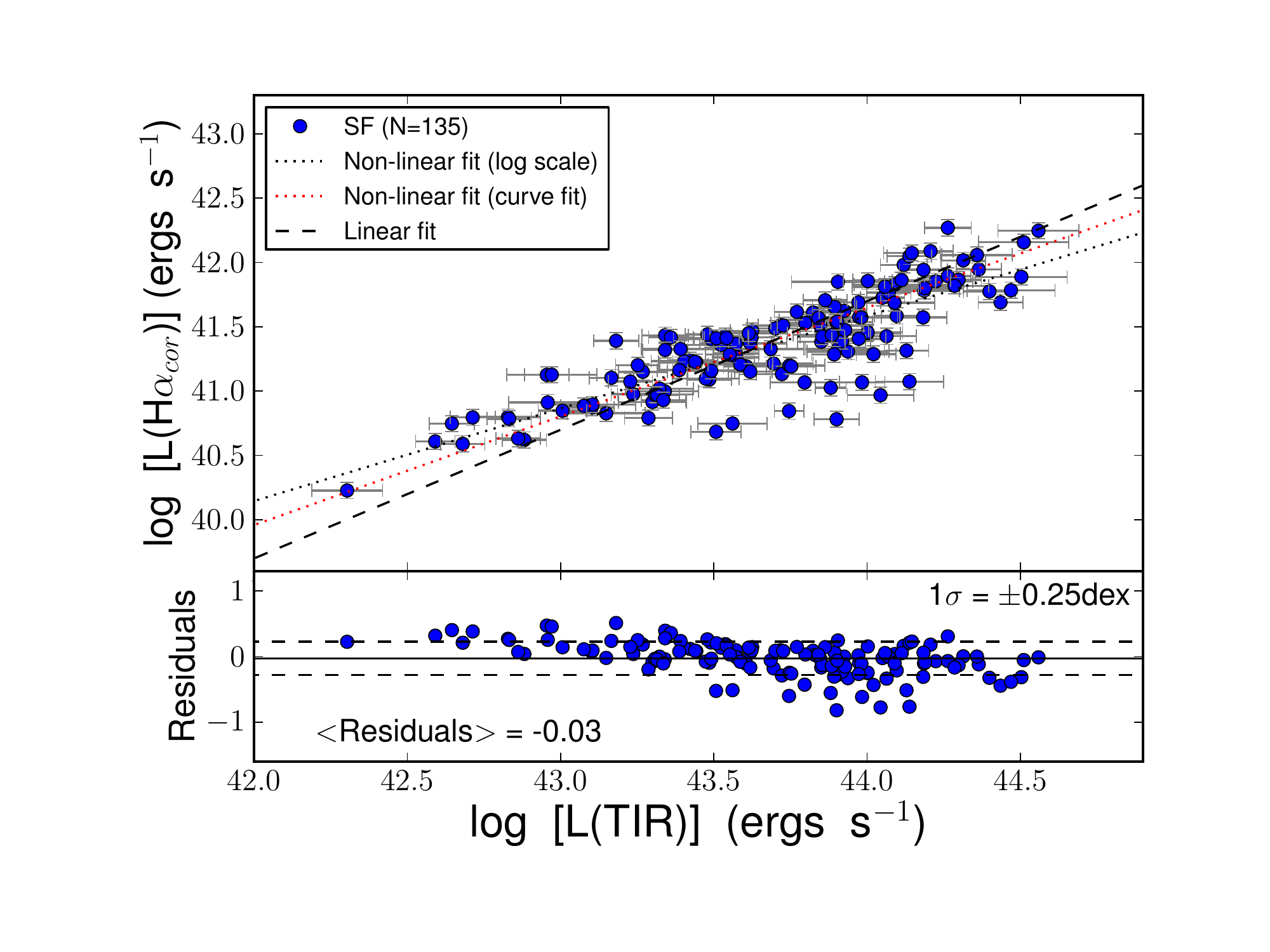} \\ 
\caption{Updated calibrations for the 22\,$\mu$m and TIR single-band SFR tracers anchoring them to extinction-corrected H$\alpha$ luminosity (Equation \ref{equation_sfr_ha}). Orange points in left panels correspond to type-2 AGN while blue points refer to star-forming galaxies. Linear fits are shown in dashed lines while non-linear fits are shown in dotted lines. The results for these calibrations appear in table \ref{single_table}. The residuals are computed as the average value of the log[5.5$\times$10$^{-42}$$\times$L(H$\alpha$$_{corr}$)/a$\times$L(IR)] where L(IR) could be 22\,$\mu$m or TIR for the case of the linear fits, after applying a 4 $\sigma$ rejection. These values are computed for all galaxies (black), star forming (blue) and type-2 AGN host galaxies (orange).}
\label{single_band_updated_v2}
\end{figure*}

\subsubsection{Single-band tracers}
\label{Single-band updated}

In this section we provide calibrations for the observed 22\,$\mu$m and TIR luminosities as tracers of the SFR anchoring them to the SFR given by the extinction-corrected H$\alpha$ luminosity according to Equation \ref{equation_sfr_ha}. Figure~\ref{single_band_updated_v2} shows the relation between L(H$\alpha$$_{corr}$) and the observed infrared luminosities. We include in these plots both non-linear\,\footnote{We normalize the luminosities to 10$^{43}$\,ergs\,s$^{-1}$ to ensure that the y-intercept for these non-linear fits is located near the values covered by our sample}, log[L(H$\alpha$$_{corr}$)]=b$\times$log\Big[$\frac{L(IR)}{10^{43}}$\Big]+log\Big[$\frac{a'}{5.5\times10^{-42}}$\Big] and linear fits, log[L(H$\alpha$$_{corr}$)]=log[L(IR)]+log\Big[$\frac{a}{5.5\times10^{-42}}$\Big]; \citep[see][]{Calzetti_2012}. We warn the reader that the use of non-linear calibrations should be restricted to studies using similar (1) selection criteria, (2) apertures and (3) corrections or the resulting SFRs could be affected by severe systematics. Linear fits are shown in dashed line while non-linear fits are shown in dotted lines. The coefficients for these fits are given in Table~$\ref{single_table}$. Note that in the case of the linear fit we name the constant {\it a} which is expressed in units of M$_{\odot}$yr$^{-1}$/ergs\,s$^{-1}$ and for the non-linear fit we use {\it a$'$} because it lacks the physical meaning of {\it a}. We are going to use $a_{IR}$ for the hybrid tracers as in this case it is dimensionless and has a different physical meaning than the previous constants (see Section \ref{Hybrid tracers} for more details). The values for {\it a$'$} and the exponent {\it b} are obtained by two different methods, a least-squares linear fit in log scale and a non-linear least squares fit using the Python task {\it curve\_fit}. Both methods yield similar values for these fitting parameters with 22\,$\mu$m and TIR luminosities. 

Figure~\ref{single_band_updated_v2} also shows the results of this analysis after including (left panels) or excluding (right panels) type-2 AGN from the sample. In all cases, with and without type-2 AGN and using either the 22\,$\mu$m or TIR luminosity, a non-linear behaviour is clearly present, especially at low luminosities (log[L(22$\mu$m)]$<$41.8 or log[L(TIR)]$<$43.3), where most galaxies are located above the best linear fit (see Figure~\ref{single_band_updated_v2}). On the other hand, galaxies with high 22\,$\mu$m luminosities (log[L(22$\mu$m)]$>$43.4) are all found below the linear fit, as also found by \cite{Rieke_2009} for 24$\mu$m luminosities above 5$\times$10$^{43}$\,ergs\,s$^{-1}$. The best-fitting global slope for our non-linear SFR calibrations based on 22\,$\mu$m luminosity, 0.733 (0.702) when type-2 galaxies are (not) included in the sample, is somewhat smaller (less linear) than the {\it local} value (500\,pc scale) of 0.885 obtained by \cite{Calzetti_2007} and than the value of 0.82 given by \cite{Cluver_2014} from the analysis of the GAMA survey. Regarding {\it a$'$}, the values can be very different from those in the literature (even their units are different, obviously) but the uncertainties are of the order of 5-18$\%$, similar to the value quoted by \citet{Calzetti_2012}, when the log is computed in normalized luminosities. It is worth noting that the 22\,$\mu$m or TIR luminosities explored by our sample are significantly lower than those of the sample studied by \cite{Rieke_2009}, which explains why these authors only needed to make use of a non-linear fit at their high-luminosity end. 

For the case of the linear fit, the difference between the {\it a} coefficients with and without type-2 AGN is very small, leading to smaller {\it a} coefficient when these objects are included by roughly 7 per cent in the case of the 22\,$\mu$m and 18 per cent in the TIR. This is likely due to the enhanced contribution of an AGN or, alternatively, obscured circumnuclear star formation to the infrared emission compared to H$\alpha$. In the former case, the use of this calibration would remove, statistically speaking at least, part of the AGN contamination, although some fraction of the H$\alpha$ could still arise from the AGN. Regarding the latter possibility, the use of a calibration anchored to the extinction-corrected H$\alpha$ luminosity would slightly underestimate the total SFR, as the star formation due to highly-obscured circumnuclear regions in type-2 AGN could be missed. In general, independently of its origin (star formation or not), the total UV light emitted in these regions is hardly recovered using the UV slope or even the Balmer decrement as a measure of its dust attenuation, especially in type-2 AGN where the BLR is completely hidden from us. The difference here is that if the emission is due to either a BLR or NLR is something that should not be accounted for in terms of the SFR anyway.

Finally, we can not rule out at this stage that, since we are dealing with single-band tracers, this difference arises from a dependence of the attenuation with the level of nuclear activity at a given SFR. 

The analysis of the hybrid tracers and the dependence of the {\it a$_{IR}$} coefficient with the attenuation presented later in this work favors the scenario where it is the contribution of the AGN itself what leads to these small changes in the single-band SFR calibrations. This will be studied in more detail in a future (spatially-resolved) analysis. Although, as mentioned before, we will still be unable to disentangle the relative contribution of AGN or circumnuclear star formation to the nuclear emission of type-2 AGN host galaxies using these data \citep[see][for alternative approaches]{alonso_herrero_2006, diaz_santos_2010, castro_2014}.

\subsubsection{Hybrid tracers}
\label{Hybrid tracers}

In the case of the hybrid indicators we assume a simple energy balance \citep[see][for more details]{Kennicutt_2009}.

\begin{equation}
\mathrm{SFR} \thickspace (M_{\odot} yr^{-1}) = 5.5 \times 10^{-42}\thickspace [L(H\alpha_{obs}) + a_{IR} \times L(IR)]
\label{halpha_calibration}
\end{equation}
\begin{equation}
\mathrm{SFR} \thickspace (M_{\odot} yr^{-1}) = 4.6 \times 10^{-44}\thickspace [L(FUV_{obs}) + a_{IR} \times L(IR)]
\label{fuv_calibration}
\end{equation}

where L(FUV$_{obs}$) and L(H$\alpha_{obs}$) are the observed luminosities in ergs s$^{-1}$ and L(IR) could be either L(22\,$\mu$m) or L(TIR), also in ergs s$^{-1}$ . 

We calculate the value of the dimensionless {\it a$_{IR}$} coefficient in the previous hybrid relations as the median of the following ratio for the L(H$\alpha$) case (Equation \ref{equation_ha_22} $\&$ \ref{equation_ha_tir}) and the same with L(FUV) (Equation \ref{equation_fuv_22} $\&$ \ref{equation_fuv_tir}): 
\begin{equation}
a_{IR} = \frac{L(H\alpha_{corr}) - L(H\alpha_{obs})} {L(IR)} 
\label{air_ha}
\end{equation} 

\begin{equation}
a_{IR} = \frac{\frac{C_{H\alpha}}{C_{FUV}}L(H\alpha_{corr}) - L(FUV_{obs})} {L(IR)} 
\label{air_fuv}
\end{equation}

Where C$_{H\alpha}$ and C$_{FUV}$ are the constants that multiply the L(H$\alpha$) and L(FUV) in Equations \ref{equation_sfr_ha} and \ref{equation_sfr_fuv} (5.5$\times$10$^{-42}$ and {4.6$\times$10$^{-44}$ [M$_{\odot}$yr$^{-1}$/ergs$^{-1}$]), respectively. 

Histograms in Figure \ref{histograms_constant} show the distribution of the {\it a$_{IR}$} coefficient for different hybrid SFR tracers. We refer the reader to the end of this section and Sections \ref{morphological dependence} through \ref{attenuation dependence} for an extensive analysis on the nature of the variation of a$_{IR}$.

In the case of the combined UV + IR SFR tracers, there are several ways of estimating the calibration. The most common methods are (1) using an energetic balance approach once we have corrected for attenuation in the UV or (2) anchoring our data to other SFRs measurements.

\begin{figure}
\includegraphics[trim=0.7cm 0.0cm 1.1cm 1.0cm, clip, width=90mm]{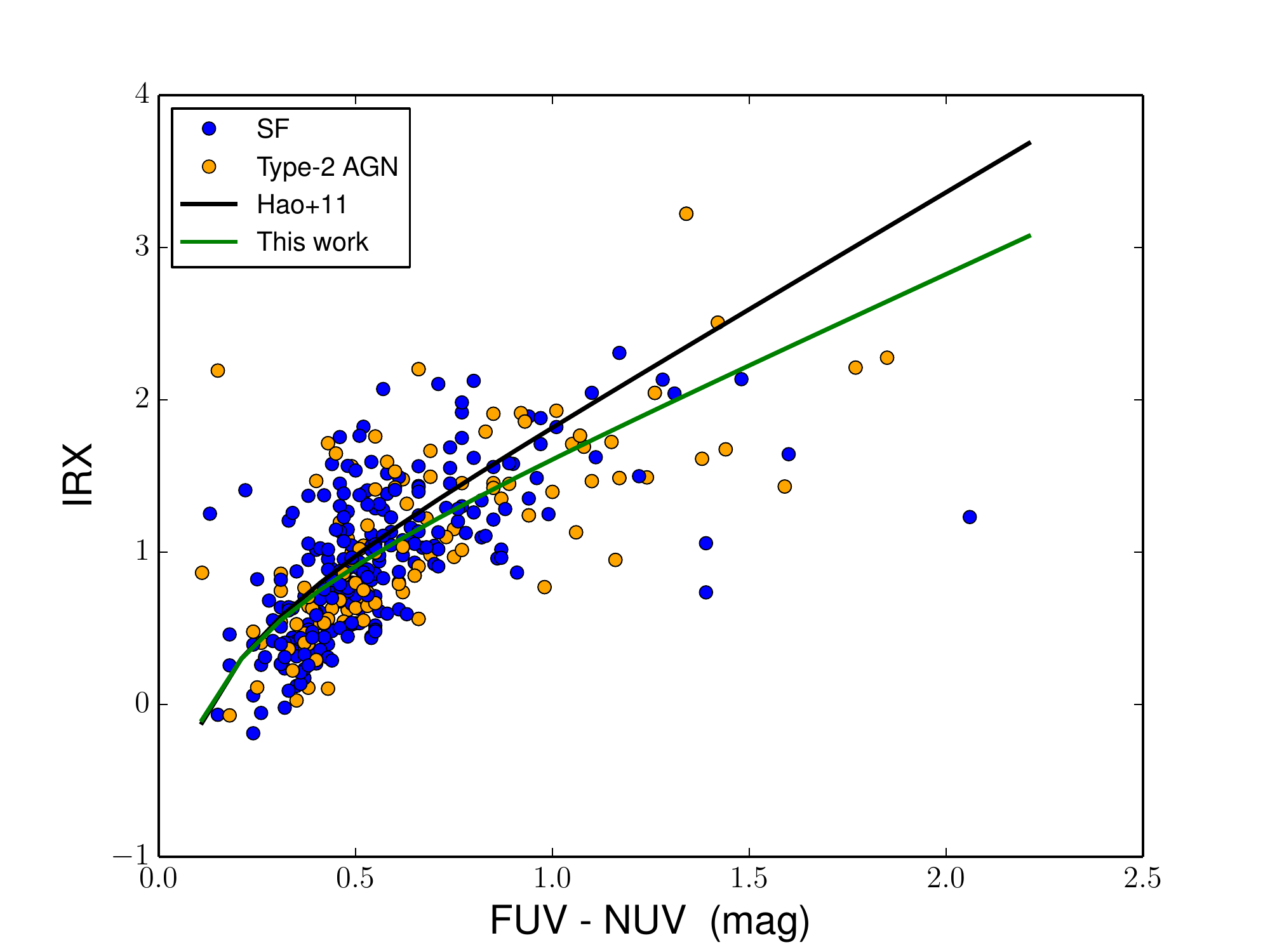}
\caption{IRX-$\beta$ relation for the galaxies that have FUV and TIR measurements in the CALIFA mother sample. Blue points represent star-forming galaxies while orange points correspond to type-2 AGN host galaxies. Black line shows the fit from \cite{Hao_2011} while green line shows our fit. We obtain a value of a$_{IR}$ $=$ 0.33 $\pm$ 0.08 for our fit.}
\label{irx_beta}
\end{figure}

\begin{figure*}
\centering
\includegraphics[trim=2.0cm 0.8cm 0cm 1cm, clip, width=90mm]{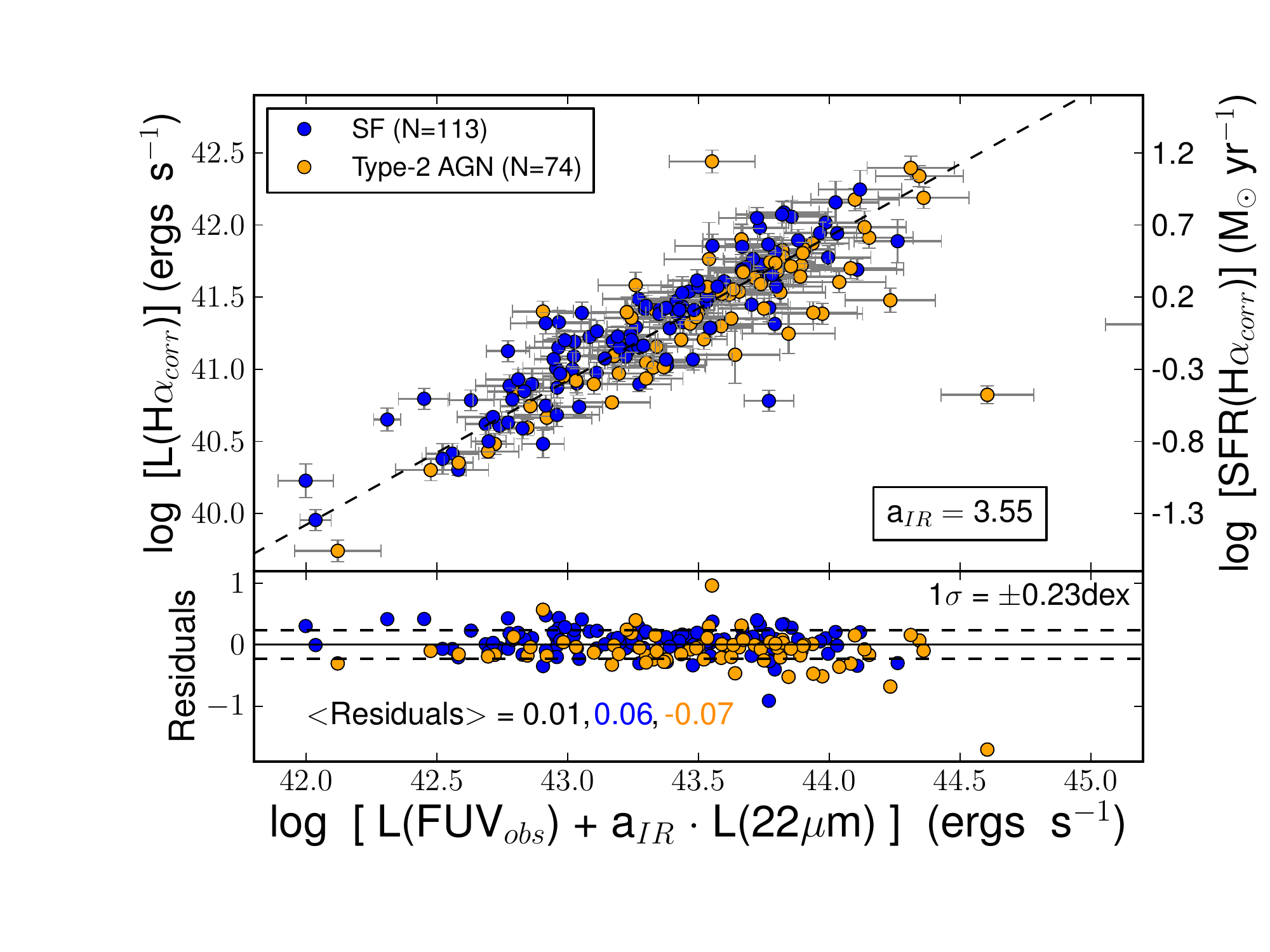} 
\includegraphics[trim=1.8cm 0.8cm 0.2cm 1cm, clip,width=90mm]{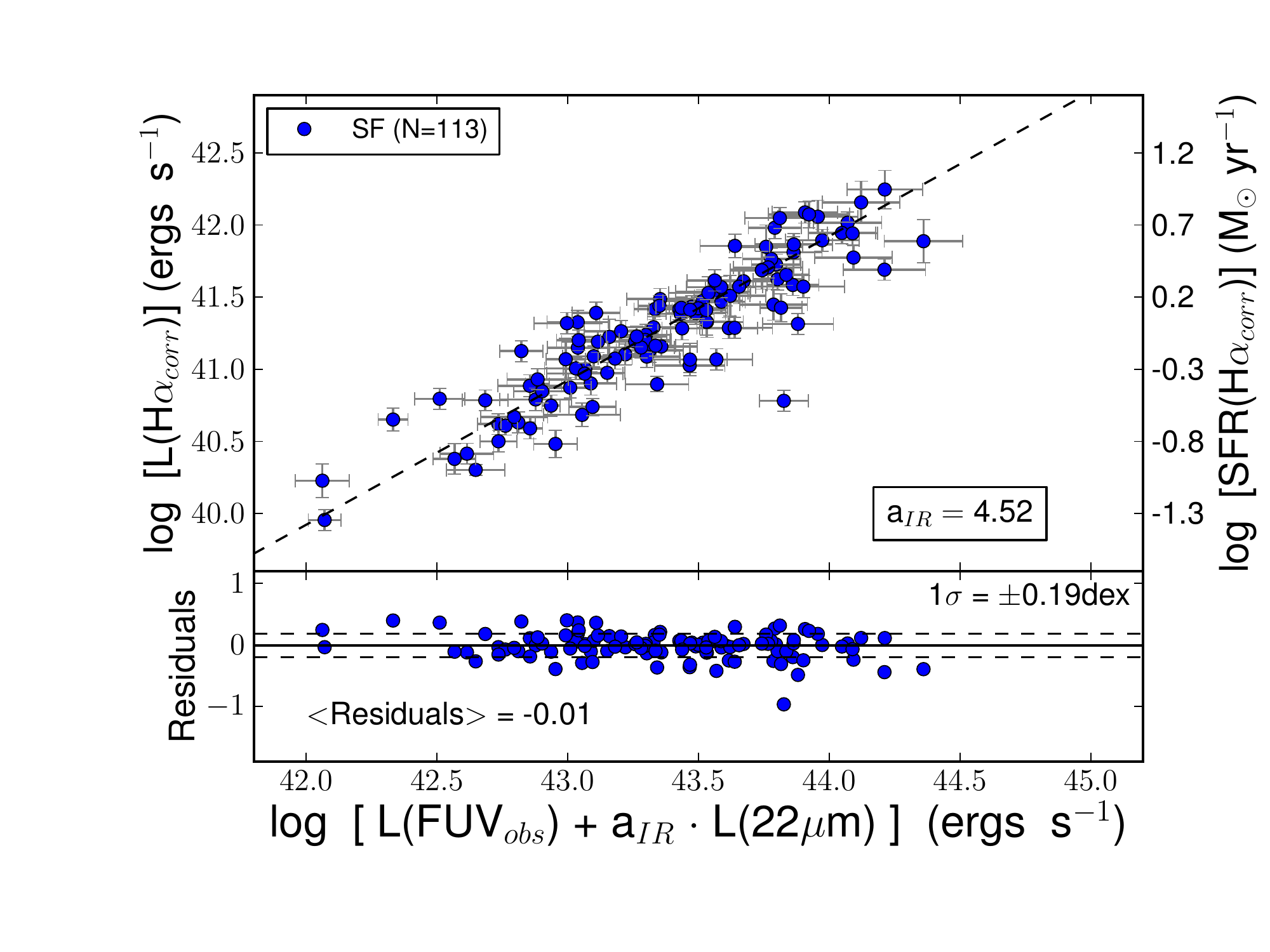} \\
\includegraphics[trim=2.0cm 1.2cm 0cm 0.5cm, clip,width=90mm]{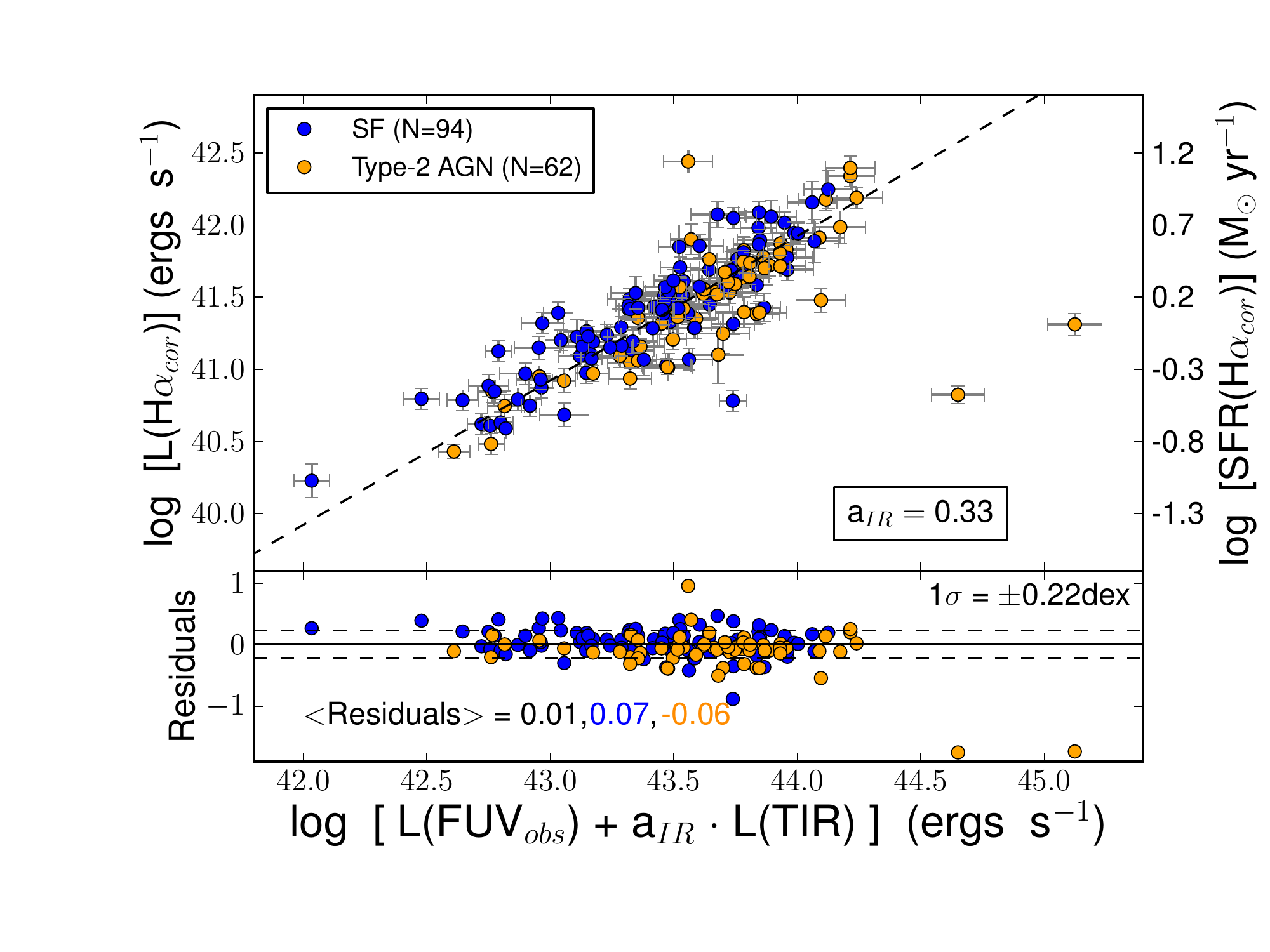} 
\includegraphics[trim=1.8cm 1.2cm 0.2cm 0.5cm,clip, width=90mm]{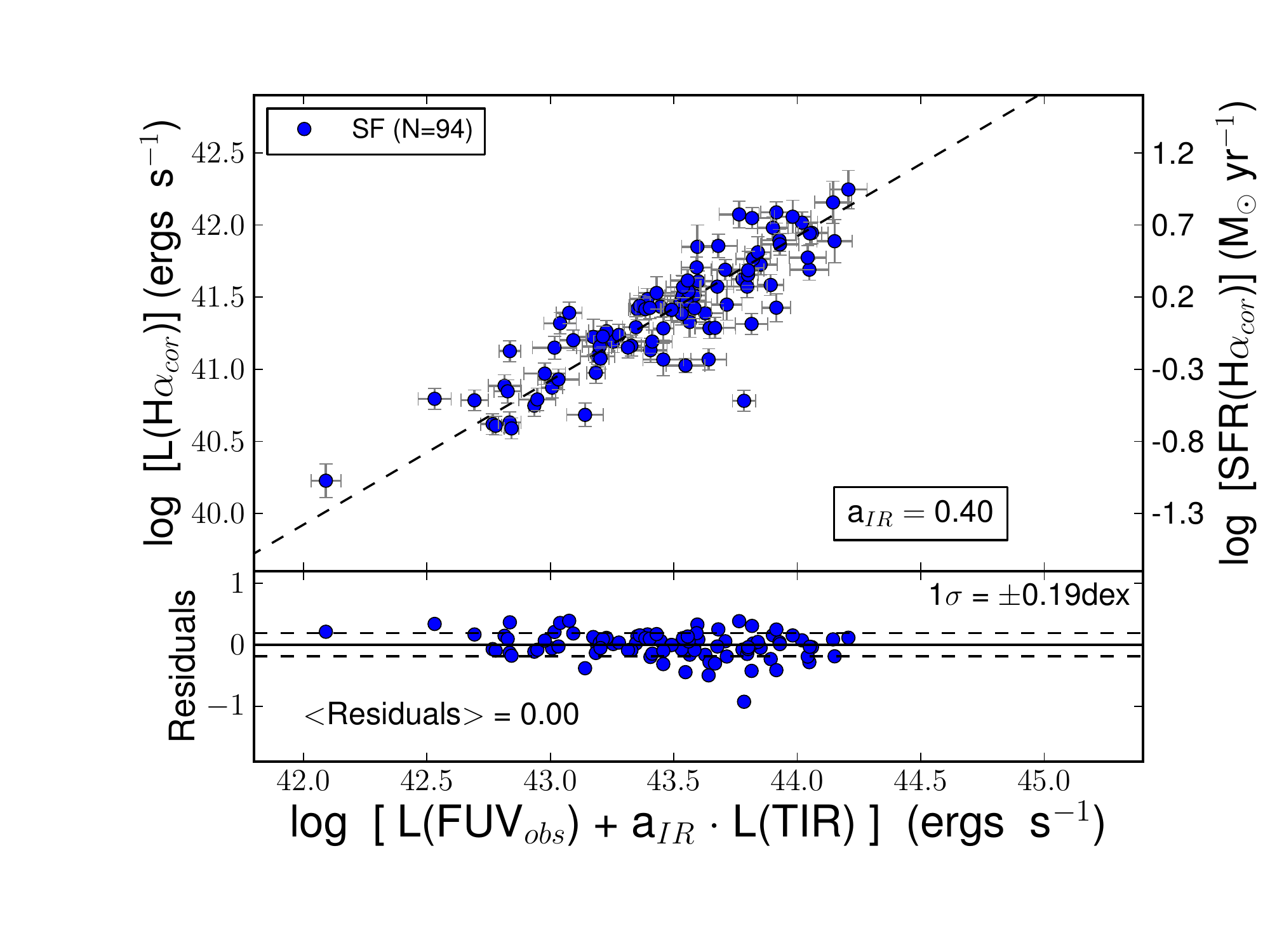} \\ 
\caption{Updated calibrations for the hybrids tracers that combine FUV observed luminosity and IR luminosity. Top panels show the FUV$_{obs}$ + 22\,$\mu$m hybrid tracer while the FUV$_{obs}$ +TIR hybrid tracers appear on the bottom panels. Galaxies hosting type-2 AGN (orange points) are included on the left panels. Blue points refer to star-forming galaxies. Dashed lines correspond to the 1:1 line taking into account the C$_{H\alpha}$ and C$_{FUV}$ constants (5.5$\times$10$^{-42}$ and 4.6$\times$10$^{-44}$) given in Equations \ref{equation_sfr_fuv} and \ref{equation_sfr_ha}, respectively. The best fitting {\it a$_{IR}$} coefficients calculated as the median value of the expression \ref{air_fuv} are shown for clarity. These {\it a$_{IR}$} values and their corresponding errors appear in table \ref{hybrids_global_table}. The residuals are computed as the average value of the log[C$_{H\alpha}$$\times$L[H$\alpha$$_{corr}$]/(C$_{FUV}$$\times$(L[FUV$_{obs}$] + a$_{IR}$$\times$L[IR]))], where L(IR) could be 22\,$\mu$m or TIR, after applying a 4 $\sigma$ rejection. These hybrid tracers show a trend with the a$_{IR}$ coefficient, so when type-2 AGN host galaxies are included the value of a$_{IR}$ decreases.}
\label{sfr_update_fuv}
\end{figure*}

\begin{figure*}
\centering
\includegraphics[trim=2.1cm 0.8cm 0cm 1cm, clip, width=90mm]{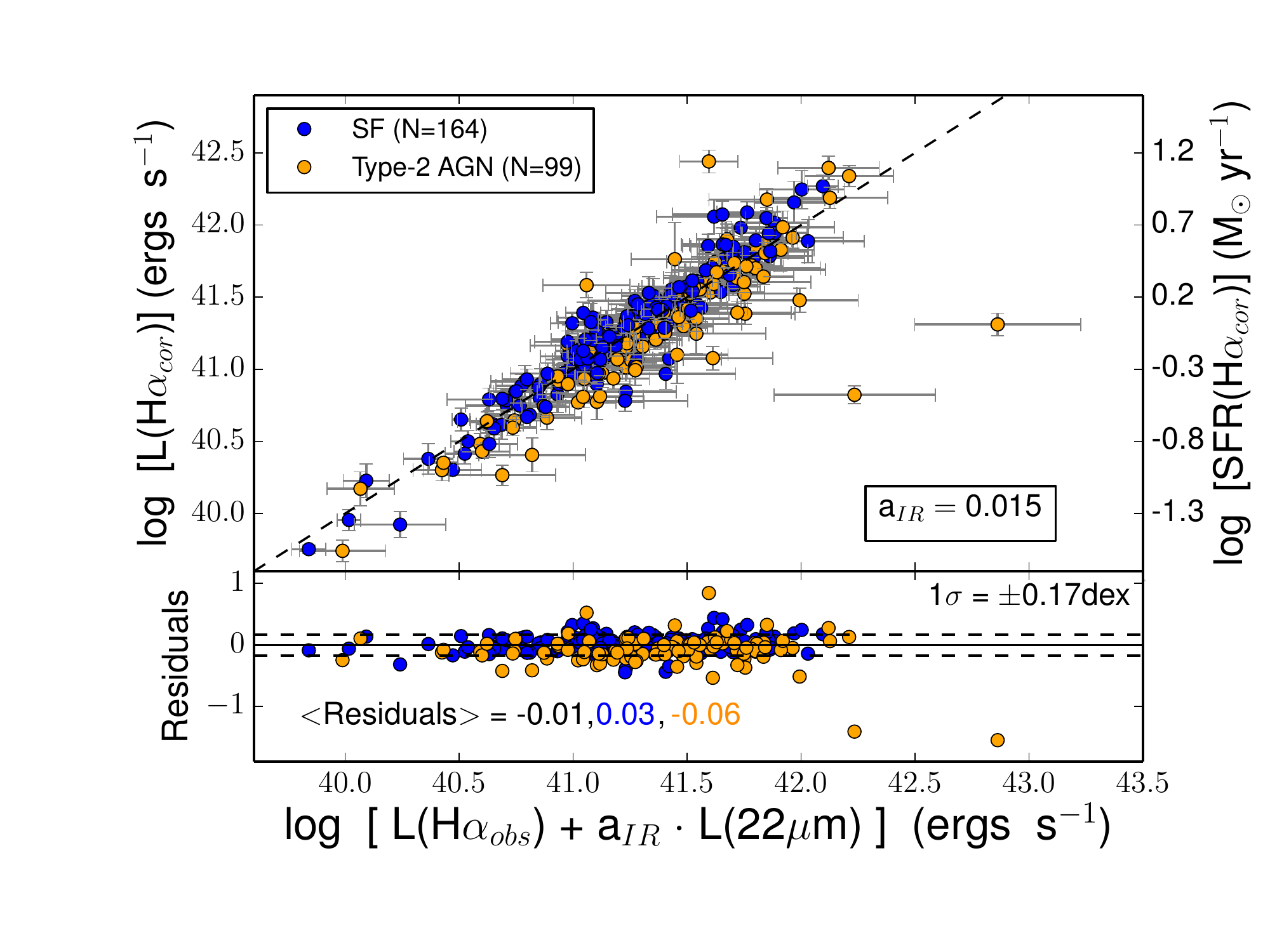} 
\includegraphics[trim=1.8cm 0.8cm 0.3cm 1cm, clip, width=90mm]{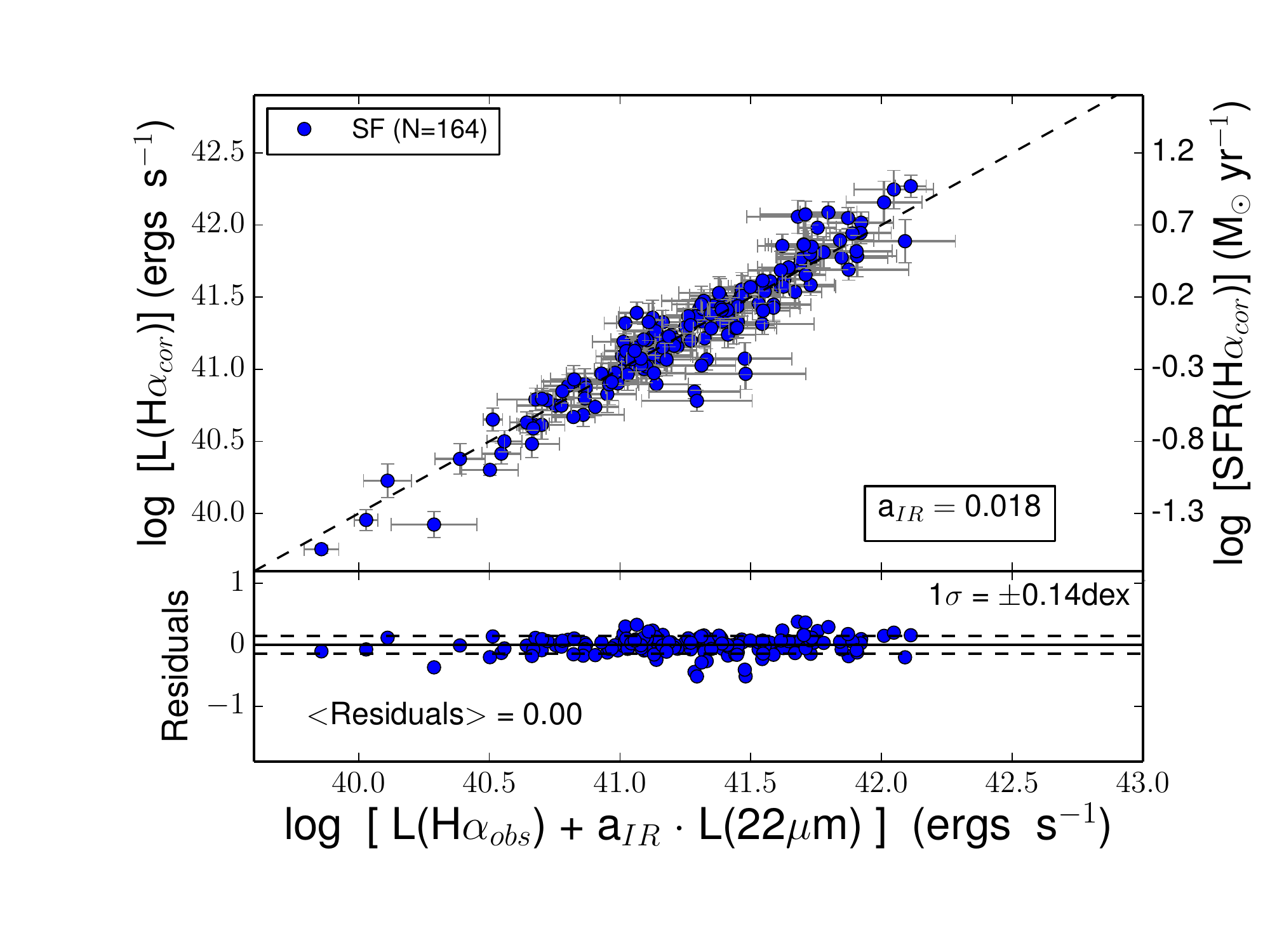} \\
\includegraphics[trim=2.1cm 1.2cm 0cm 0.5cm, clip, width=90mm]{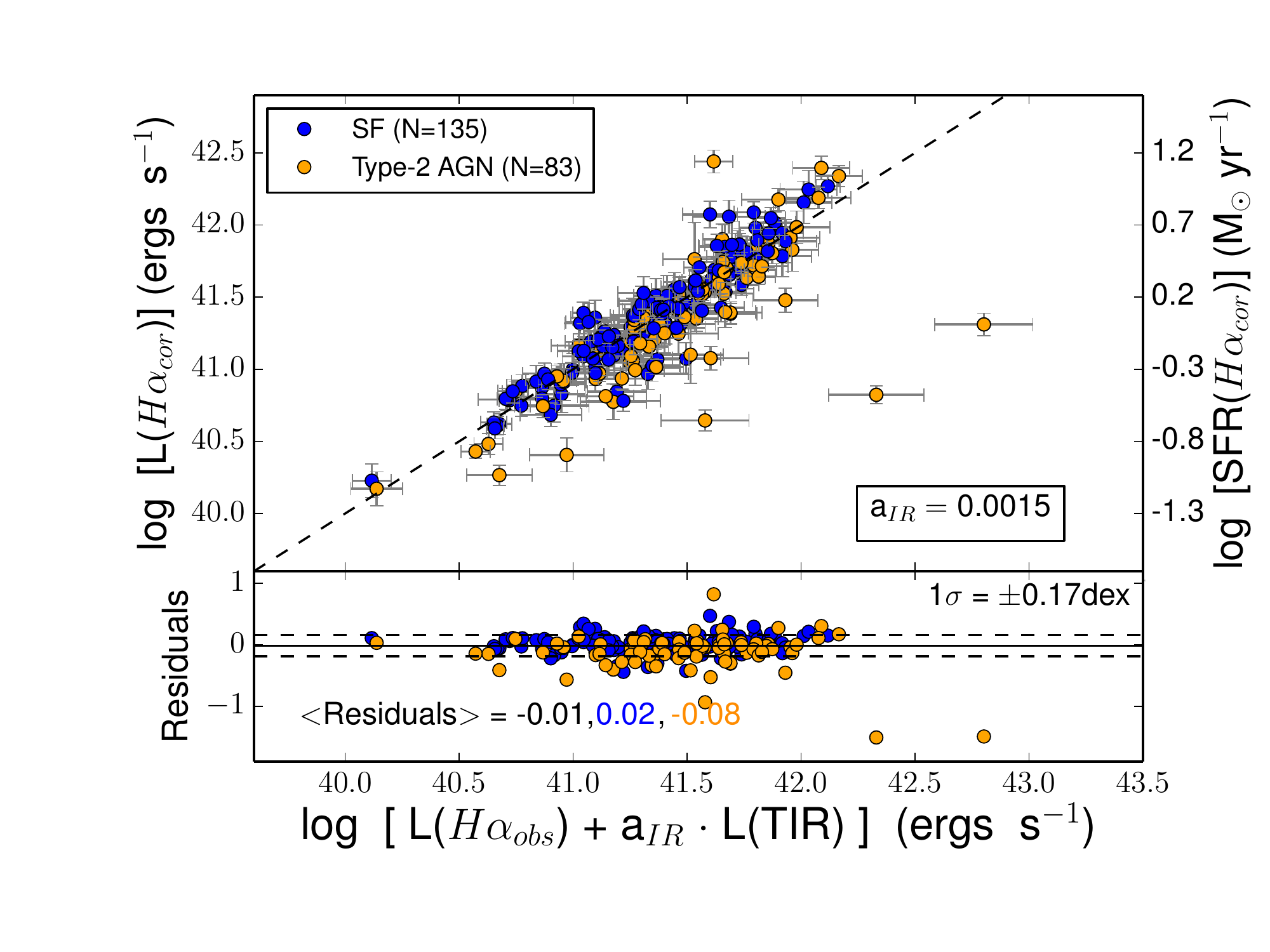} 
\includegraphics[trim=1.8cm 1.2cm 0.3cm 0.5cm,clip, width=90mm]{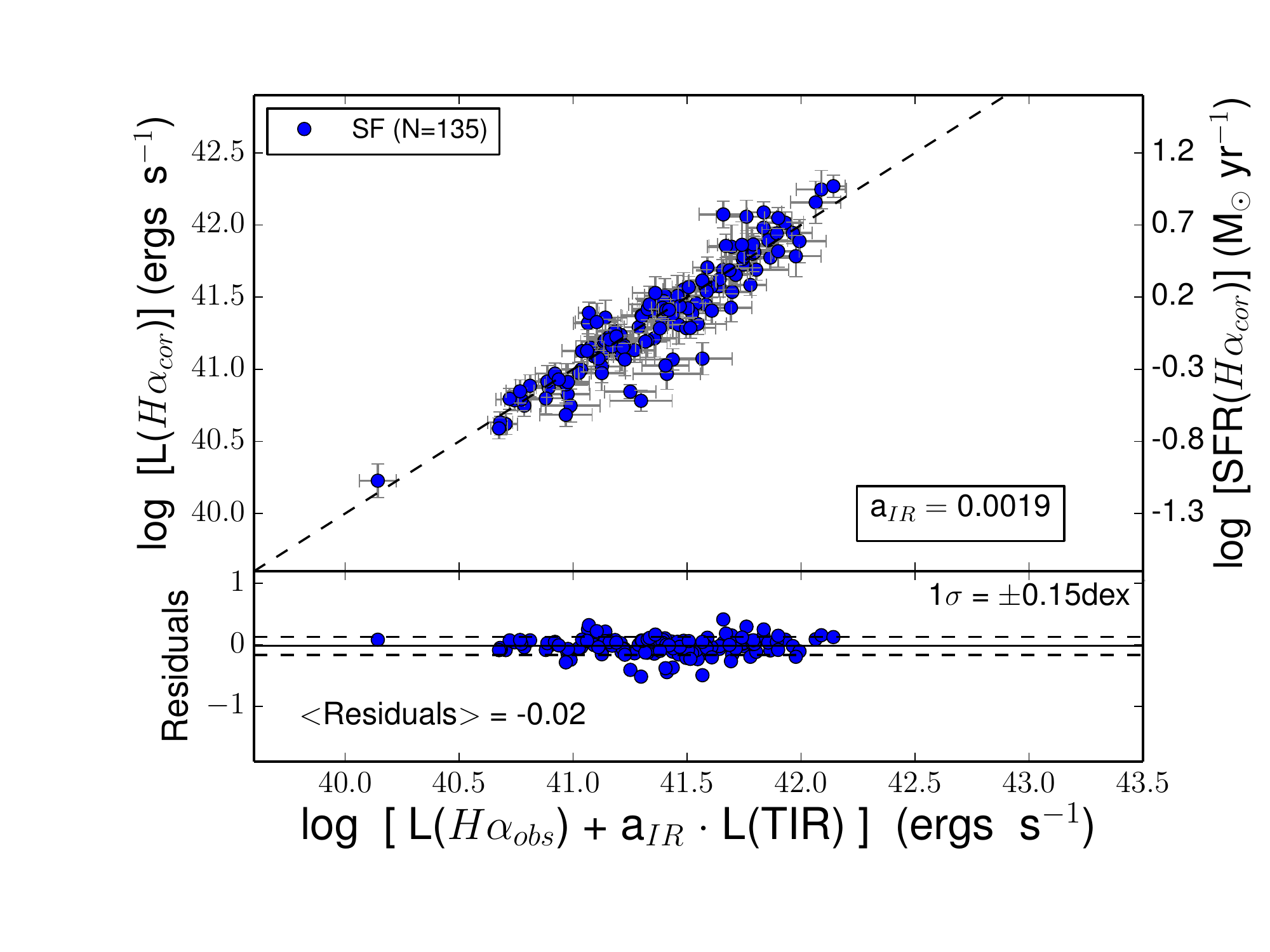} \\
\caption{Updated calibrations for the hybrid tracers that combine H$\alpha$ observed luminosity and infrared emission. The ones with 22\,$\mu$m luminosity appear at the top while the tracers that used TIR luminosity are shown at the bottom. Galaxies hosting type-2 AGN (orange points) are included on the left panels. Blue points refer to star-forming galaxies. Dashed lines correspond to the 1:1 line. The best fitting {\it a$_{IR}$} coefficients calculated as the median value of the expression \ref{air_ha} are shown for clarity. These {\it a$_{IR}$} values and their corresponding errors appear in table \ref{hybrids_global_table}. The residuals are computed as the average value of the log[L(H$\alpha$$_{corr}$)/(L(H$\alpha$$_{obs}$) + a$_{IR}$$\times$L(IR))], where L(IR) could be 22\,$\mu$m or TIR, after applying a 4 $\sigma$ rejection. The hybrid tracers show dispersions lower than in the case of the single-band tracers. These hybrid tracers show the same pattern as the ones in Figure \ref{sfr_update_fuv}, the a$_{IR}$ coefficient decreases when galaxies hosting type-2 AGN are considered. These calibrations and the ones in Figure \ref{sfr_update_fuv} show that applying an energy balance approximation is a good approach to obtain reliable SFR tracers for integrated measurements of nearby galaxies. Here, we use IFS data for the first time to achieve this goal.}
\label{sfr_update_ha}
\end{figure*}

With regard to (1), we must keep in mind that the estimation of the FUV attenuation is challenging so several methods have been put forward for that purpose. The most common approaches are the use of the $\beta$ slope of the UV continuum (similar but not identical to the FUV$-$NUV color) or the ratio of IR to UV luminosity. We obtained the FUV$-$NUV-corrected FUV luminosity (using the $\beta$ slope) in Section \ref{Single-band SFR tracers}, but the results show that with this method recovering the SFR is complicated, especially for the highest values of the SFR where the attenuation appears to be underestimated. Other expressions for the FUV attenuation using the FUV$-$NUV color can be found in the literature \citep[e.g.,][]{Kong_2004,Seibert_2005,Salim_2007,Hao_2011}. As an example, \cite{Hao_2011} use their own expression for the A(FUV) (equation 16 in their paper) to calibrate the TIR + FUV hybrid tracer obtaining that the FUV$-$NUV-corrected FUV luminosity also underestimates the highest SFRs. As explained before, the other way of deriving the attenuation is using the ratio of IR to UV luminosity. We can use the IRX-corrected FUV luminosity to calibrate the TIR + FUV and the 22\,$\mu$m + FUV hybrid tracers as done in sections 4.1 and 4.3 in \cite{Hao_2011}. The authors obtain a relation (equation 13 in their paper) between the IRX and the FUV$-$NUV observed color.  An important parameter that appears in the former equation is the a$_{IR}$ value linked directly to the IRX by A(FUV) $=$ 2.5log(1+a$_{IR}$ $\times$ 10$^{IRX}$). \citep[See][equation 2, where the authors name it a$_{FUV}$ instead of a$_{IR}$; we used a$_{IR}$ for consistency along this work]{Hao_2011}. For the IRX, they use the definition of \cite{Meurer_1995}: IRX = log[{L(TIR)}/{L(FUV)$_{obs}$}]. In order to see the differences from Hao et al.'s sample and this work, we have derived our own IRX$-$$\beta$ relation for the galaxies that have FUV and TIR measurements in the CALIFA mother sample including SF and type-2 AGN host galaxies (see Figure~\ref{irx_beta}). We use the intrinsic FUV$-$NUV color obtained by \citet{Armando_2007}, 0.025 $\pm$ 0.049 mag, which is very similar to the one obtained by \citet{Hao_2011}, 0.022 $\pm$ 0.024 mag. Black line shows Hao et al.'s fit and the green line is our own fit. This fit gives us a value for the coefficient a$_{IR}$ of 0.33 $\pm$ 0.08, in comparison with their value a$_{IR}$ $=$ 0.46 $\pm$ 0.12. Notice that there is a large dispersion in the previous figure, so even taking into account that our measure of the a$_{IR}$ coefficient is in good agreement with the one found by \cite{Hao_2011}, we trust more on method (2) as we discuss next.

Finally, method (2) relies on anchoring the data we want to calibrate to other SFRs measurements, i.e. we establish a reference SFR against which we can compare the hybrid tracers. One possibility would be to use the SFR provided by the extinction-corrected Pa$\alpha$ line emission \citep[see][]{Calzetti_2007}. This line is only moderately influenced by dust extinction and gives us a good measure of the current SFR. The problems related with this emission line are due to its faintness and the difficulty of observing a large number of nearby galaxies as it is only accessible from space.

In our case, we are going to use the extinction-corrected H$\alpha$ SFR tracer measurements obtained for the first time from IFS data as a reference. These data are required to obtain a proper estimation of the stellar continuum and therefore, to estimate a reliable measurement of the ionized-gas dust attenuation via the Balmer decrement. Besides, we count with an homogeneous large survey that provides us with good statistics on the properties of nearby star-forming galaxies. For these reasons (and others explained in Section \ref{Origin of the discrepancies among SFR tracers}), we consider this tracer as a robust estimator of the SFR. Using this method we obtain the updated calibrations for FUV + 22\,$\mu$m, FUV + TIR, H$\alpha$ + 22\,$\mu$m and H$\alpha$ + TIR hybrid tracers that appear in Table \ref{hybrids_global_table}.

The resulting hybrid-tracer calibrations obtained using FUV and H$\alpha$ as observed luminosities are shown in Figures~\ref{sfr_update_fuv} and \ref{sfr_update_ha}. The dispersions found for FUV $+$ 22\,$\mu$m, FUV + TIR, H$\alpha$ $+$ 22\,$\mu$m and H$\alpha$ + TIR tracers are 0.23 (0.19), 0.22 (0.19), 0.17 (0.14) and 0.17 (0.15) dex when type-2 AGN are (not) included, respectively. The single best-fitting parameter in each of these plots is the median of the distribution of the coefficients that multiply the corresponding infrared luminosity in each galaxy (a$_{IR}$) to match the SFR based on the extinction-corrected H$\alpha$ luminosity. As explained before, a$_{IR}$ has been obtained using expressions \ref{air_ha} and \ref{air_fuv} and could be found in these plots. The line shown in these figures corresponds to the 1:1 relation in SFR. This line corresponds to the 1:1 line also in luminosity in the case of the H$\alpha$ + IR tracers (Figure~\ref{sfr_update_ha}) but takes into account the different constant for FUV and H$\alpha$ given in Equations \ref{equation_sfr_fuv} and \ref{equation_sfr_ha}, respectively (Figure~\ref{sfr_update_fuv}). 
An interesting result found is a nearly constant difference ($\sim$ 9$\,\%$) in the coefficients of the infrared term, a$_{IR}$, 3.55 (4.52) in the FUV $+$ 22\,$\mu$m and 0.33 (0.40) in FUV + TIR tracers with (and without) type-2 AGN host galaxies being included. A $\sim$ 10$\,\%$ difference also appears when we compare the a$_{IR}$ coefficients between the H$\alpha$ $+$ 22\,$\mu$m, 0.015 (0.018), and H$\alpha$ + TIR, 0.0015 (0.0019), calibrators with (and without) type-2 AGN, respectively. If we compare the ratio between these a$_{IR}$ coefficients for the combinations of 22\,$\mu$m and TIR data with the luminosity ratio expected for infrared SEDs with different interstellar radiation fields, starlight intensities, dust chemical composition, etc., we would predict the ratio of the energy absorbed by dust at $\lambda$ $<$ or $>$ 4000 \citep[see Figure 2 of][]{Cortese_2008}. The most optimal models for carrying out such study, those by \citet{Draine_Li_2007}, assume a specific and fixed shape for the interstellar radiation field (the local one) so the effect of optical photons is hidden in the variation of the factor $\gamma$, which parameterize the fraction of dust heated by intense radiation fields. The comparison of the a$_{IR}$ for 22\,$\mu$m and a$_{IR}$ for TIR coefficients yields a factor of 0.1 between L(22) and L(TIR), which \citep[according to Figure 19 of][]{Draine_Li_2007} corresponds to $\gamma$ $=$ 0.02, quite independently of the fractional abundance of PAHs.

\begin{figure*}
\centering
\includegraphics[width=76mm]{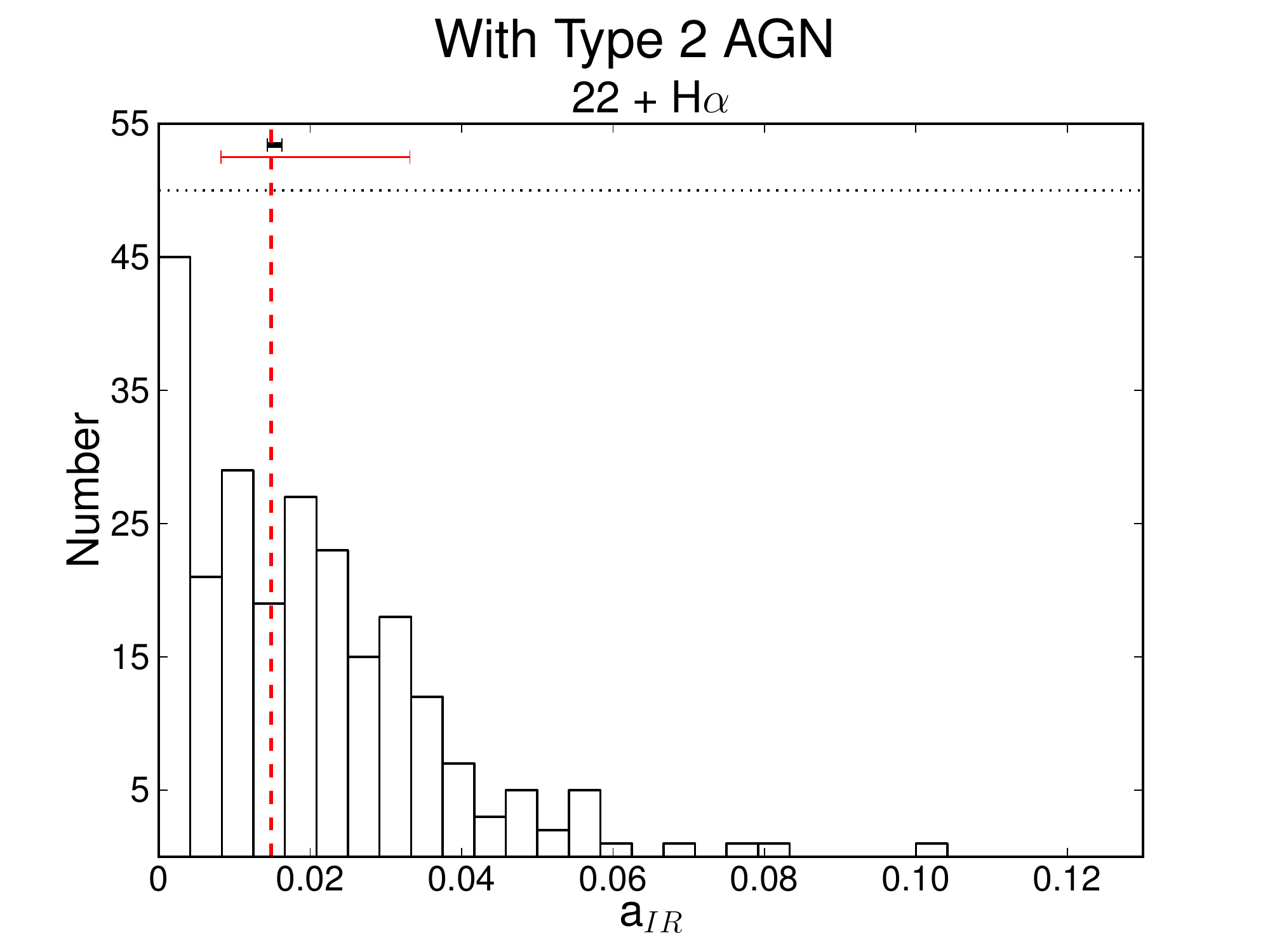} 
\includegraphics[width=76mm]{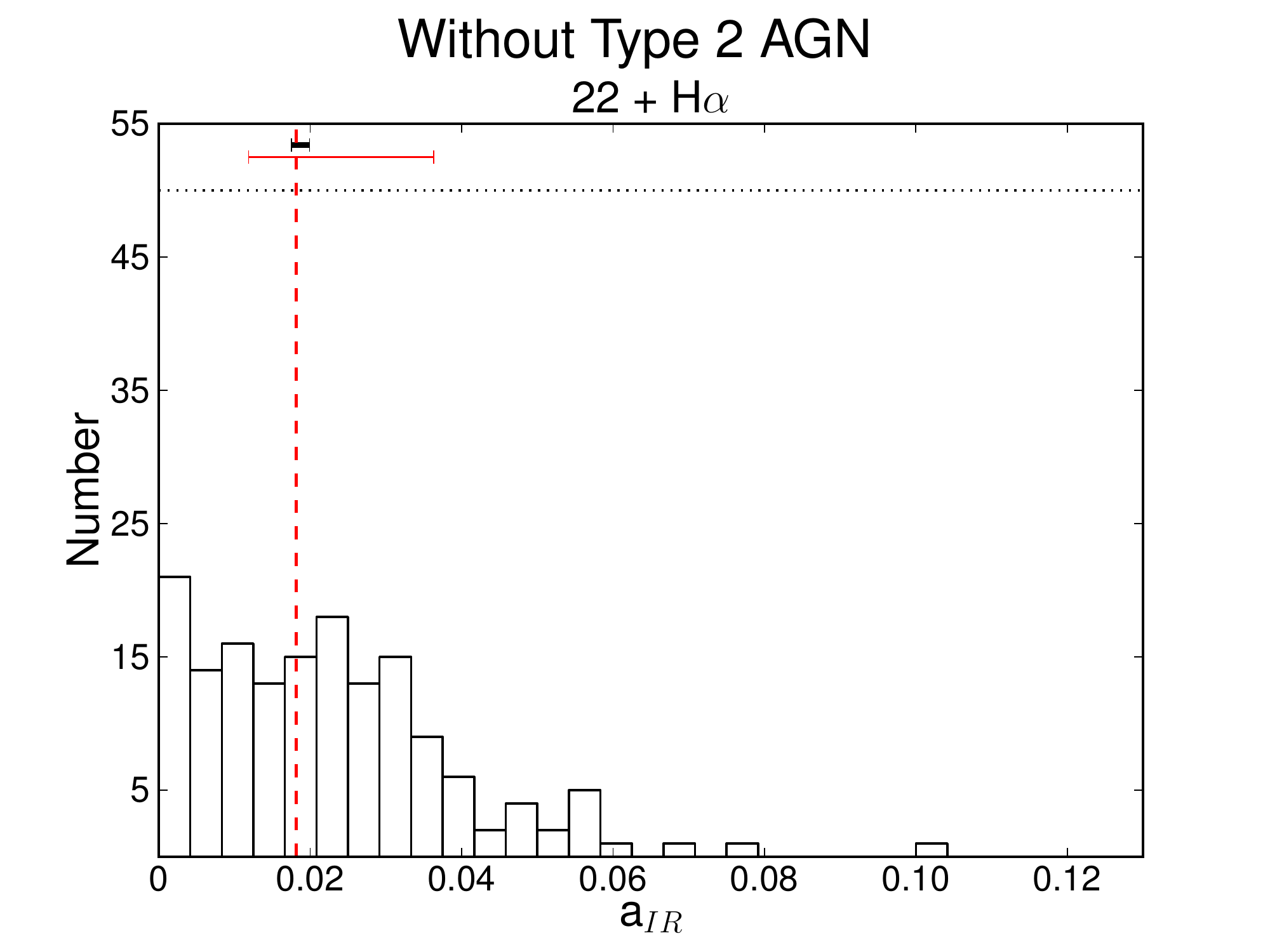} \\
\includegraphics[width=76mm]{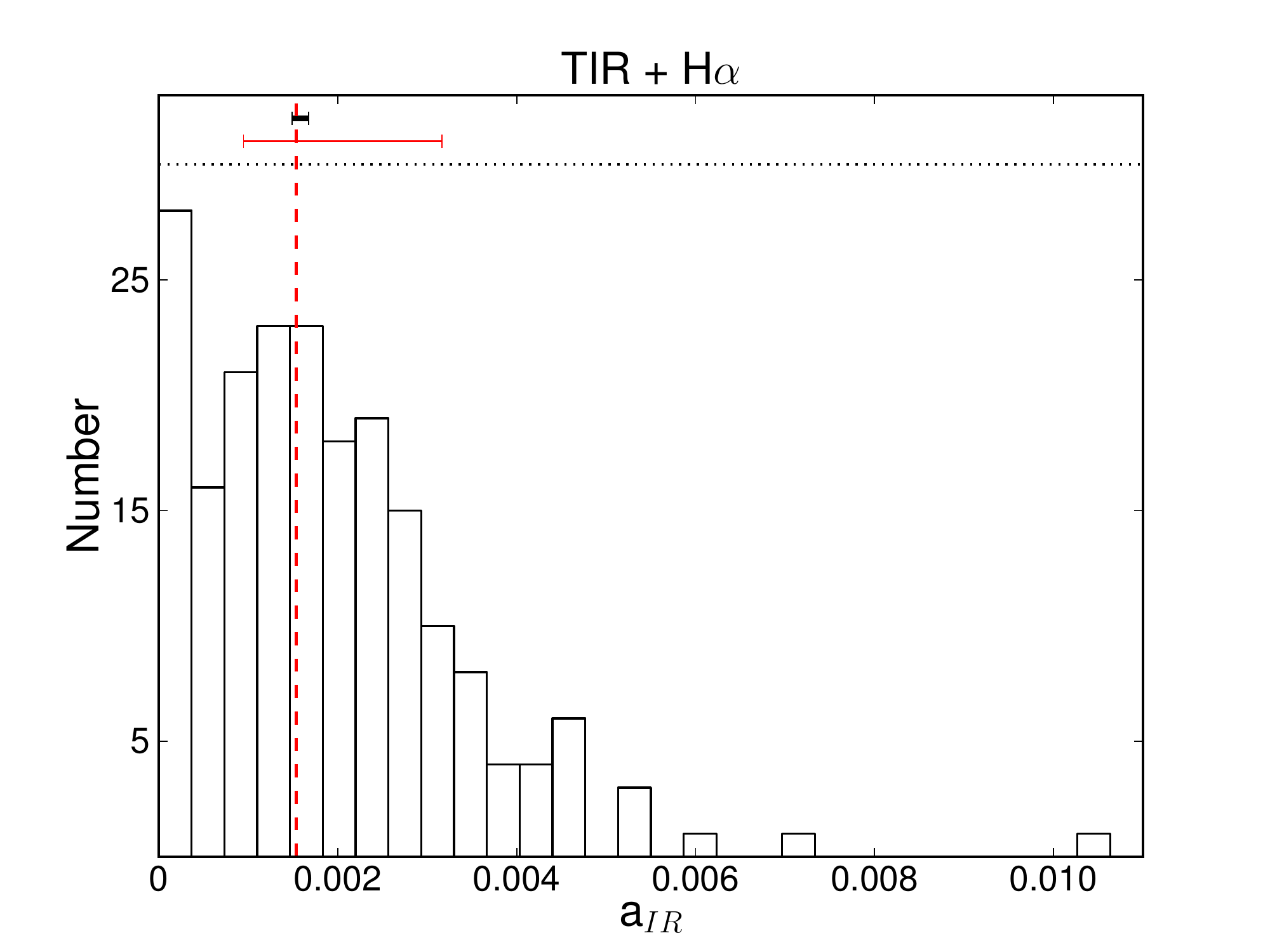} 
\includegraphics[width=76mm]{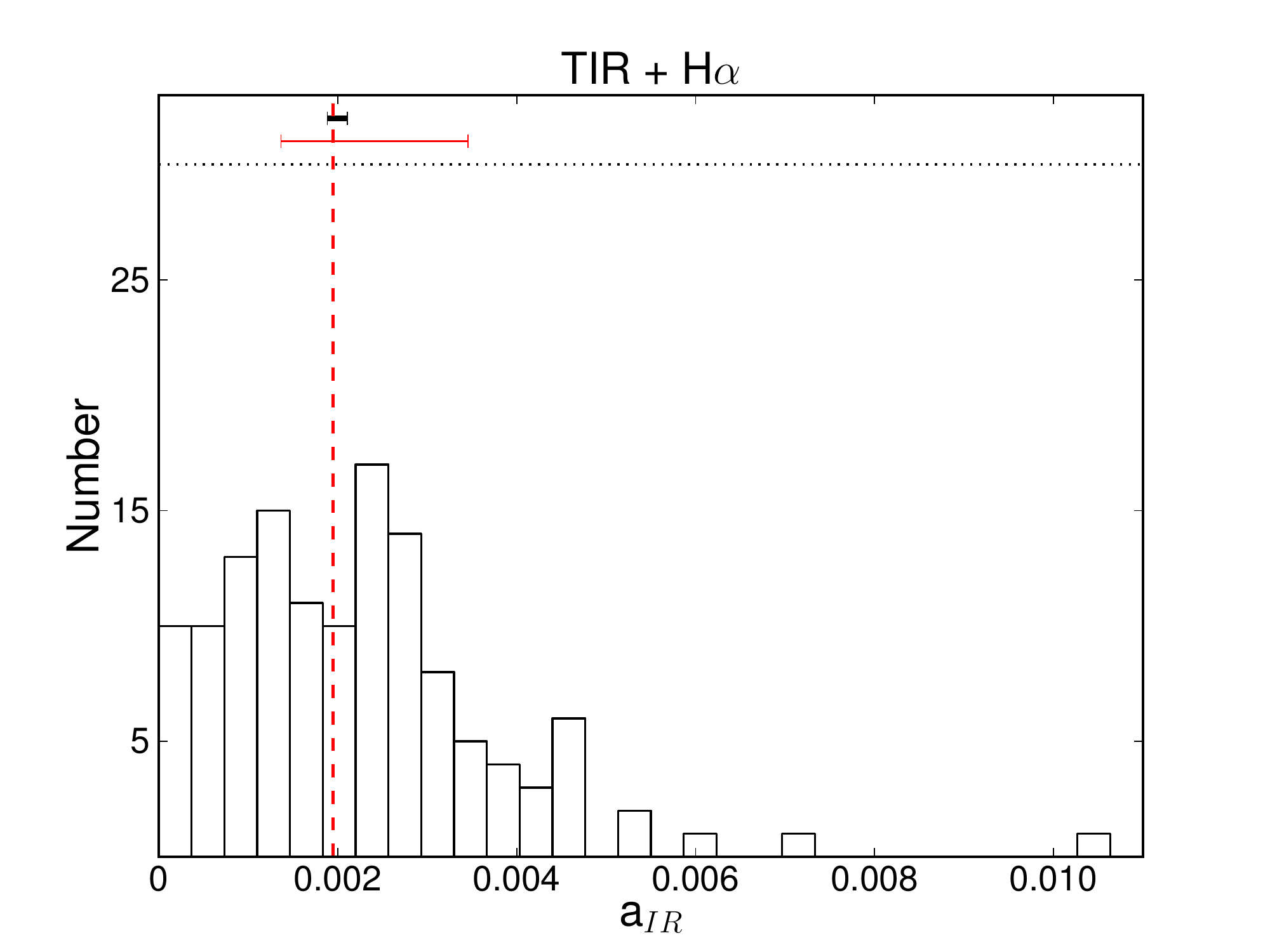} \\
\includegraphics[width=76mm]{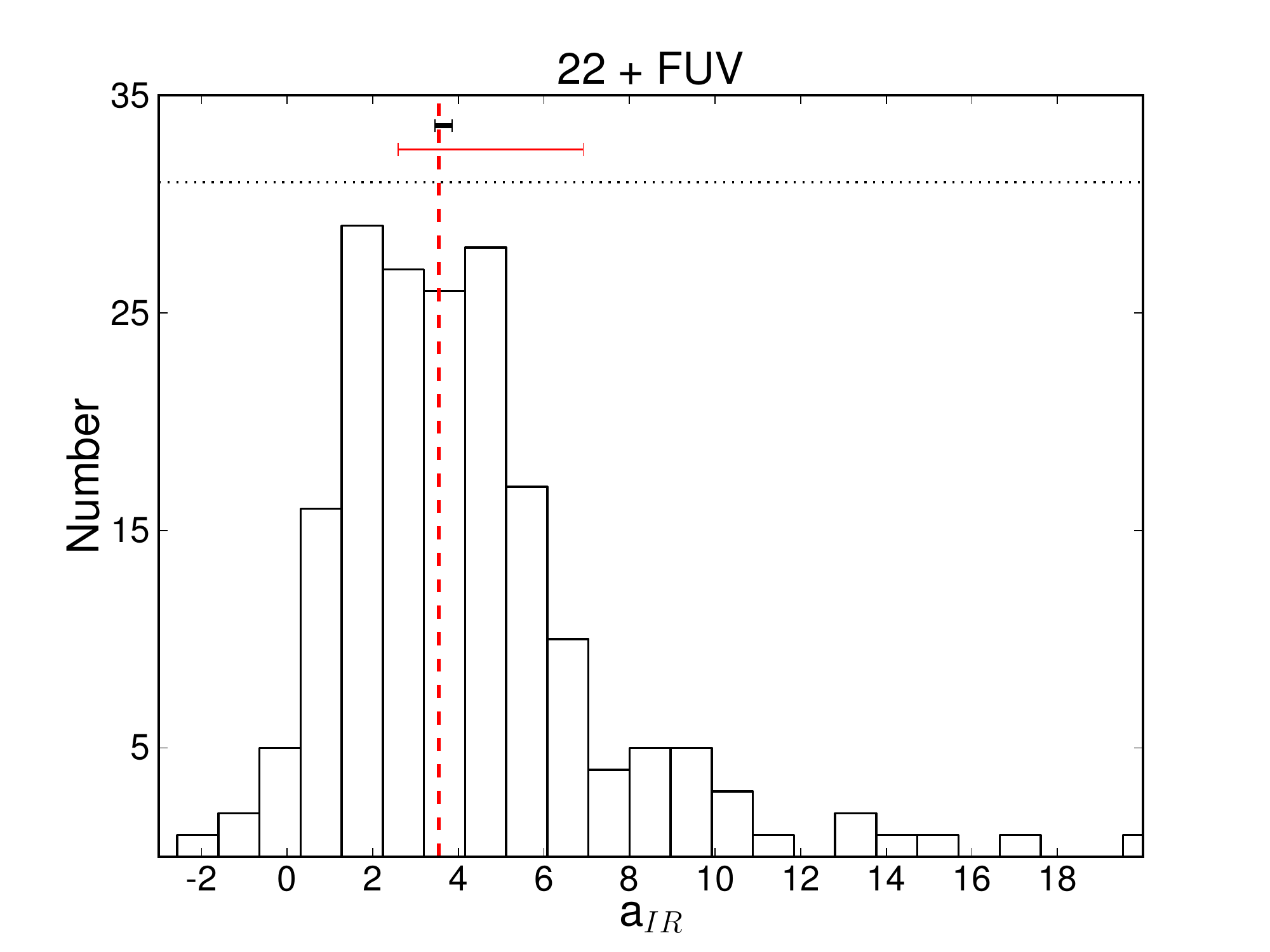} 
\includegraphics[width=76mm]{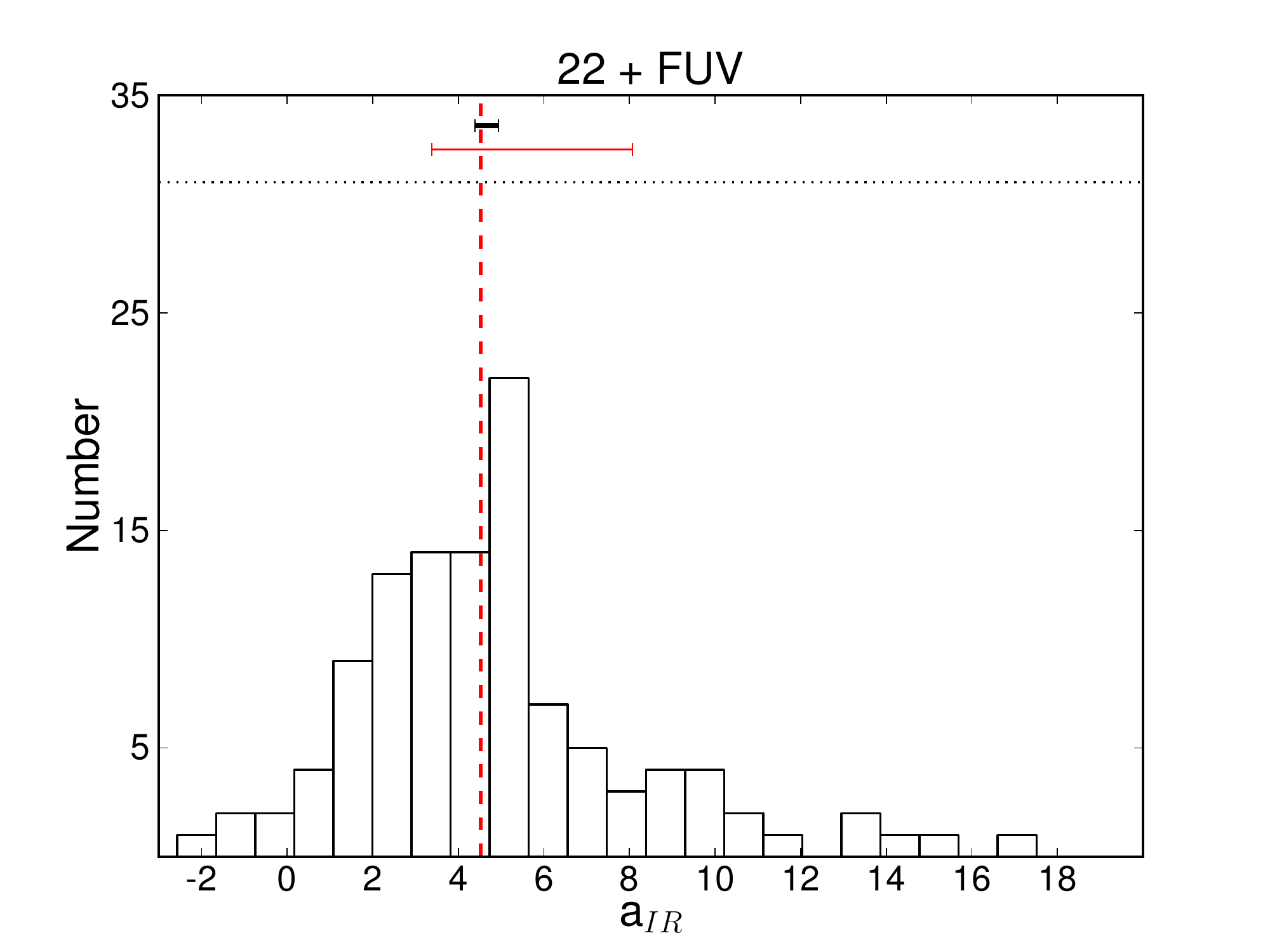} \\
\includegraphics[width=76mm]{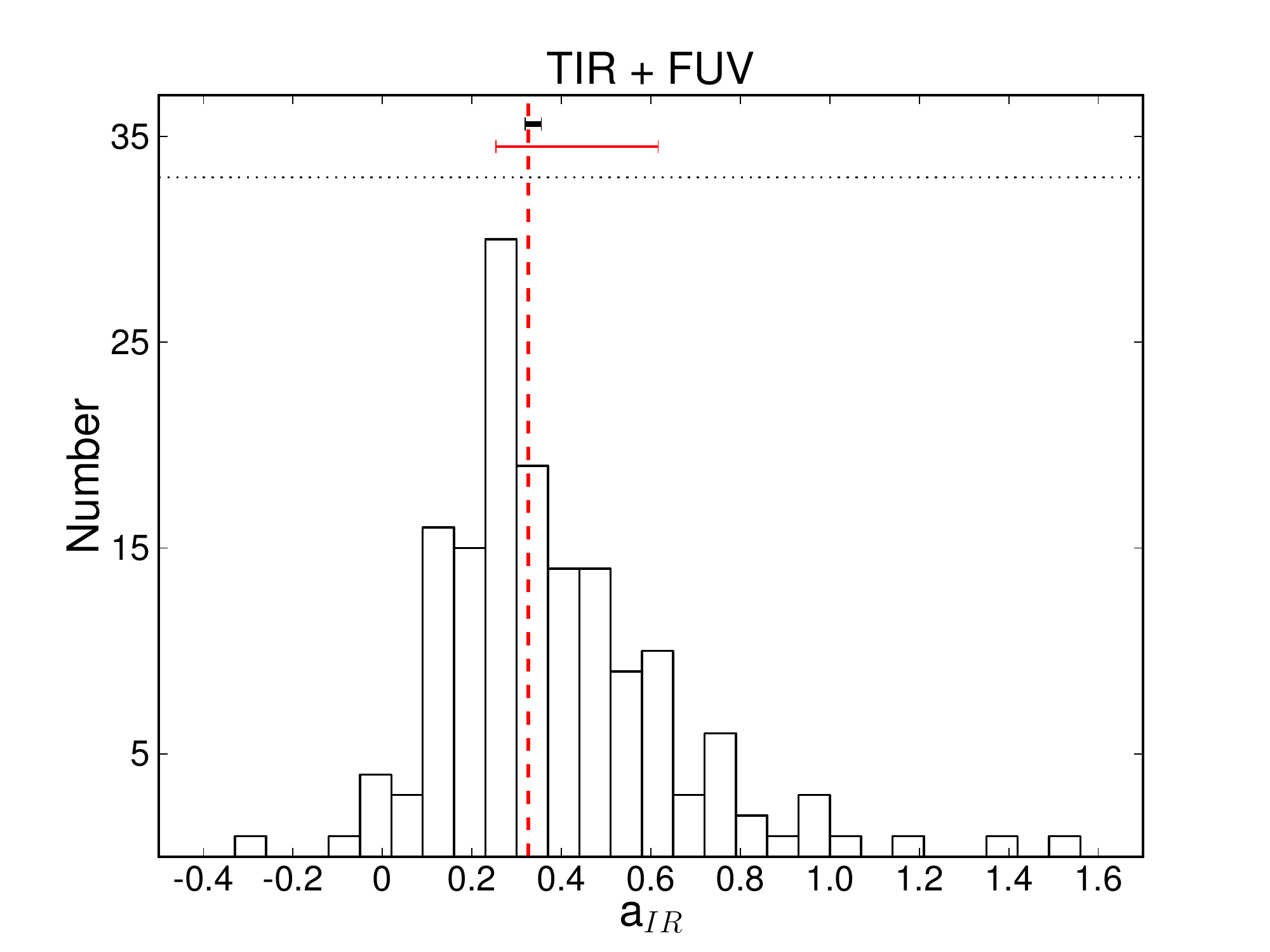} 
\includegraphics[width=76mm]{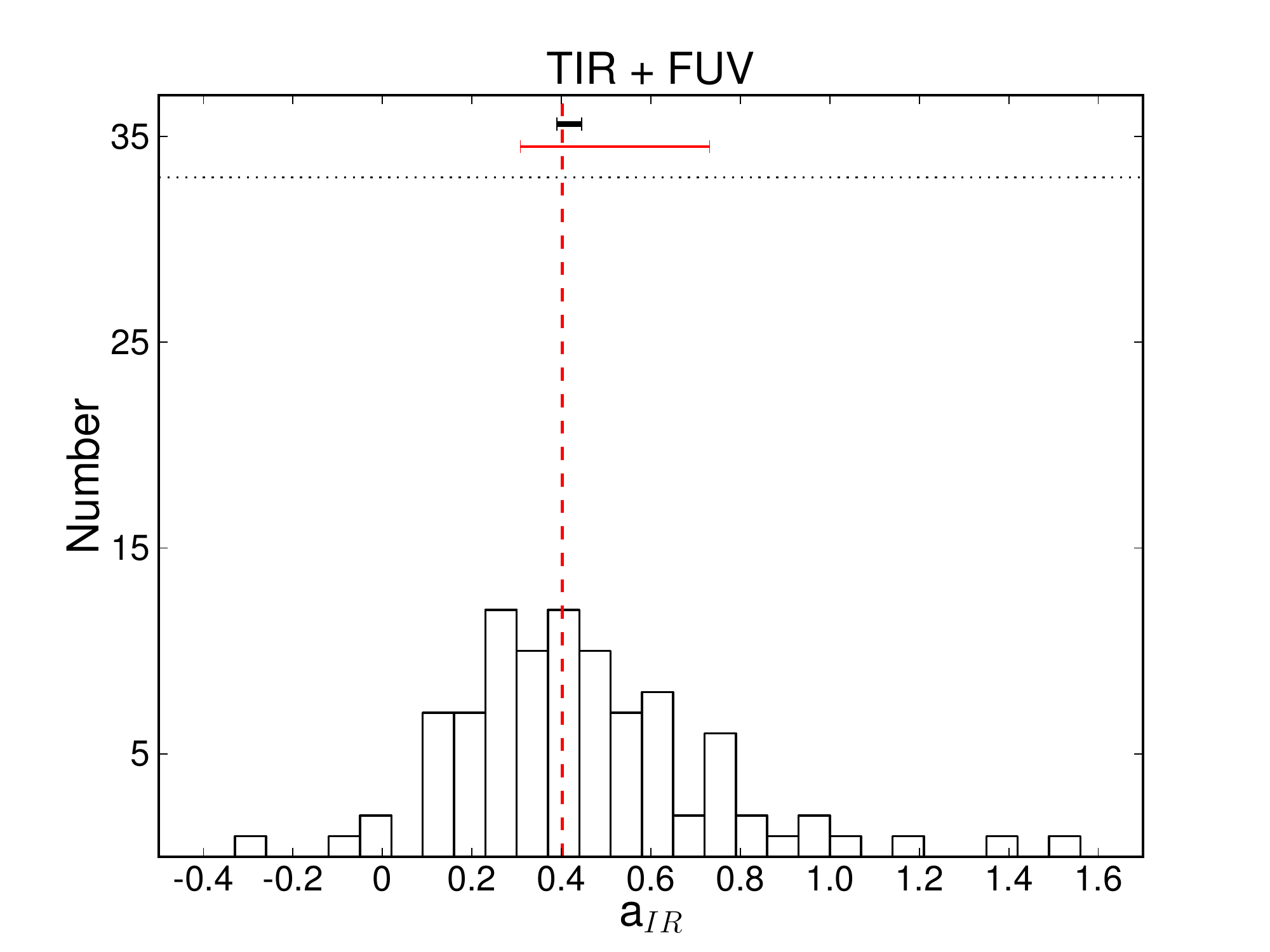} \\
\caption{Histograms showing the distribution of the a$_{IR}$ coefficient values obtained for the different hybrid tracers. a$_{IR}$ is computed using the expressions \ref{air_ha} and \ref{air_fuv}. The red dashed line corresponds to the median value of this coefficient. The red tick-marks shown at the top refers to the 1 $\sigma$ dispersions measured as the interval that includes 68$\%$ of the data points around the median quoted in Table \ref{hybrids_global_table} while black tick-marks indicate the standard error of the median computed from the asymptotic variance formula using these 1 $\sigma$ dispersions. Type-2 AGN galaxies are excluded from the histograms at right.}
\label{histograms_constant}
\end{figure*}

\begin{figure*}
\centering
\includegraphics[width=74mm]{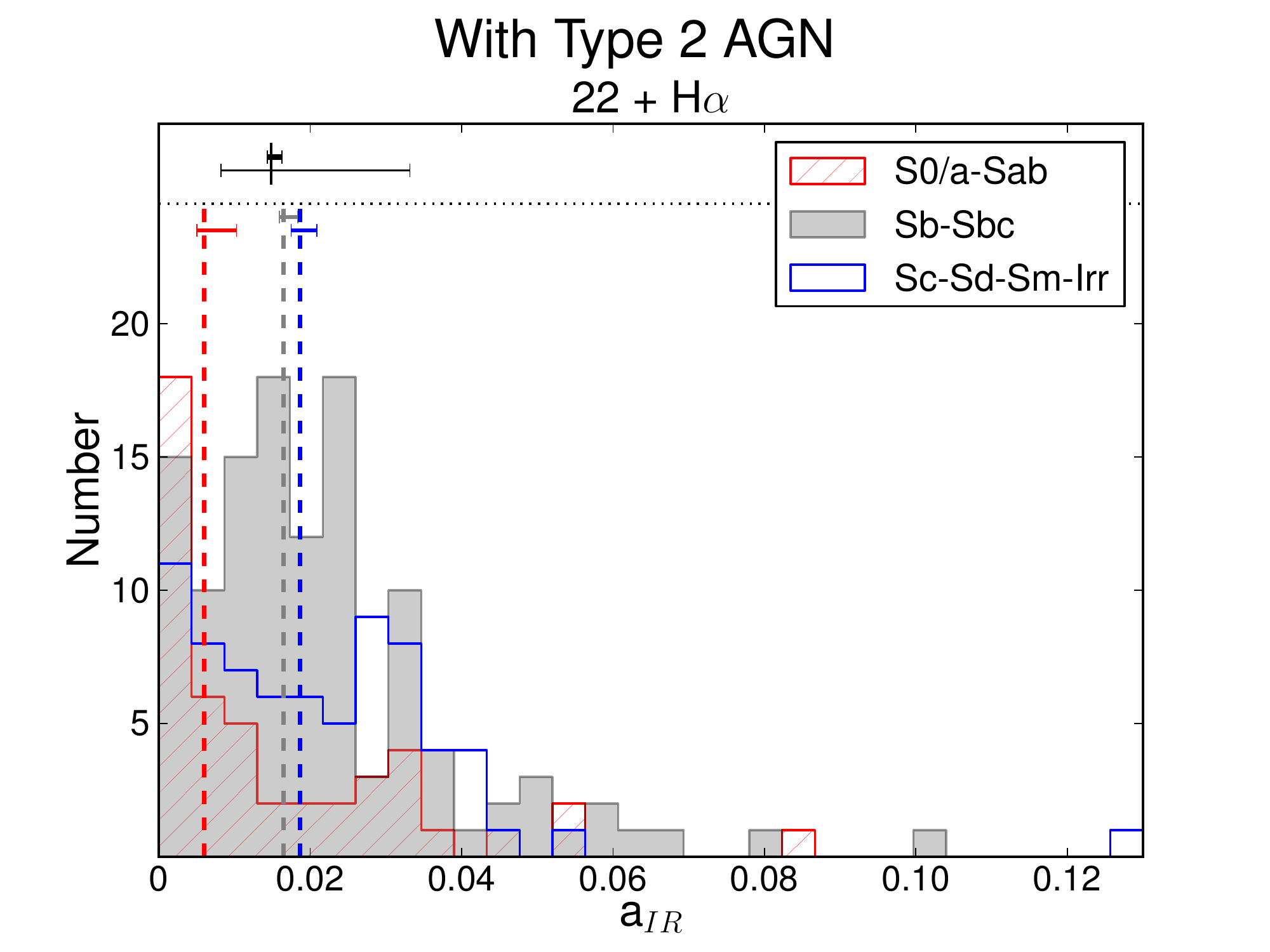}
\includegraphics[width=74mm]{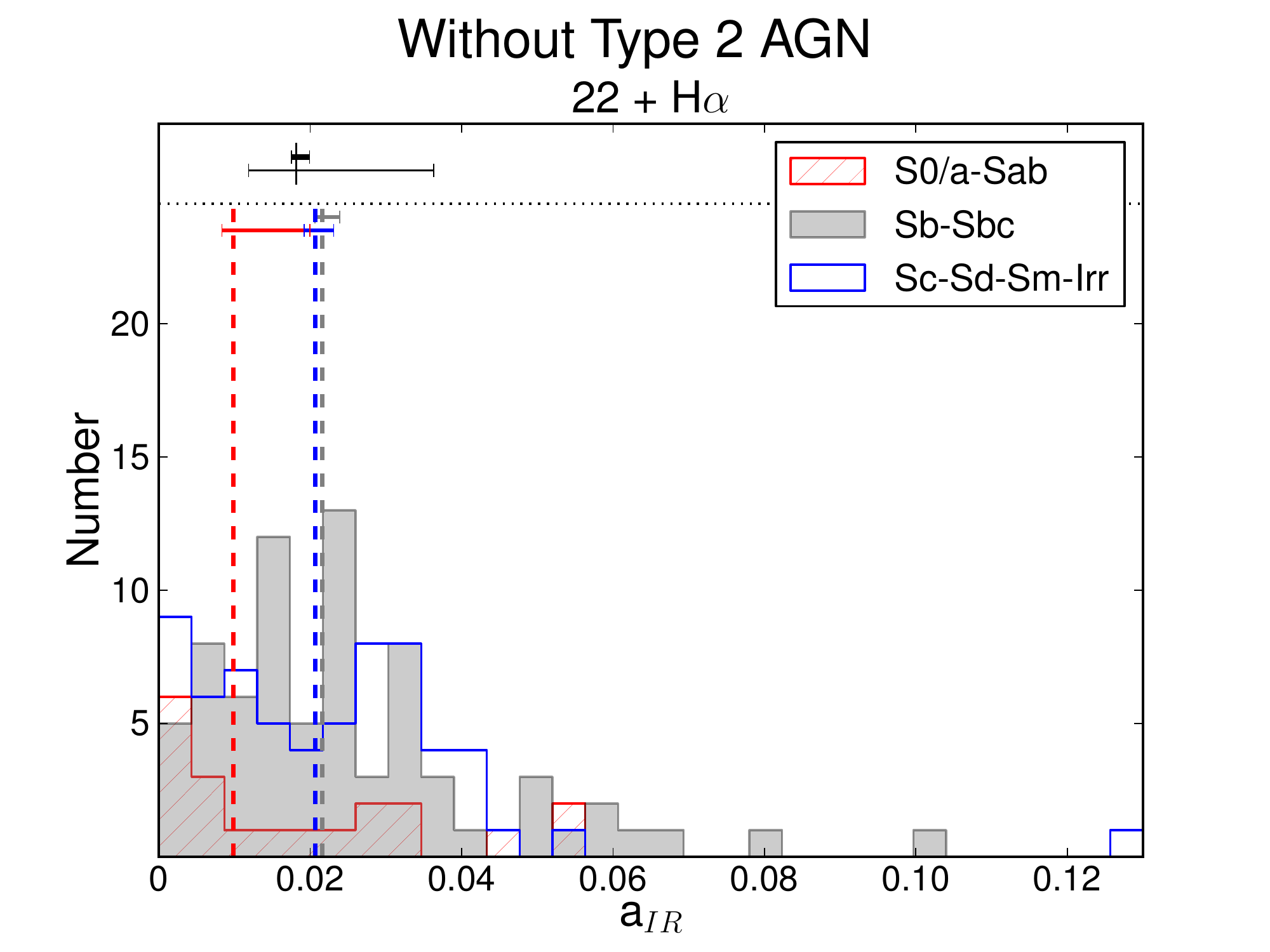} \\
\includegraphics[width=74mm]{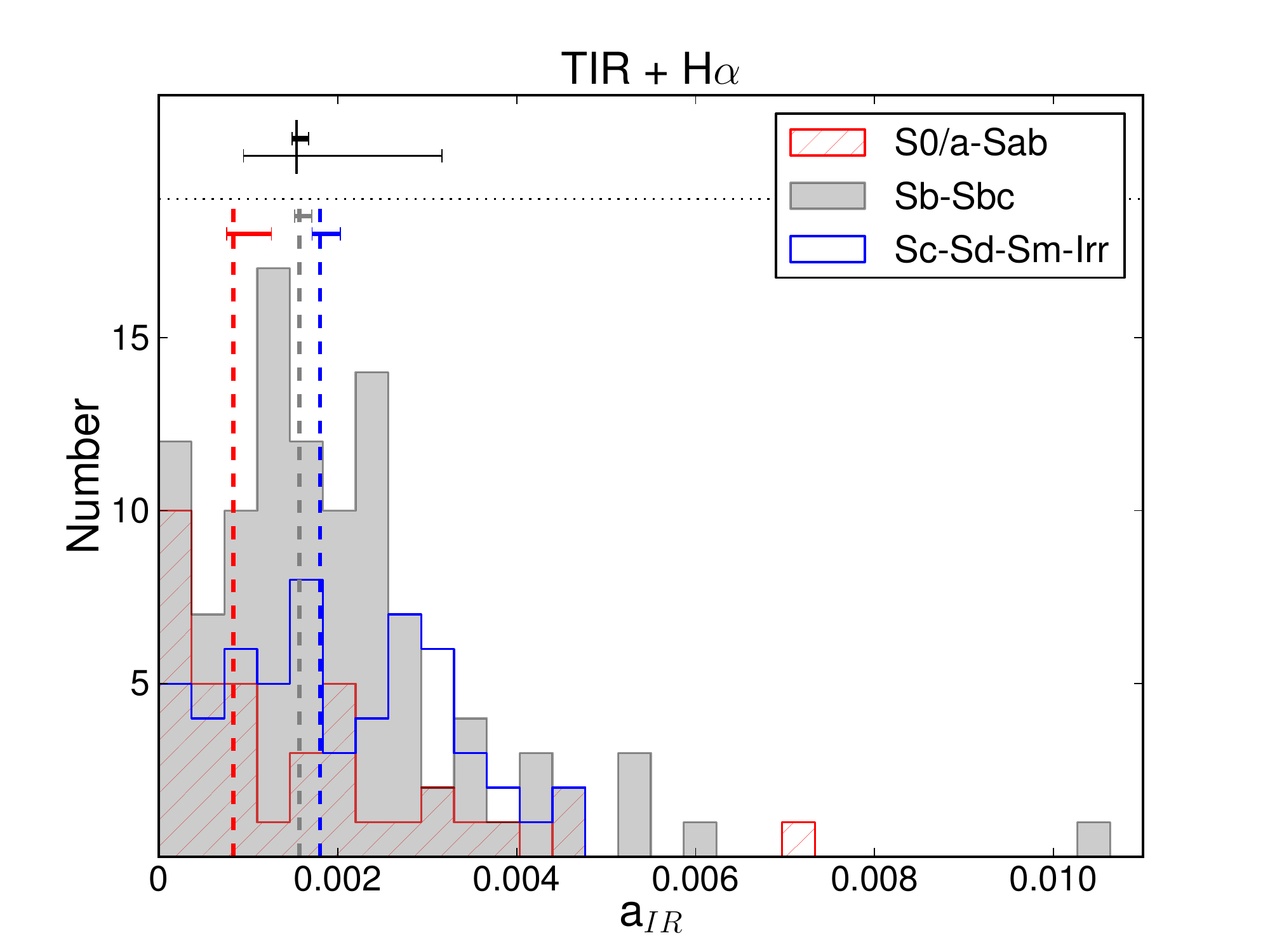}
\includegraphics[width=74mm]{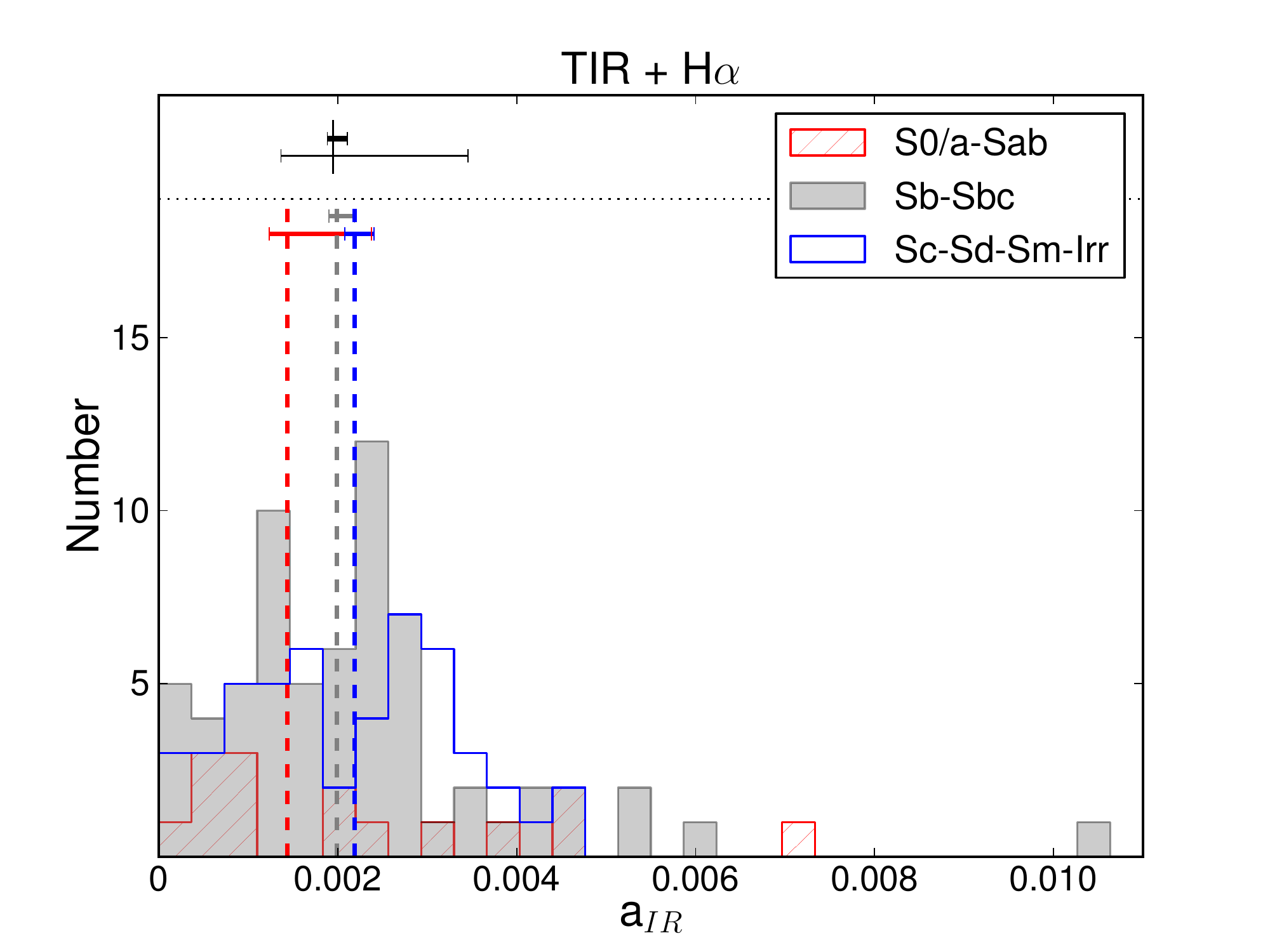} \\
\includegraphics[width=74mm]{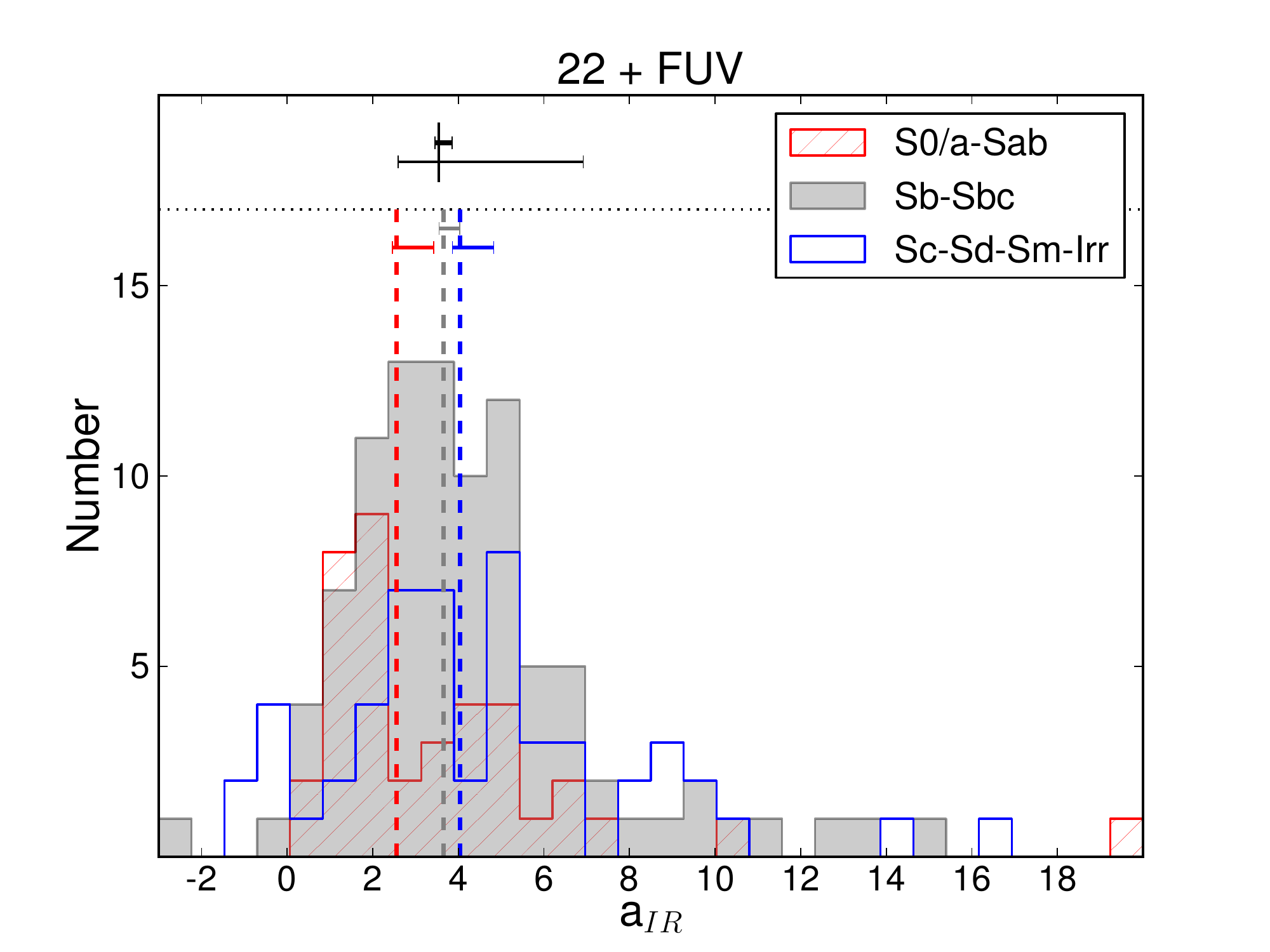}
\includegraphics[width=74mm]{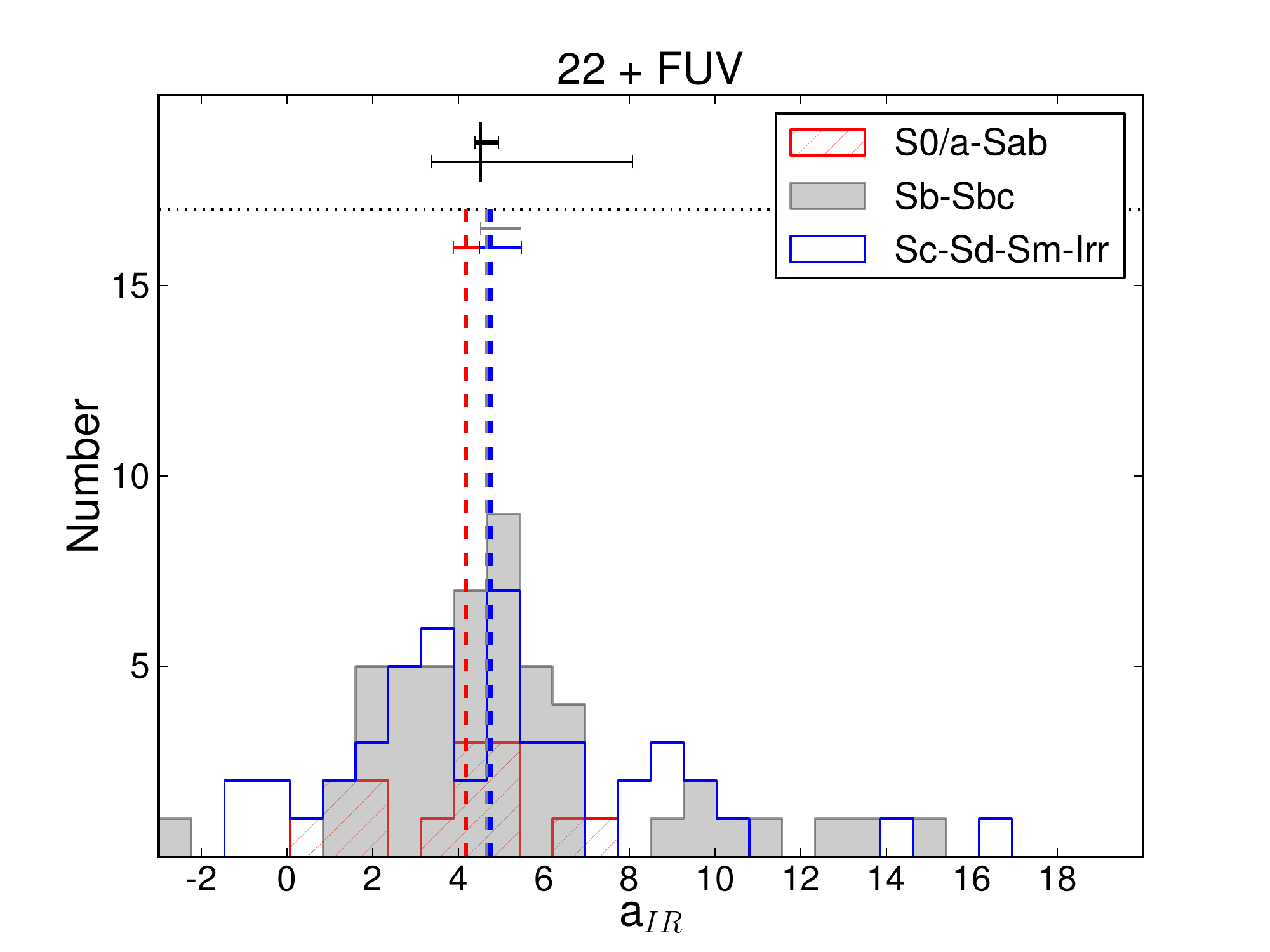} \\
\includegraphics[width=74mm]{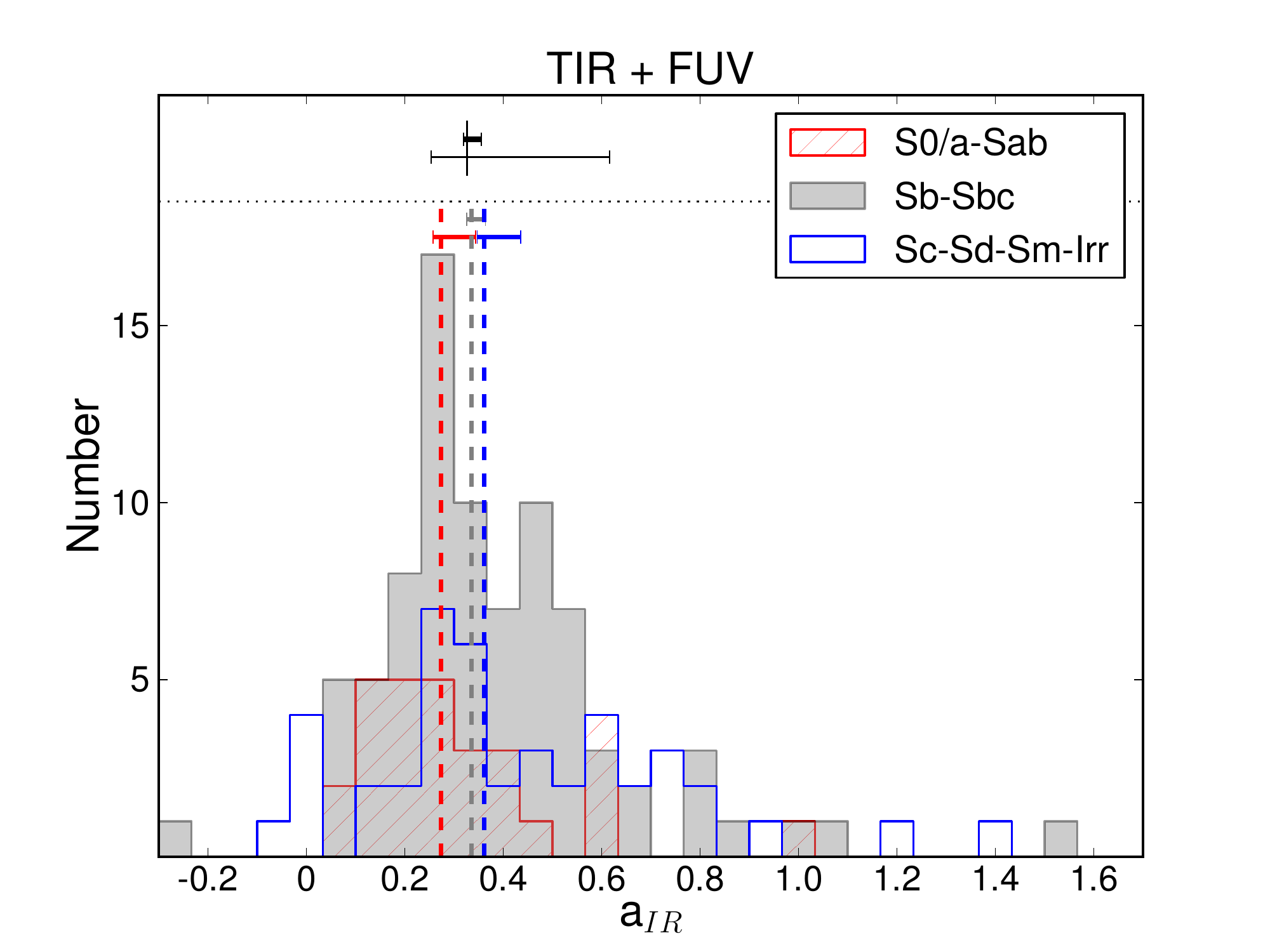}
\includegraphics[width=74mm]{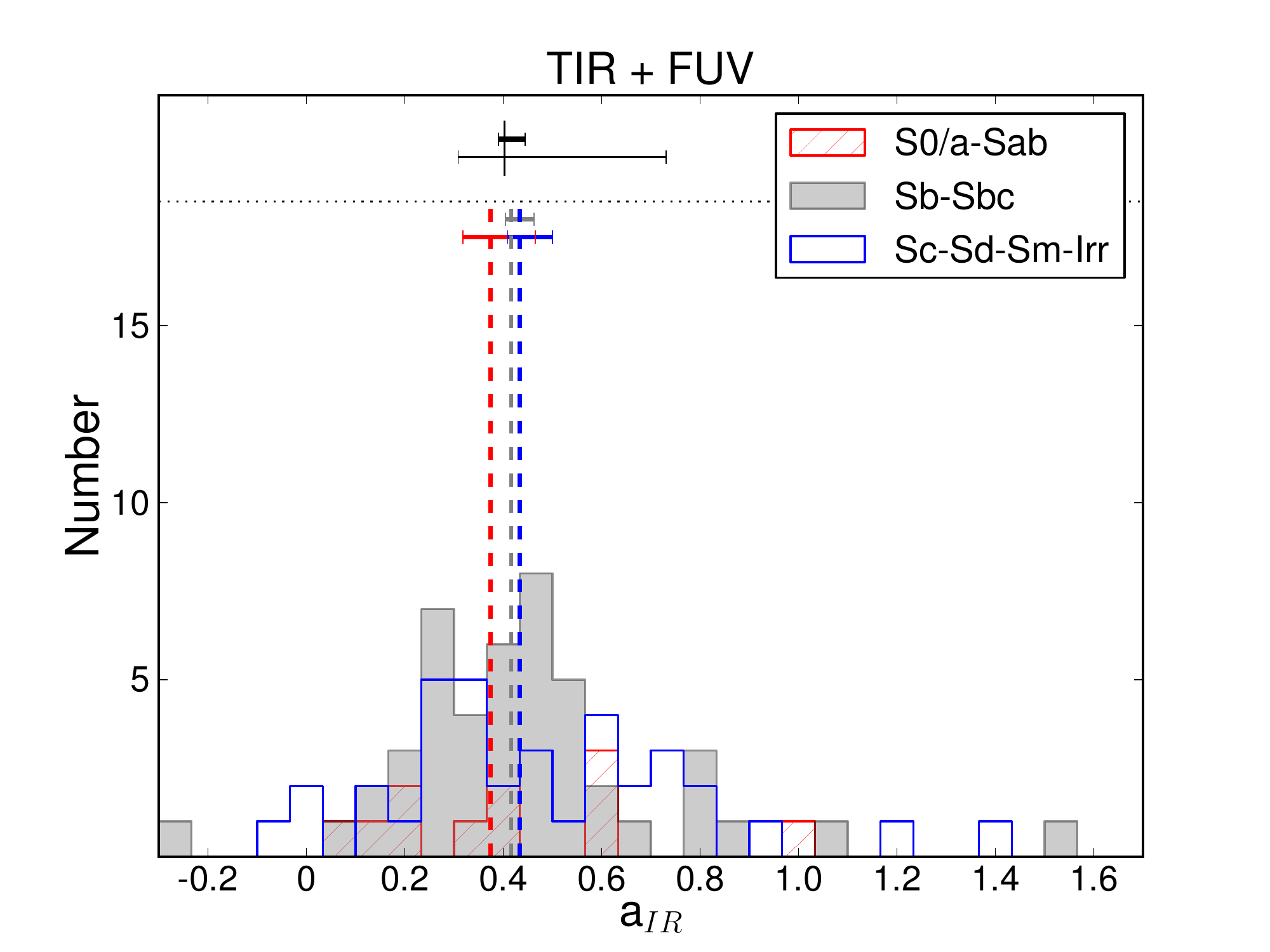} \\
\caption{Four left panels: Histograms showing the values of the coefficient that multiply the IR luminosity, {\it a$_{IR}$}, in the hybrid tracers using bins of morphological types. Early-type star-forming galaxies, considered here as S0/a, Sa and Sab, are shown in red, intermediate-type spirals such as Sb and Sbc appear in grey and Sc-Sd-Sm-Irr galaxies are represented in blue. Vertical dashed lines correspond to the median value of each galaxy group. Black top marks show the median value for all the galaxies as in Figure \ref{histograms_constant}. There is a clear trend with the morphological type, late-type galaxies need a higher value of the {\it a$_{IR}$} coefficient than early-type galaxies. This trend could be explained in terms of the contribution of an obscured AGN, a missing fraction of the H$\alpha$ extinction-corrected SFR or the heating by optical photons. Four right panels: Same histograms as before but removing the type-2 AGN. There is still a trend with the morphological type although less obvious that in the previous case.}
\label{histograms_morphological_type}
\end{figure*}

\begin{figure*}
\centering
\includegraphics[width=73.50mm]{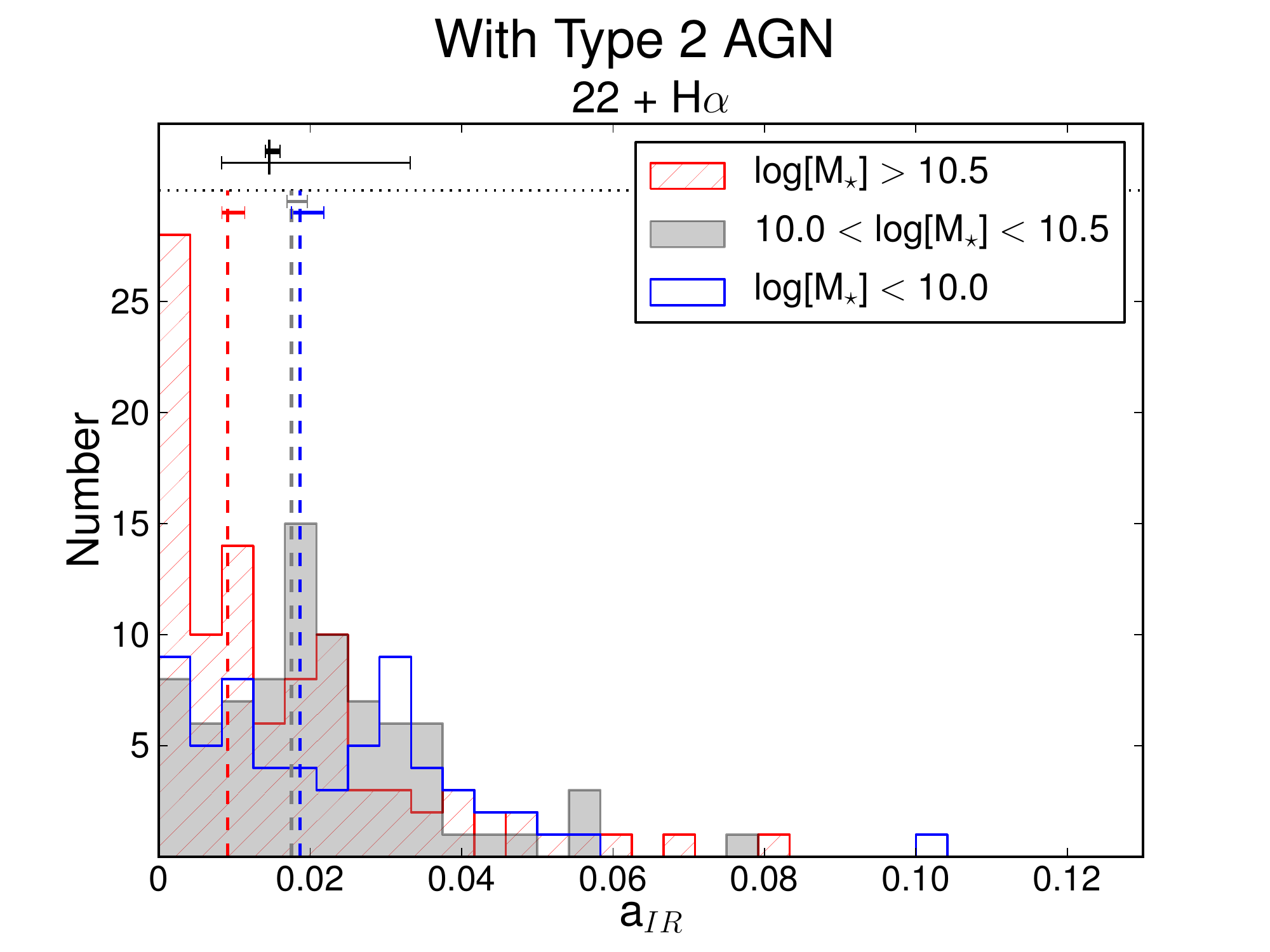} 
\includegraphics[width=73.50mm]{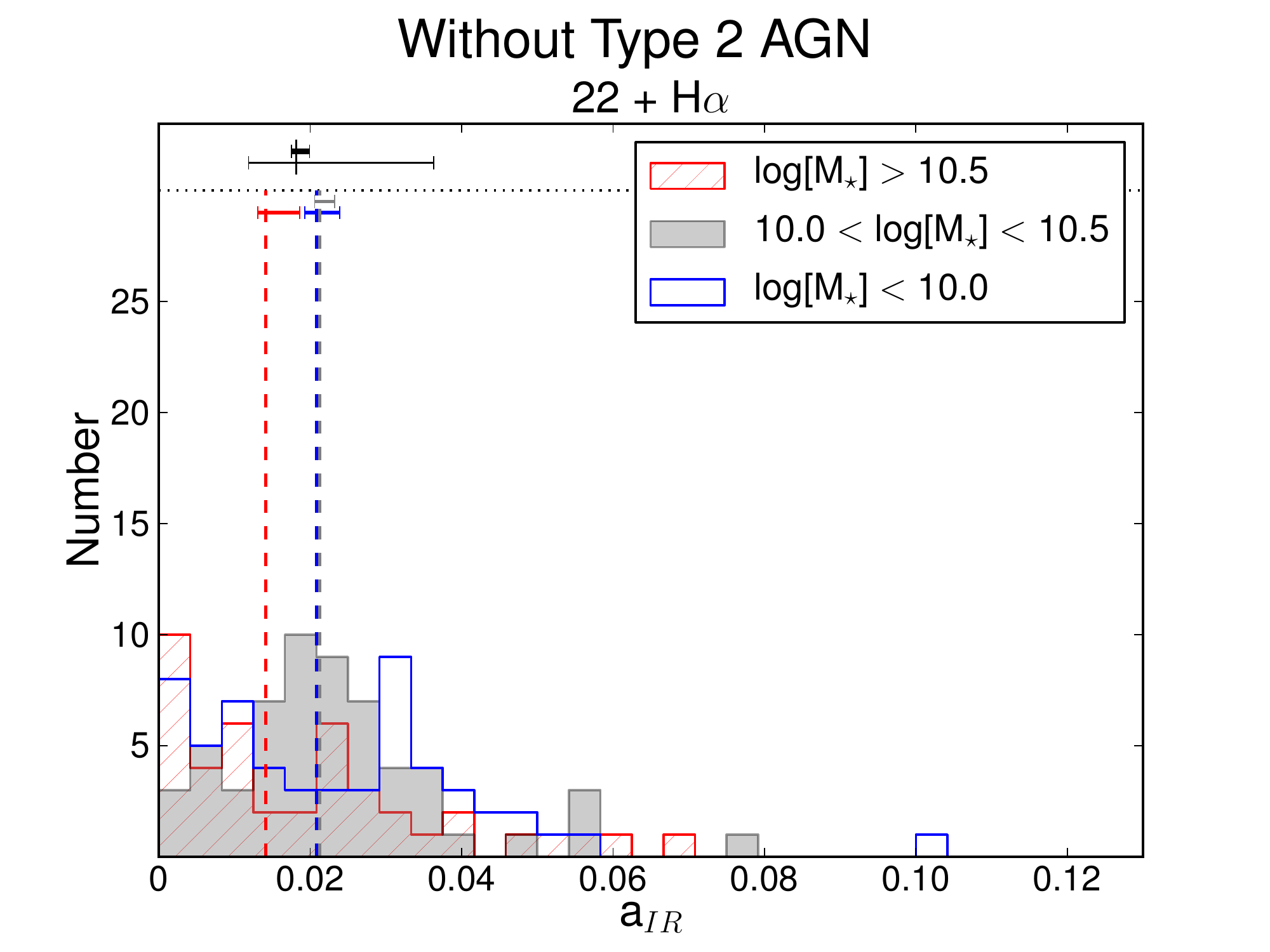} \\
\includegraphics[width=73.50mm]{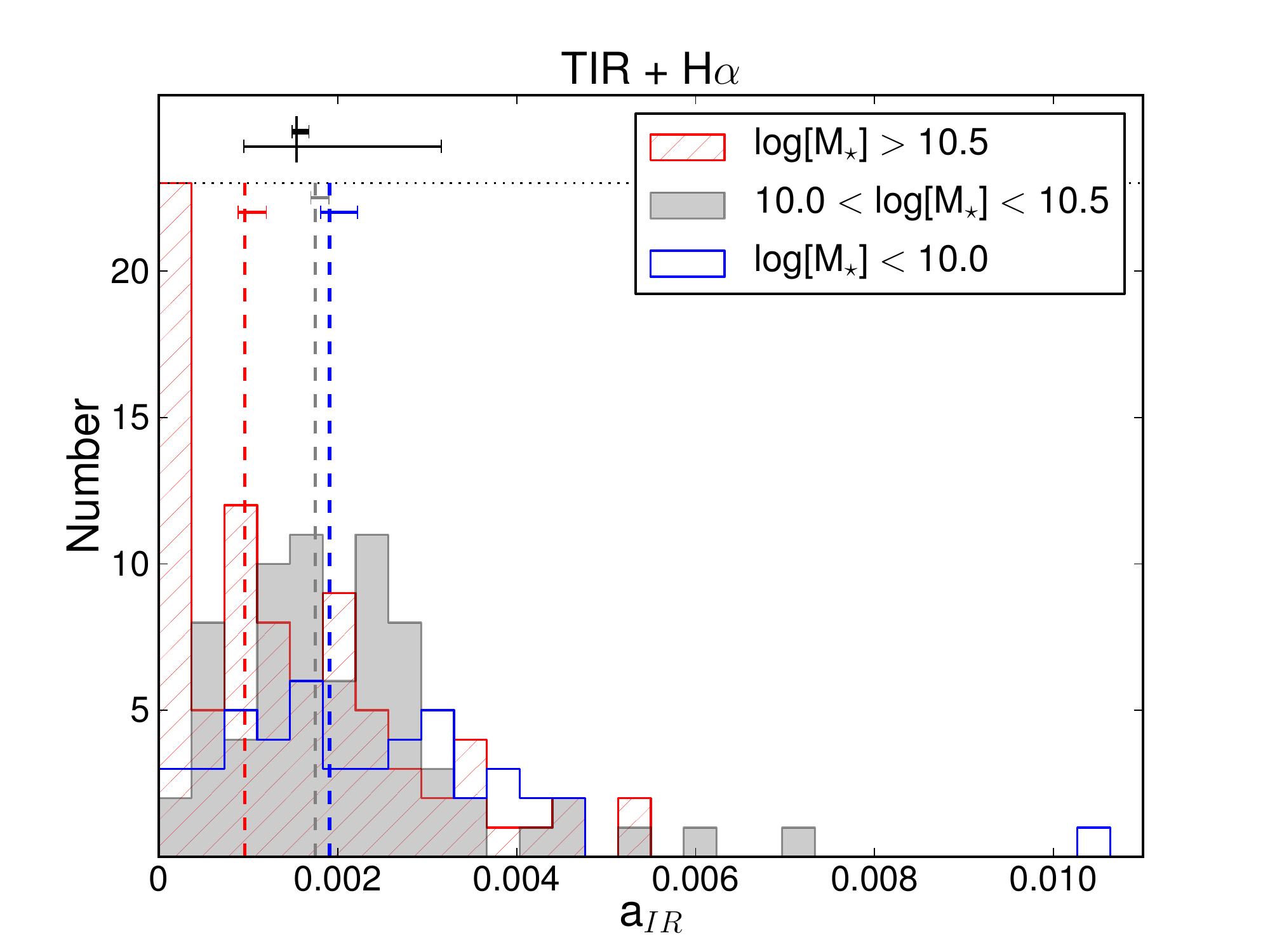} 
\includegraphics[width=73.50mm]{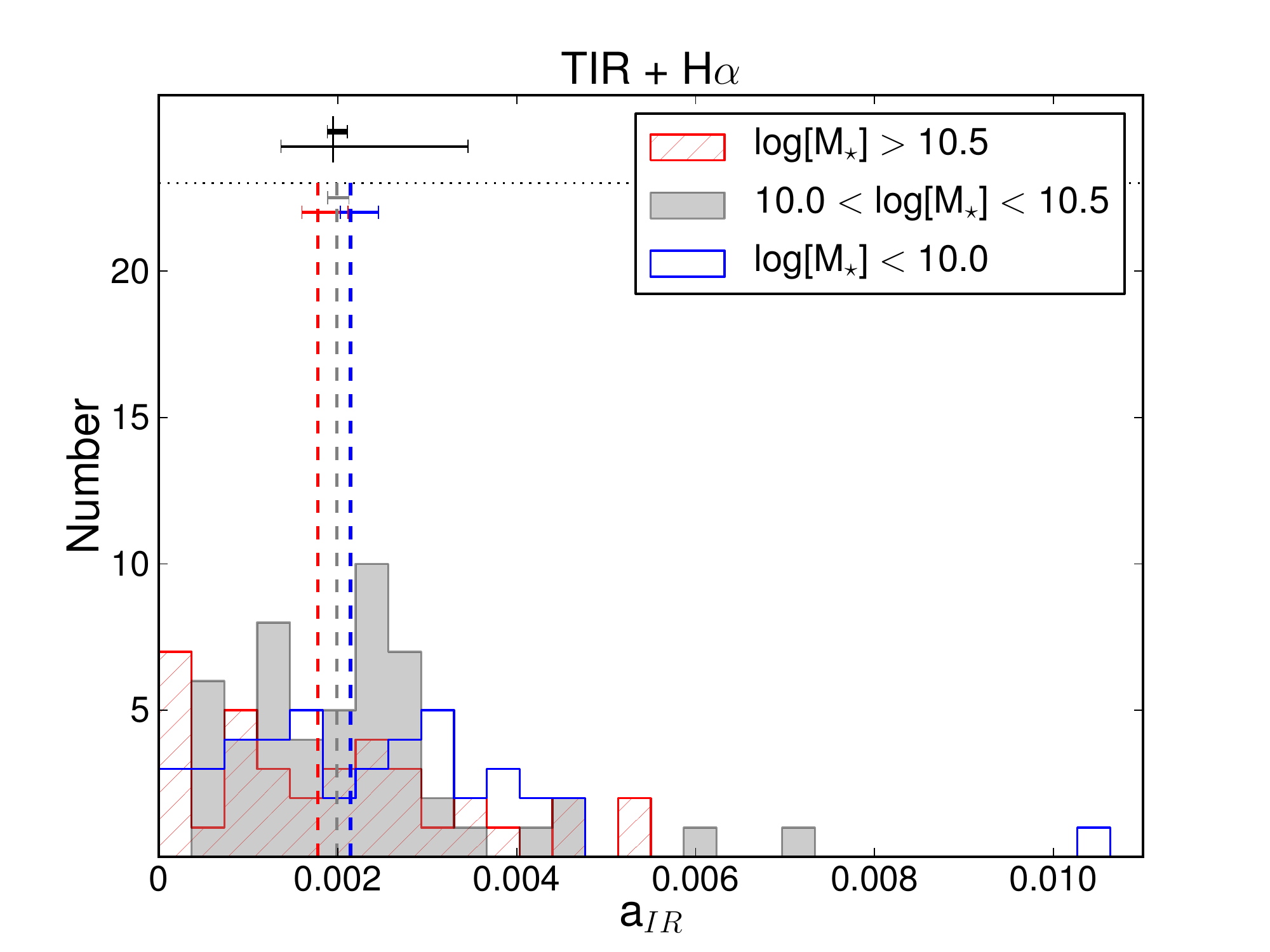} \\
\includegraphics[width=73.50mm]{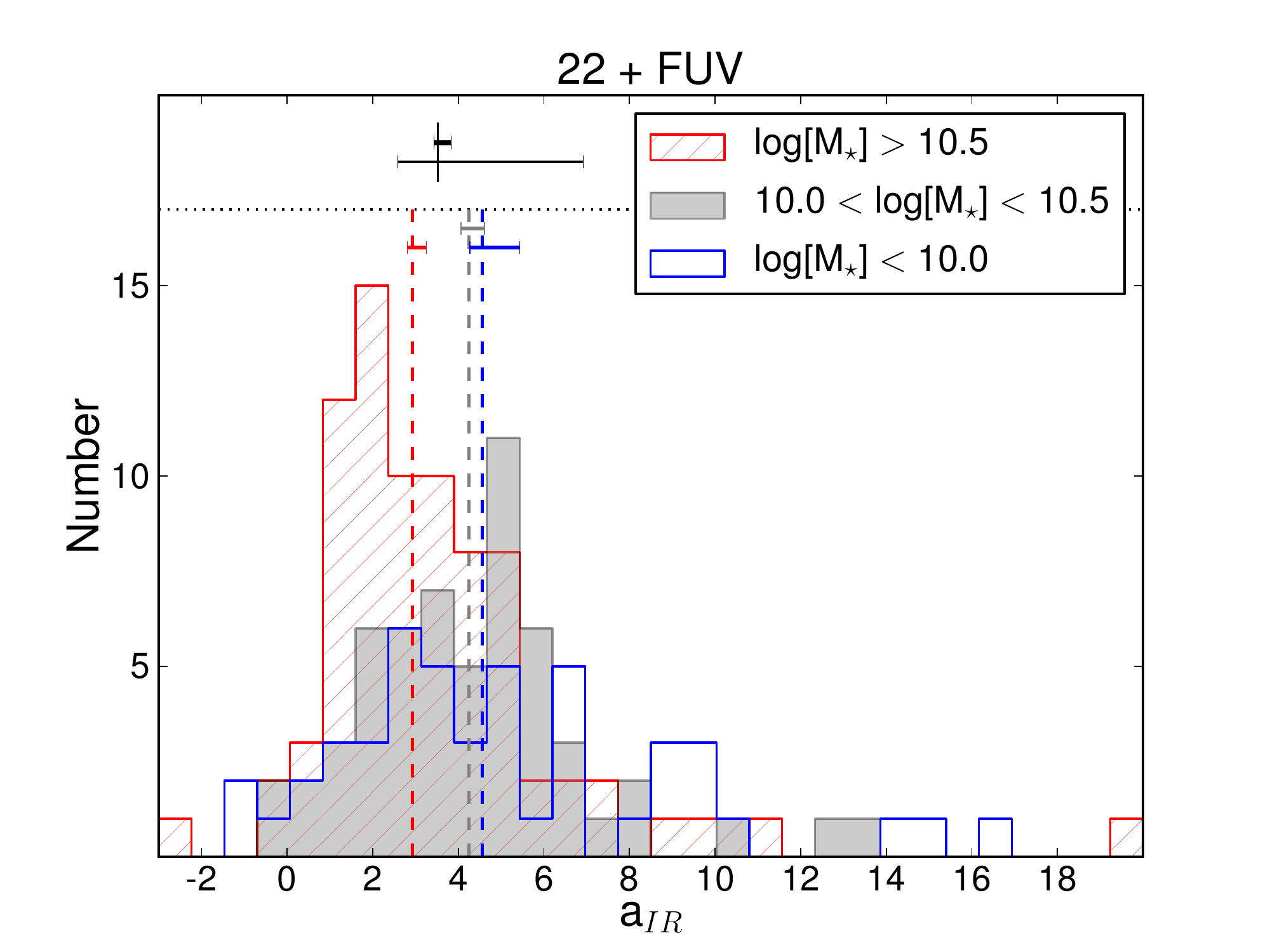} 
\includegraphics[width=73.50mm]{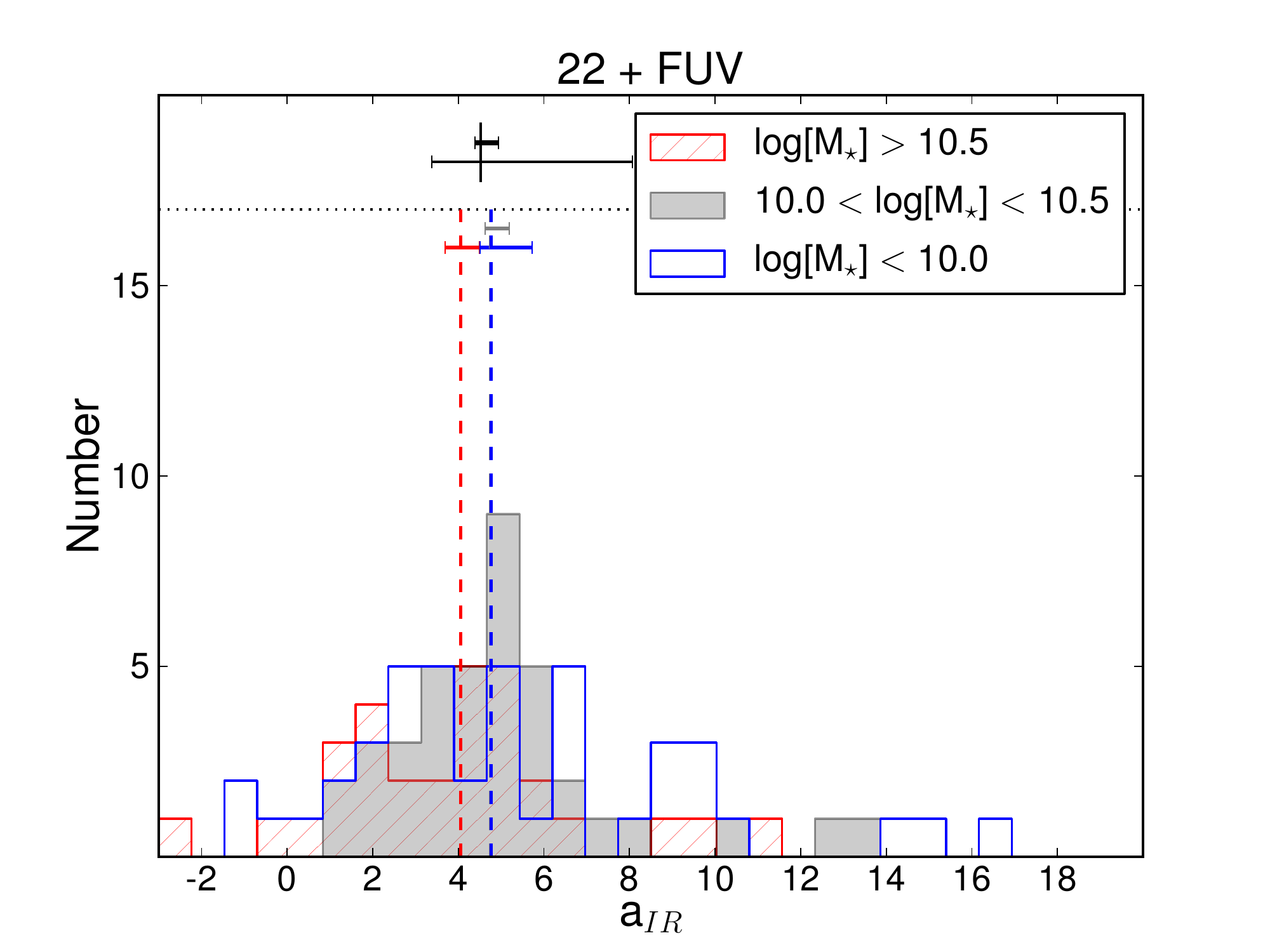} \\
\includegraphics[width=73.50mm]{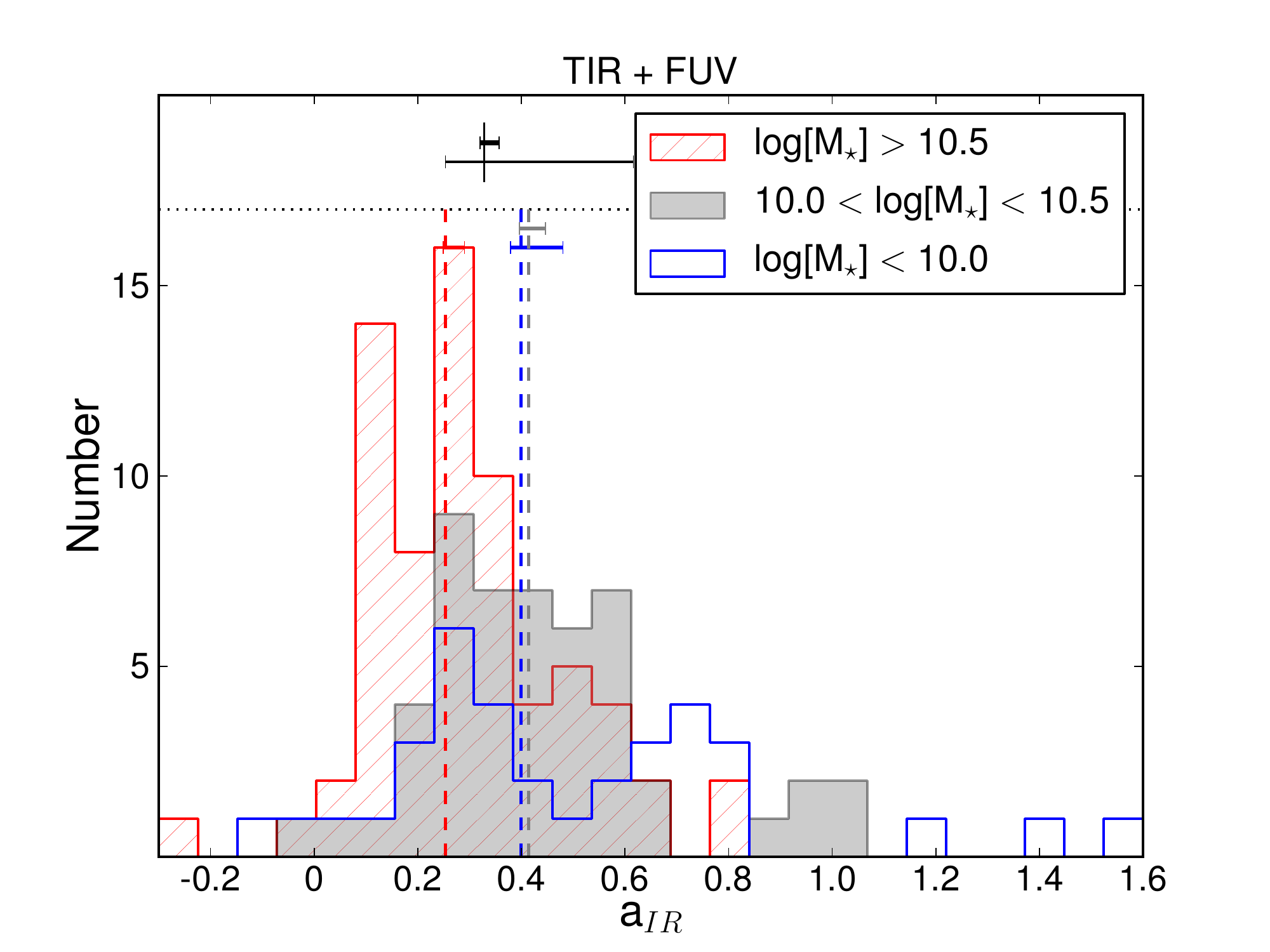} 
\includegraphics[width=73.50mm]{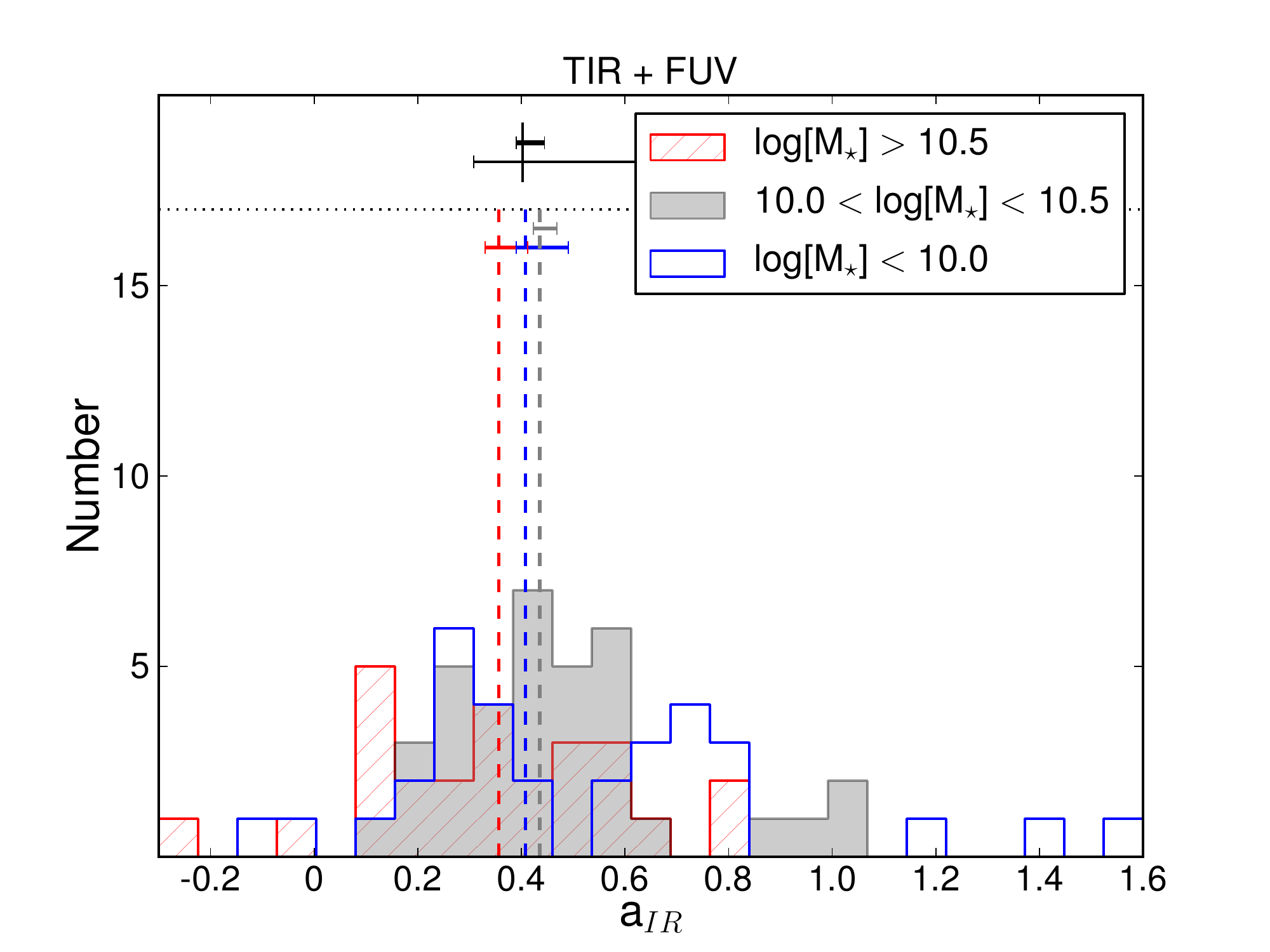} \\
\caption{Left four panels: Frequency histograms of {\it a$_{IR}$} for different hybrid tracers as a function of stellar mass. Massive galaxies (log[M$_*$/M$_{\odot}]$$>$10.5) appear in red, intermediate-mass galaxies (10$<$log[M$_*$/M$_{\odot}]$$<$10.5) are shown in grey and low-mass galaxies (log[M$_*$/M$_{\odot}]$$<$10) in blue. Dashed vertical lines correspond to the median value of each galaxy group. Black top marks show the median value for all the galaxies as in Figure \ref{histograms_constant}. There is a clear trend with the stellar mass, less massive galaxies need a higher value of the a$_{IR}$ coefficient compared with massive ones. Right four panels: Same as in left panels but this time type-2 AGNs are excluded for the sample. The distribution of the {\it a$_{IR}$} coefficient with the stellar mass and morphological type (Figure \ref{histograms_morphological_type}) allow us to provide, for the first time, a set of hybrid calibrations in terms of these galaxy properties. If the sample to be analyzed is biased towards morphology or, more commonly, luminosity or stellar mass, these tracers would be particularly useful (see Tables \ref{hybrids_type_table} and \ref{hybrids_mass_table}).}
\label{histograms_stellar_mass}
\end{figure*}

In Figures \ref{sfr_update_fuv} and \ref{sfr_update_ha} we also show the results of this analysis after including (left panels) or excluding (right panels) type-2 AGN from the sample in order to establish whether the behavior of the hybrid calibrators changes in each case. The four hybrid tracers show the same pattern, the {\it a$_{IR}$} coefficient decreases when galaxies hosting type-2 AGN are considered. As for the case of the single-band tracers (see Section \ref{Single-band updated}) this decrease in the value of the {\it a$_{IR}$} coefficient implies that we need to slightly reduce the contribution of the infrared emission in type-2 AGN to match that measured in H$\alpha$. This implication means that either (1) galaxies hosting type-2 AGN are emitting more light in the infrared that is not associated to the sites or processes that lead to the H$\alpha$ emission, both at 22\,$\mu$m and TIR luminosities, than normal star-forming galaxies or (2) the Balmer-corrected H$\alpha$ luminosity underestimates the actual SFR in these galaxies. 

The distribution of the {\it a$_{IR}$} coefficient appears in the histograms of Figure \ref{histograms_constant} where red dashed lines are referred to its median value. This coefficient has a large dispersion even when only star-forming galaxies are studied. In Table \ref{hybrids_global_table} we give the resulting median values of {\it a$_{IR}$} and the corresponding dispersions (measured as the interval that includes 68\% of the data points around the median). These dispersions appear as red tick-marks at the top panels in Figure \ref{histograms_constant} while black tick-marks indicate the standard error of the median computed from the asymptotic variance formula (which assumes that the underlying distribution is Gaussian) using the previous 1 $\sigma$ dispersions. These values are in good agreement with the ones reported in the literature for integrated measurements of galaxies. \citet{Kennicutt_2009} found 0.020 $\pm$ 0.005 and 0.0024 $\pm$ 0.0006 for L(H$\alpha$) + a$_{IR}$ $\times$ L(24\,$\mu$m) and L(H$\alpha$) + a$_{IR}$ $\times$ L(TIR), respectively. For the case of the UV luminosity, \citet{Hao_2011} found 3.89 $\pm$ 0.15 and 0.46 $\pm$ 0.12 for L(FUV) + a$_{IR}$ $\times$ L(25\,$\mu$m) and L(FUV) + a$_{IR}$ $\times$ L(TIR), respectively.

In the rest of this section we study the value of the {\it a$_{IR}$} coefficient as a function of galaxy properties in order to get insights on the origin of this spread. As we show below, the change in the {\it a$_{IR}$} coefficient with galaxy properties does not appear only when studying nuclear activity but also galaxy morphology, stellar mass, color, axial ratio and attenuation.


\subsubsection{Morphological-type dependence of {\it a$_{IR}$} in hybrid tracers}
\label{morphological dependence}

Given the large number of galaxies in our sample, we can now explore the origin of the differences between the various SFR tracers. In particular, we analyze the origin of the variation of the {\it a$_{IR}$} with different galaxy properties. Here we focus on the study of its dependence with galaxy morphology \citep[see][]{Walcher_2014}. Figure \ref{histograms_morphological_type} shows the distribution of the {\it a$_{IR}$} coefficient in bins of morphological type. In the four top plots of this figure we can see a trend for the median value of the {\it a$_{IR}$} coefficient (vertical dashed lines) with the galaxy morphology for H$\alpha$ + IR tracers. Star-forming galaxies of early-type, considered here as S0/a, Sa and Sab, have lower median values for {\it a$_{IR}$} (red dashed line) than intermediate-type spirals such as Sb and Sbc (grey dashed line). The last group of galaxies, Sc-Sd-Sm-Irr, shows the largest median value for the {\it a$_{IR}$} coefficient (blue dashed line). When type-2 AGN galaxies are excluded (right panels in Figure \ref{histograms_morphological_type}) the trend is less obvious, mainly because of a drastic increase in the median {\it a$_{IR}$} of early-type spirals.

Regarding the FUV+IR hybrid tracers (four bottom panels in Figure \ref{histograms_morphological_type}), we find that the median values for {\it a$_{IR}$} are more similar between S0/a-Sab and Sb-Sbc galaxies. However, the Sc-Sd-Sm-Irr galaxies still show the highest value for the {\it a$_{IR}$} coefficient. Table~\ref{hybrids_type_table} lists the resulting median values and their corresponding errors.

These trends are likely the combination of multiple effects (especially given the large dispersion in the value of  {\it a$_{IR}$}  within a given subsample), namely:

\begin{itemize}
\item[(1)] The contribution of obscured AGN to the IR luminosity (both at 22\,$\mu$m and in the TIR). This partly explains the fact that the average {\it a$_{IR}$} decreases when type-2 AGN are included in the sample. The fraction of type-2 AGN is larger within early-type galaxies, so part of the IR luminosity (without an equivalent extinction-corrected H$\alpha$ luminosity counterpart) is arising from the (obscured) AGN itself. 

\item[(2)] A fraction of the SFR (that assumed to be in this case accurately measured using an hybrid tracer with a nominal -- large -- value of the {\it a$_{IR}$} coefficient) is missed when using the extinction-corrected H$\alpha$ luminosity. This happens, especially, in early-type spirals so the {\it a$_{IR}$} coefficient decreases in these objects to compensate for the reduced amount of SFR derived from H$\alpha$. When the H$\alpha$ emission missed is exclusively due to an obscured AGN we will be in case (1) and H$\alpha$ would be a fair measure of the SFR.

\item[(3)] There is a fraction of the infrared emission that is due to heating by optical photons. One would expect that this effect would be more notorious when the {\it a$_{IR}$} coefficient refers to the TIR band, as optical photons are expected to heat the dust at low temperatures, where the emission at 22\,$\mu$m should be small. As discussed above, the value of  {\it a$_{IR}$} is smaller for S0/a-Sab galaxies, which are galaxies that have older stellar populations and optically bright bulges. \citet{Li_2013} also found that the coefficient that multiply the IR luminosity in the L(H$\alpha$) $+$ a $\times$ L(70\,$\mu$m) hybrid tracer is smaller when larger apertures around star-forming regions are used. The authors attribute this effect to the larger associated star formation timescale and the consequent dust heated by old stellar populations.
\end{itemize}

The fact that by removing type-2 AGN we reduce but not eliminate completely the morphological-type dependence of {\it a$_{IR}$} indicates that while (1) appears to have some role, the other possibilities are also at play. Disentangling the contribution of different mechanisms listed above is not easy. In particular, the change observed in {\it a$_{IR}$} when type-2 AGN are excluded from the sample could be also due to a decrease in the number of red massive star-forming galaxies in each morphological-type bin, galaxies which are expected to suffer from mechanism (3) as well. The analysis of the variation of {\it a$_{IR}$} with other properties, mass, color, axial ratio and ionized-gas attenuation will help us understanding the relative contribution of these mechanisms and, therefore, the specific limitations of the different SFR hybrid tracers. 

\begin{figure}
\centering
\includegraphics[trim=0.6cm 0.1cm 0.3cm 1.0cm, clip, width=95mm]{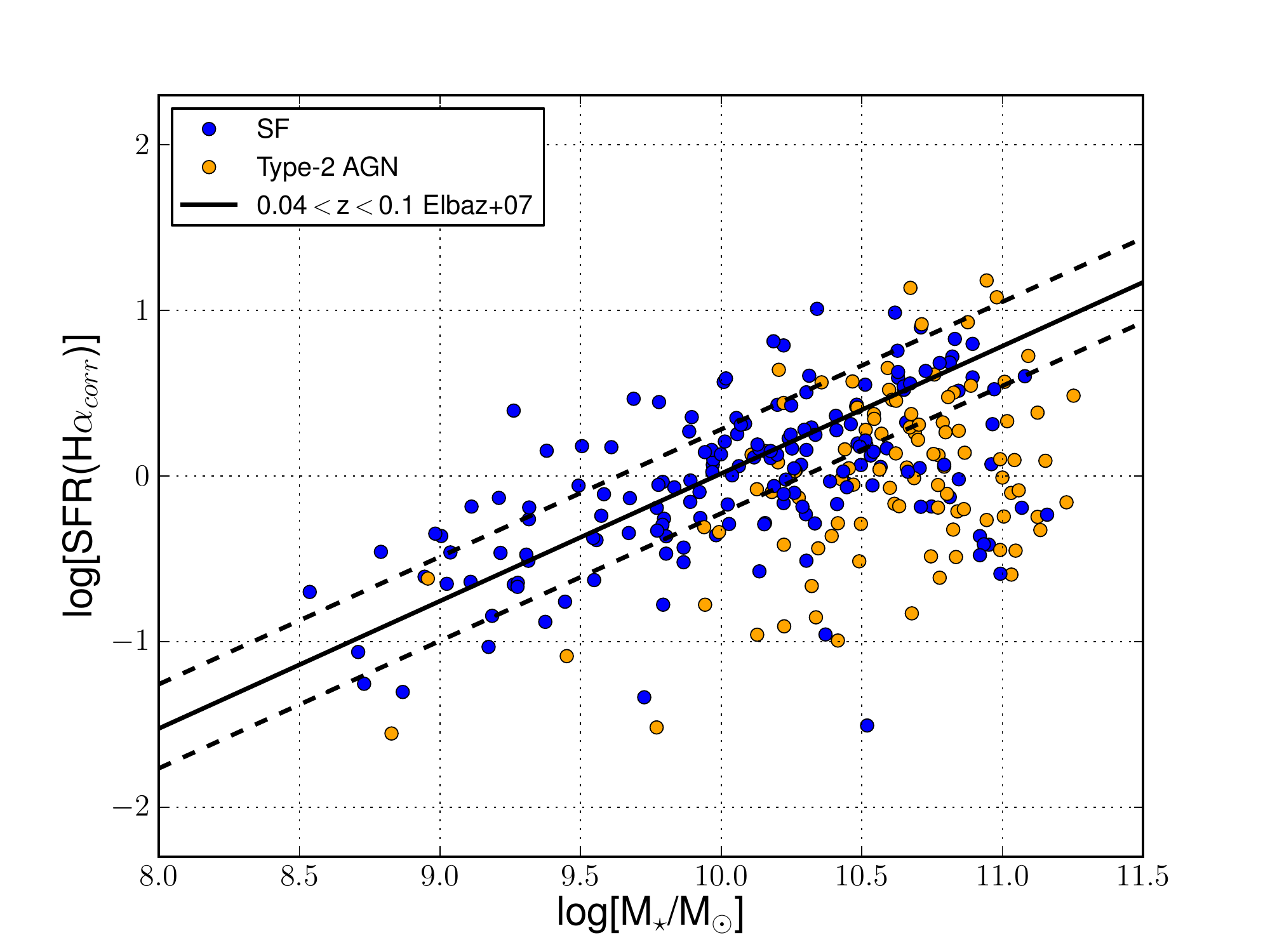}
\caption{Correlation between the extinction-corrected H$\alpha$ SFR and the total stellar mass of the galaxies. Solid line represents the fit of \citet{elbaz_2007} with a slope of 0.77 for galaxies in the 0.04$<$$z$$<$0.1 range while dashed lines correspond to the dispersion of this fit. Type-2 AGN host galaxies dominate the high-mass end in the Main Sequence plot for our galaxies. They show somewhat smaller SFR values for the same stellar mass. This fact could be due to a fraction of H$\alpha$ emission absorbed by the AGN or in the circumnuclear region or, alternatively, the presence of the type-2 AGN might impact the internal evolution of the galaxy quenching the SF.}
\label{main_sequence}
\end{figure}

\subsubsection{Stellar mass dependence of {\it a$_{IR}$} in hybrid tracers}
\label{mass dependence}

Since morphology alone is not able to establish the origin of the variation of  {\it a$_{IR}$} from galaxy to galaxy and within subsamples, we now explore its dependence with stellar mass. We use the total stellar masses for the CALIFA galaxies from \citet{Walcher_2014}, Section 6.3. (J. Walcher, priv$.$ comm$.$). The masses are publicly available on the CALIFA DR2 webpage\footnote{http://www.caha.es/CALIFA/public$\_$html/?q=content/califa-2nd-data-release}. The procedure for determining them is based on the fitting of UV-optical-NIR SEDs as described in detail in \citet{Walcher_2014}.

\begin{figure*}
\centering
\includegraphics[trim=1.1cm 0.2cm 1.0cm 3.5cm, clip, width=90mm]{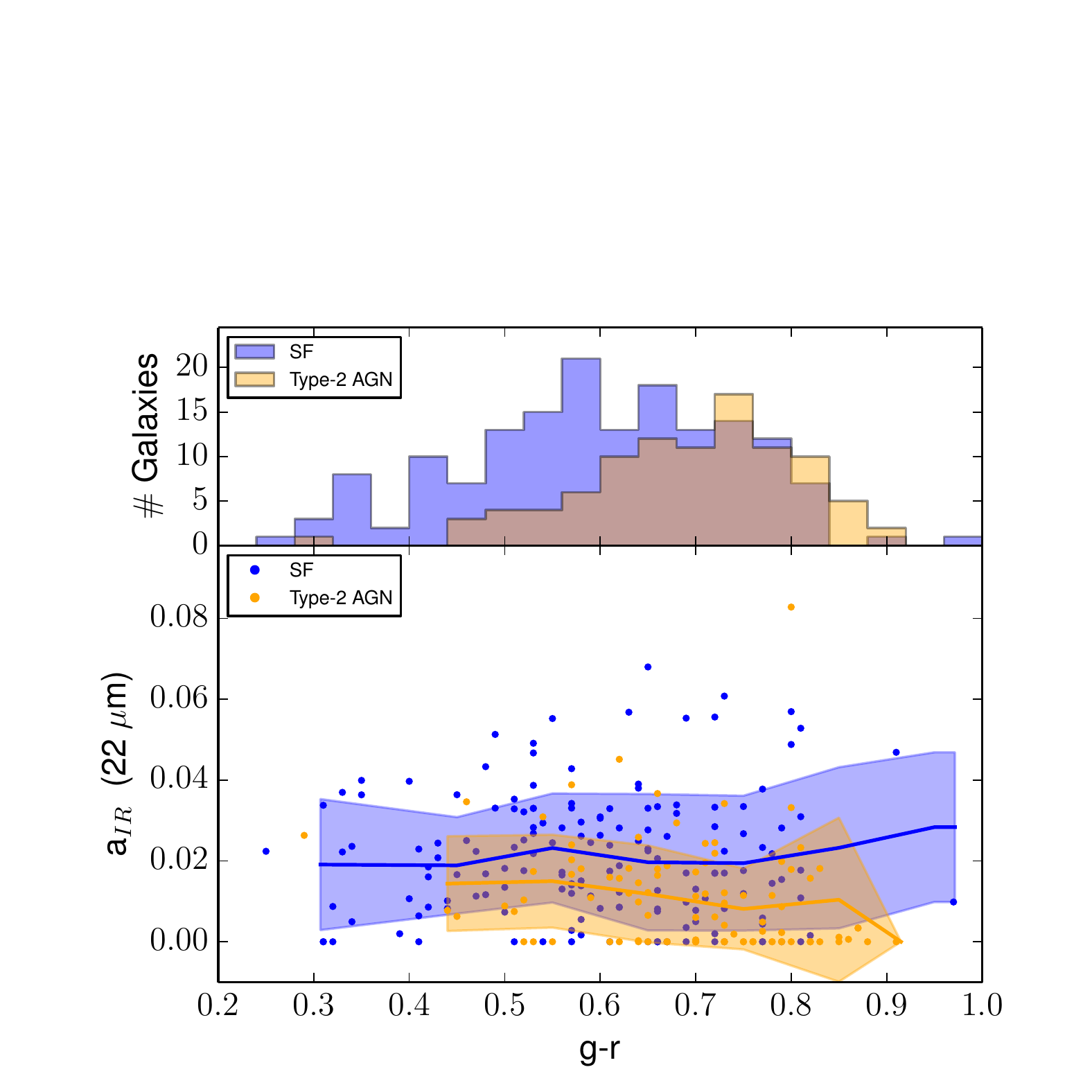} 
\includegraphics[trim=1.1cm 0.2cm 1.0cm 3.5cm, clip, width=90mm]{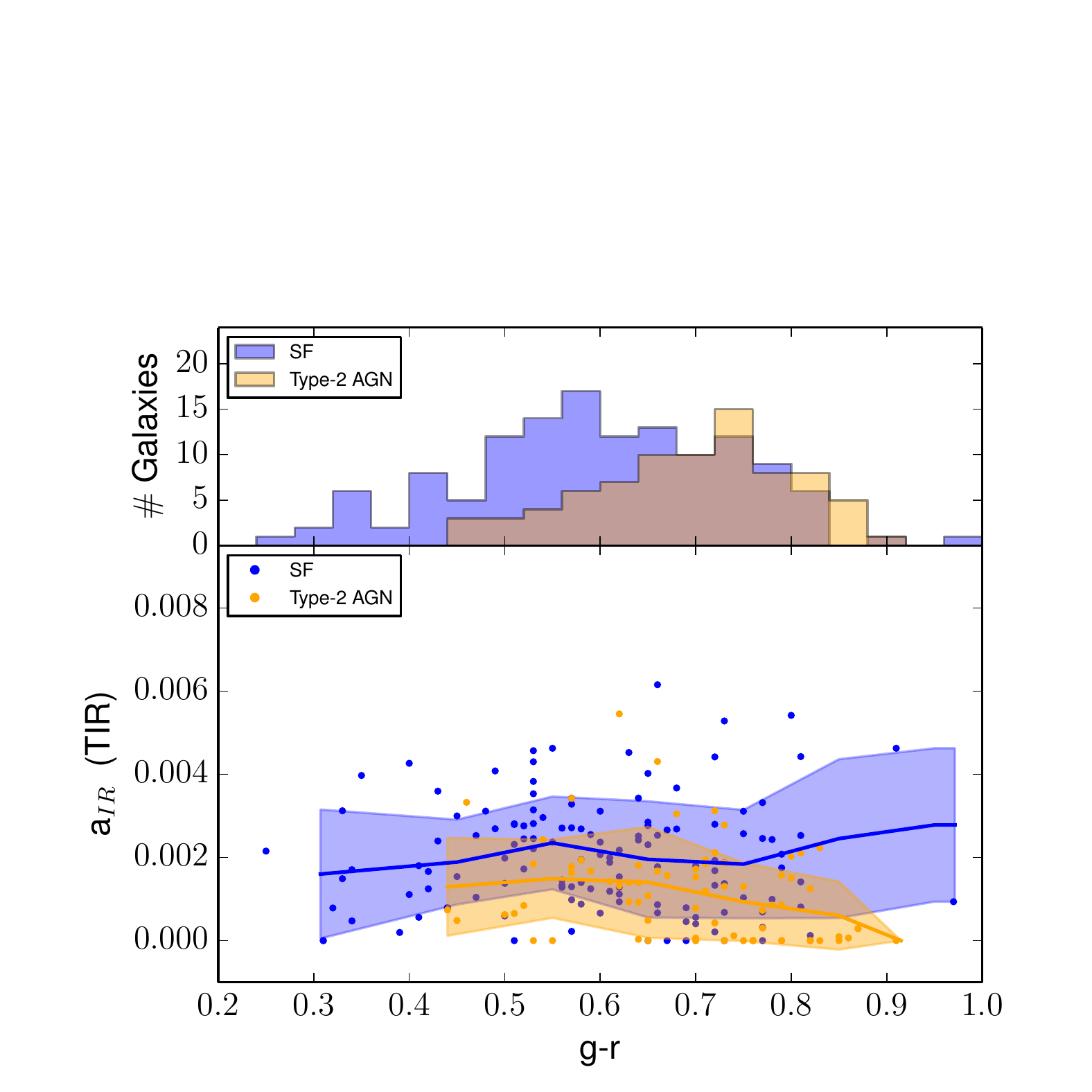} 
\caption{Left panel: Distribution of {\it a$_{IR}$} coefficients as a function of the galaxies $g-r$ {\it SDSS} color for the H$\alpha$+22\,$\mu$m hybrid tracer. Orange points show type-2 AGN galaxies while blue points represent star-forming galaxies. Filled contours represent the 1 $\sigma$ dispersion after applying a 5 $\sigma$ rejection around the mean value expressed as a blue (orange) solid line for the SF (type-2 AGN) galaxies. The corresponding histogram with the distribution of the number of galaxies for each $g-r$ {\it SDSS} color is plotted on the top for reference applying the same color-coding. Right panel: Same as the left panel showing the H$\alpha$+TIR hybrid tracer instead. There is a clear offset between the star-forming and type-2 AGN host galaxies with the a$_{IR}$ coefficient at any $g-r$ {\it SDSS} color.}
\label{g-r}
\end{figure*}

\begin{figure*}
\centering
\includegraphics[trim=1.1cm 0.2cm 1.0cm 4.5cm, clip, width=90mm]{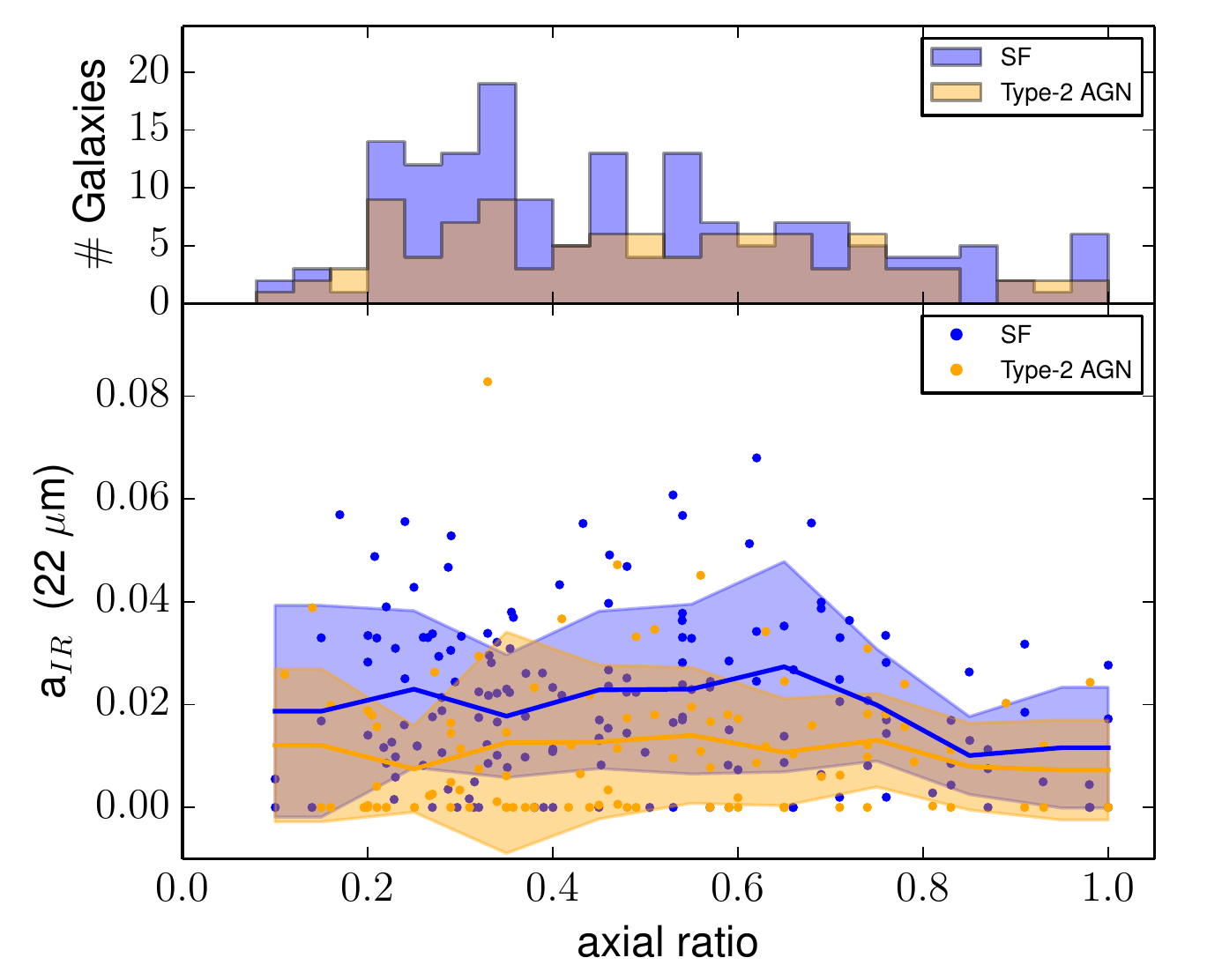}
\includegraphics[trim=1.1cm 0.2cm 1.0cm 4.5cm, clip, width=90mm]{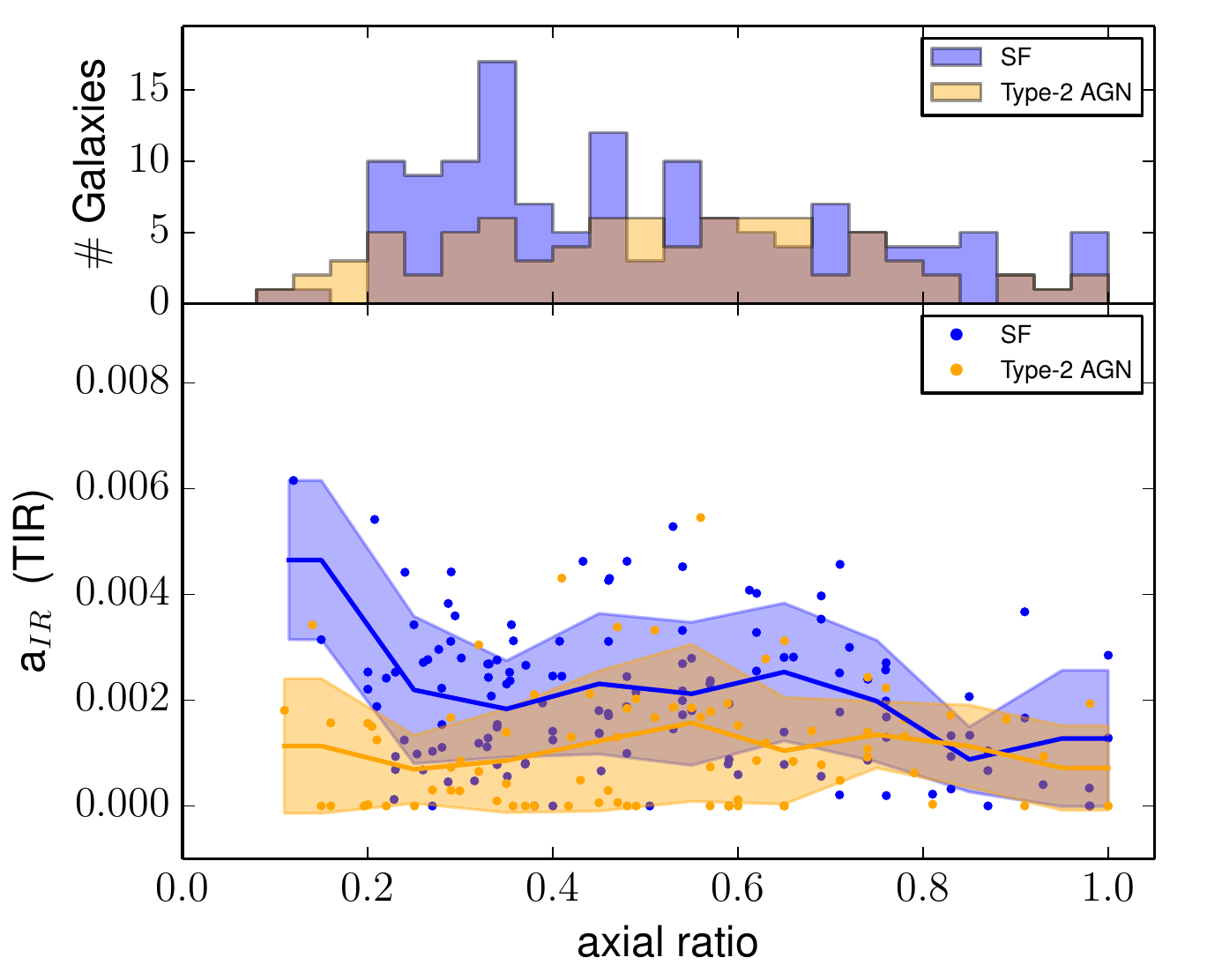}\\
\caption{Left panel: Distribution of {\it a$_{IR}$} coefficients as a function of the galaxies axial ratio for the H$\alpha$+22\,$\mu$m hybrid tracer. Orange points show type-2 AGN galaxies while blue points represent star-forming galaxies. Filled contours represent the 1 $\sigma$ dispersion after applying a 5 $\sigma$ rejection around the mean value expressed as a blue (orange) solid line for the SF (type-2 AGN) galaxies. The corresponding histogram with the distribution of the number of galaxies for each axial ratio is plotted on the top for reference applying the same color-coding. Right panel: Same as the left panel showing the H$\alpha$+TIR hybrid tracer instead. SF objects with low axial ratios (where highly-inclined disk galaxies would be located) show similar {\it a$_{IR}$} values as the rest of the galaxies. Lower {\it a$_{IR}$} would be expected if a fraction of the H$\alpha$ emission is completely obscured.}
\label{inclination_dependence}
\end{figure*}

Figure~\ref{histograms_stellar_mass} shows the frequency histograms of  {\it a$_{IR}$} for different hybrid tracers (with and without type-2 AGN in the sample) as a function of stellar mass. As for Figure~\ref{histograms_morphological_type}, we find a large dispersion within each mass bin, so clearly mass is not the only driver behind the variation of  {\it a$_{IR}$} from galaxy to galaxy. We find that most massive galaxies (log[M$_*$/M$_{\odot}]$$>$10.5) have lower median values of {\it a$_{IR}$} (red dashed line) than intermediate-mass (10$<$log[M$_*$/M$_{\odot}]$$<$10.5) (grey dashed line). In general, the low-mass galaxies (log[M$_*$/M$_{\odot}]$$<$10) show the largest median value for {\it a$_{IR}$} (blue dashed line). Table~\ref{hybrids_mass_table} compiles the resulting median values and their corresponding errors.

It should be noted that besides the relation between mass and color or attenuation, also the presence of intense nuclear star formation \citep[such as that found in the classical starburst nuclei, SBN; e.g.][]{gonzalez_delgado_1995,Gallego_1996} is far more common among massive star-forming systems than in low mass ones \citep[see][]{Perez_Gonzalez_2003}. It is precisely in these objects where complete obscuration effects in H$\alpha$ (that would reduce the value {\it a$_{IR}$}) might occur.

Figure~\ref{main_sequence} compares SFR (derived from the extinction-corrected H$\alpha$ luminosity) with the total stellar mass of the galaxies in the sample, the so-called 'main sequence' of galaxies \citep{Brinchmann_2004}. This figure shows that type-2 AGN host galaxies (orange dots) dominate the high-mass end of those in our diameter-limited sample. Besides, for the same stellar mass, active galaxies show somewhat smaller star formation rates. Some simulations show that when including the AGN feedback, most massive galaxies show a decrease in the specific SFR \citep{taylor_2015}. This could be due to a fraction of H$\alpha$ emission being completely absorbed either at the AGN or in circumnuclear star formation or to correlations between nuclear activity and other properties, besides mass, such as morphological type or environment. The latter is related with the fact that the presence of an AGN might impact the internal evolution of the galaxy quenching the SF by feedback mechanisms \citep[for a complete review on this topic see][]{alexander_hickox_2012}. The analysis of potential effects of the AGN on the current level of star formation at fixed mass \citep[e.g. preference of type-2 AGN for the Green Valley; see][]{Kauffmann_2003,sebastian_2004} is beyond the scope of this paper.

\subsubsection{Color dependence of {\it a$_{IR}$} in hybrid tracers}
\label{color dependence}

We address here the dependence of the {\it a$_{IR}$} coefficient with the color of the integrated stellar population \citep[as traced by the global SDSS $g-r$ color; see][]{Walcher_2014}. Figure \ref{g-r} shows the distribution of {\it a$_{IR}$} coefficients as a function of the galaxies $g-r$ color in the case of the H$\alpha$+22\,$\mu$m (left panel) and H$\alpha$+TIR (right panel) hybrid tracers. In these plots, type-2 AGN galaxies are shown as orange points and star-forming objects as blue. 
A clearer picture is obtained when looking separately at star-forming and type-2 AGN galaxies, as traced by the blue and orange-shaded areas in the bottom panels (mean $\pm$1$\sigma$ curves computed after an initial 5$\sigma$ rejection). We see here that most of the decrease in $a_{IR}$ with color is driven by type-2 AGN host galaxies that appeared to be a little redder than SF galaxies  in the top histogram. We find a trend for redder type-2 AGN host galaxies to show a lower value of {\it a$_{IR}$} especially at colors $g-r$ $>$ 0.6, although with a large scatter. This trend could be due to the fact that redder colors are likely related to galaxies with more massive bulges, and these with systems where the IR emission of a (luminous) obscured AGN could effectively dominate over that due to star formation alone. With regard to the pure star-forming galaxies in the sample, we find a relatively flat trend considering the scatter.

\subsubsection{Axial ratio dependence of {\it a$_{IR}$} in hybrid tracers}
\label{Axial ratio dependence}

Since highly-inclined systems might be subject to important obscuration effects in the derivation of the SFR, we have explored the dependence between the $a_{IR}$ coefficient and the axial ratio, as a proxy for the galaxy inclination. Figure \ref{inclination_dependence} shows the histograms of both star-forming and type-2 AGN host galaxies as a function of the axial ratio as given by the RC3 catalog, i.e. measured in the D25 B$-$band isophote. In addition to a clear offset between the two samples at any axial ratio, we find a nearly flat distribution within each sample at axial ratios below $\sim$0.65. An apparent decrease of $a_{IR}$ appear for face-on SF systems, although statistics are poor in this case. This is true both for the H$\alpha$+22$\mu$m and the H$\alpha$+TIR hybrid tracers (left and right panel in Figure \ref{inclination_dependence}, respectively). We do not find star-forming objects with low axial ratios (where highly-inclined disk galaxies would be located) to show lower $a_{IR}$. One would expect it if a fraction of the dust-absorbed H$\alpha$ emission will not be recovered by our Balmer decrement based extinction correction (in other words, the H$\alpha$ emission will be completely obscured.) Therefore, should H$\alpha$ be missing a fraction of the SFR in some galaxies, these are not necessarily the most inclined systems. Alternatively, that missed SFR (if present) could arise from dense nuclear regions, such as (circum)nuclear starbursts. 

\begin{figure}[ht]
\centering
\includegraphics[trim=0.2cm 0.0cm 0.2cm 1.0cm, clip, width=95mm]{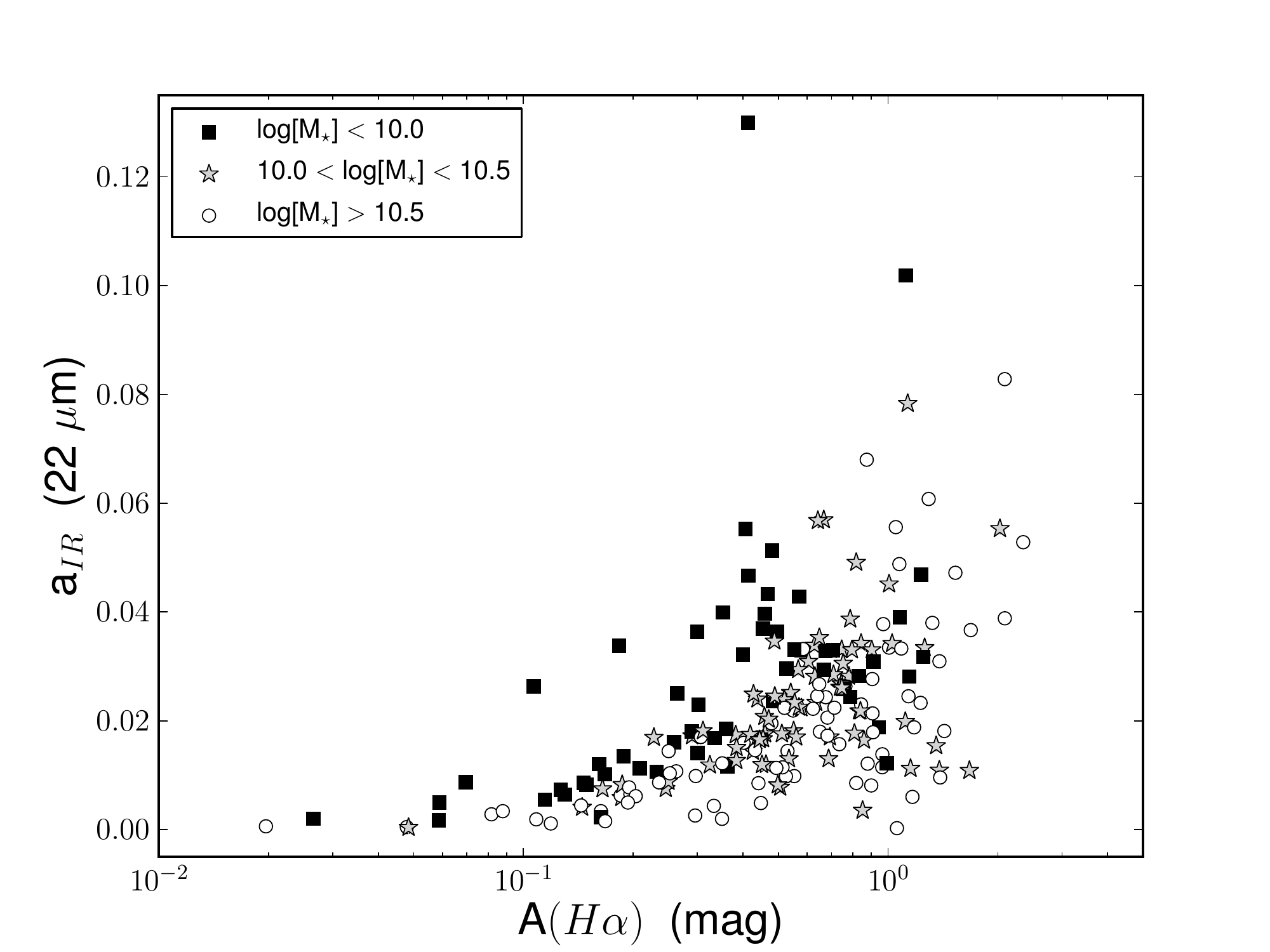}
\includegraphics[trim=0.2cm 0.0cm 0.2cm 1.0cm, clip, width=95mm]{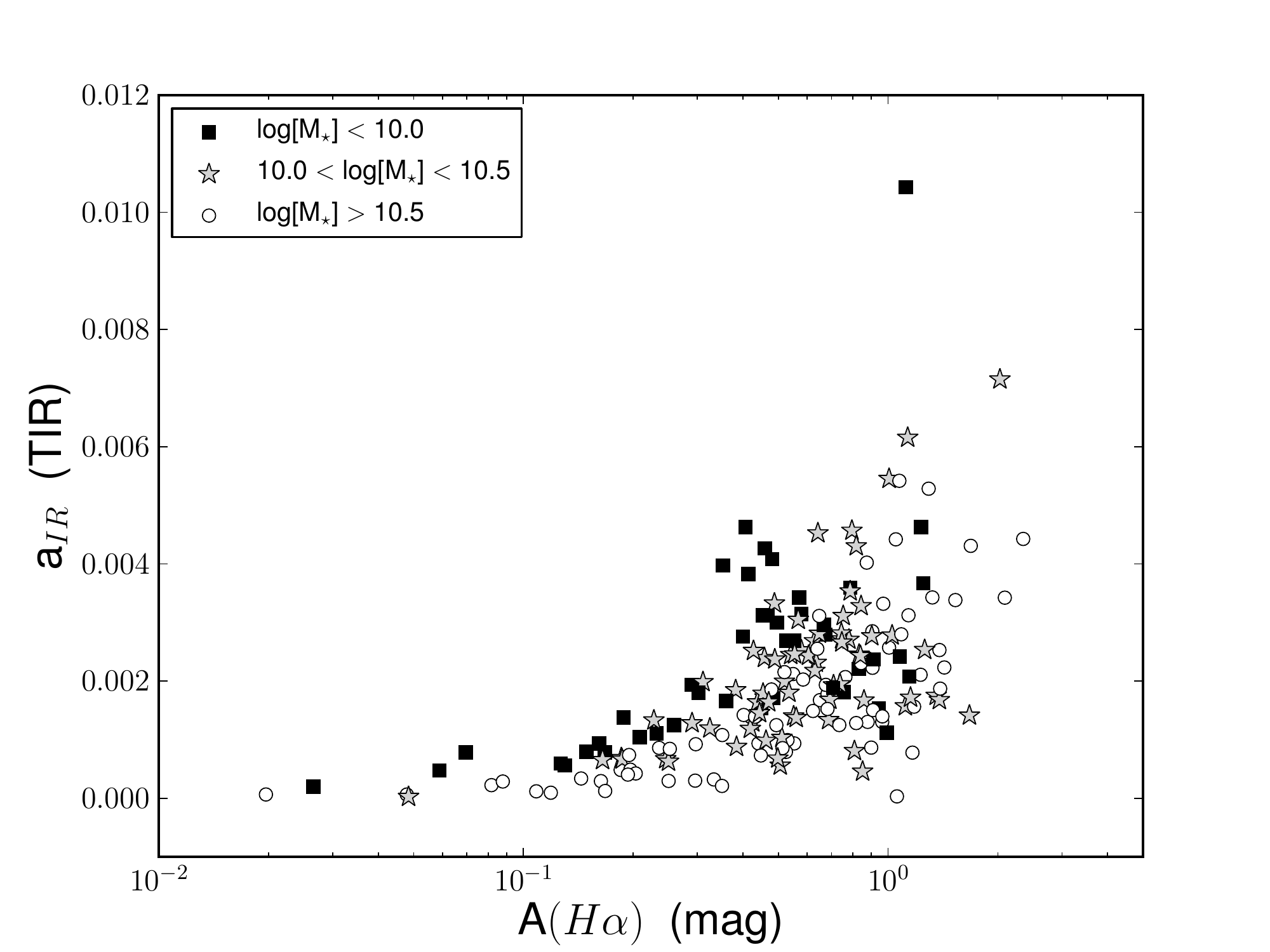}
\caption{Top panel: Variation of the {\it a$_{IR}$} coefficient with the H$\alpha$ attenuation derived using the Balmer decrement for the H$\alpha$+22\,$\mu$m hybrid tracer. Black squares show galaxies with stellar masses lower than log[M$_{\star}$] $<$ 10.0, grey stars represent galaxies with stellar masses in the range of 10.0 $<$ log[M$_{\star}$] $<$ 10.5 and, finally, open circles are for the most massive galaxies with log[M$_{\star}$] $>$ 10.5. Bottom panel: Same as previous panel this time for the {\it a$_{IR}$} coefficient that corresponds to the H$\alpha$+TIR hybrid tracer.} 
\label{constant_attenuation_ha}
\end{figure}

\begin{figure}[ht]
\centering
\includegraphics[trim=0.2cm 0.0cm 0.2cm 1.2cm, clip, width=95mm]{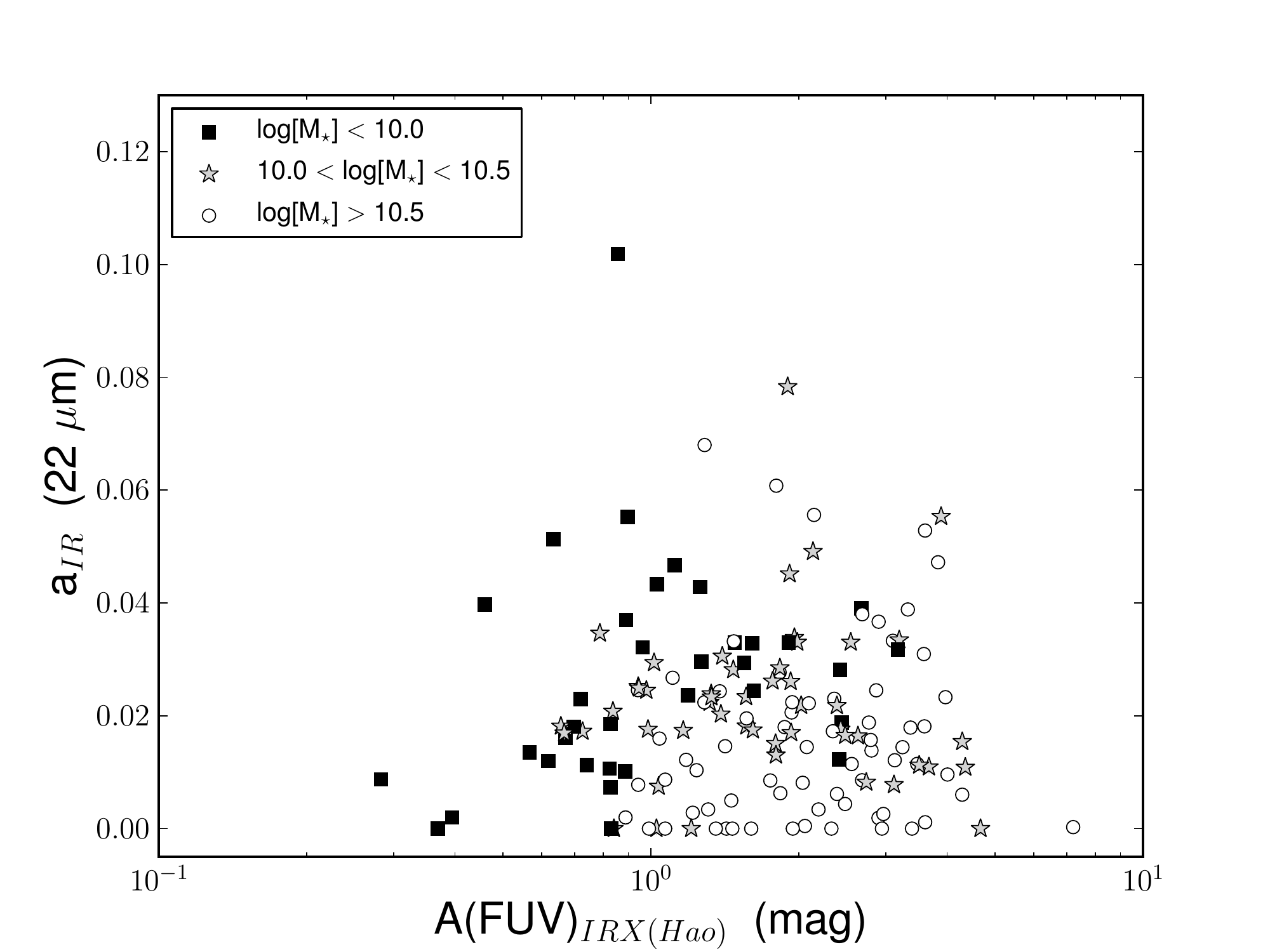}
\includegraphics[trim=0.2cm 0.0cm 0.2cm 1.2cm, clip, width=95mm]{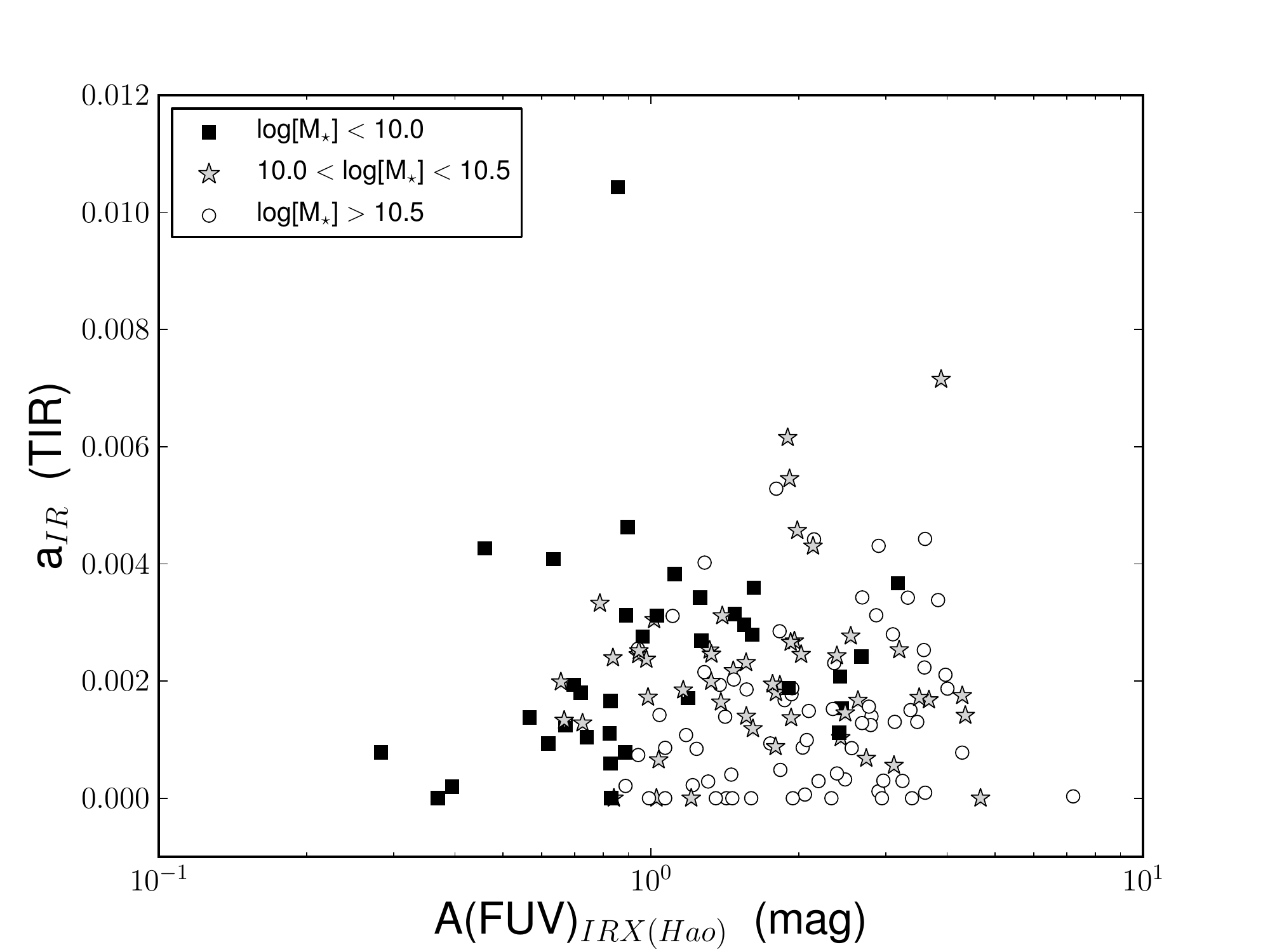}
\caption{Top panel: Variation of the {\it a$_{IR}$} coefficient with the FUV attenuation derived using the IRX from \cite{Hao_2011} for the H$\alpha$+22\,$\mu$m hybrid tracer. Same symbol-coding as in Figure \ref{constant_attenuation_ha}. Bottom panel: Same as previous panel this time for the {\it a$_{IR}$} coefficient that corresponds to the H$\alpha$+TIR hybrid tracer.} 
\label{constant_attenuation_uv}
\end{figure}

\citet{Prescott_2007} studied the incidence of obscured SF in a large sample of infrared-selected star forming regions in normal galaxies. They used the 24 $\mu$m flux as a tracer of the obscured emission due to SF and the uncorrected H$\alpha$ flux as a tracer of the unobscured portion (the same way as we use our 22 $\mu$m + H$\alpha$ hybrid tracer along this work, but we compute integrated measurements of galaxies). These authors conclude that the fraction of highly obscured regions in normal, star forming disk galaxies is small on 500 pc scales. They are more luminous and tend to be closer to the center of the host galaxy. The analysis of obscuration effects is the subject of Section \ref{attenuation dependence} below in which efforts had been made to further explore this issue.

\subsubsection{Attenuation dependence of {\it a$_{IR}$} in hybrid tracers}
\label{attenuation dependence}

In order to determine whether local obscuration effects in H$\alpha$ might be behind the decrease of {\it a$_{IR}$} in galaxies of specific types, masses or colors (as this coefficient should be reduced to compensate by the SFR missed in H$\alpha$) we finally analyze its variation as a function of ionized-gas attenuation. This attenuation is derived from the H$\alpha$/H$\beta$ Balmer decrement as described in Section~\ref{Flux corrections}. Figure \ref{constant_attenuation_ha} shows the variation of the {\it a$_{IR}$} coefficient with the attenuation measured in magnitudes in H$\alpha$ (the use of log scale in the abscissa is justified by the large concentration of points at low attenuations). The most remarkable feature in this plot is that there is a number of galaxies with low global ionized-gas attenuations that show very small values of {\it a$_{IR}$}. We interpret this as consequence of dust emission that is caused by the heating of photons different from those arising in sites of current star formation. In galaxies where the attenuation derived in H$\alpha$ is compatible with no attenuation even a small amount of dust emission would lead to a null value for {\it a$_{IR}$}, which results in the number of galaxies with low attenuations and low values of {\it a$_{IR}$} seen in Figure \ref{constant_attenuation_ha}. These galaxies with very small values of {\it a$_{IR}$} indicate that, at this level, we are in the limit where A(H$\alpha$) can be properly derived, given the low global ionized-gas attenuations found. The variation of the {\it a$_{IR}$} coefficient using the FUV attenuation applying the IRX given by \cite{Hao_2011} is shown in Figure \ref{constant_attenuation_uv} for comparison.

Except for this tail at low attenuation, A(H$\alpha$)$<$0.2\,mag, we find no correlation between the two parameters. Should a significant fraction of the SFR being missed when the extinction corrected H$\alpha$ luminosity is used, one would expect to find a clear decrease in $a_{IR}$ as the ionized-gas attenuation gets larger. Only when low-mass galaxies are analyzed separately they seem to show a decline in their $a_{IR}$ values above A(H$\alpha$)$=$0.4 mags, although with some discrepant points at A(H$\alpha$) $>$ 1 mag. This decline in $a_{IR}$ might be due to the fact that only in some of these naturally low-metallicity galaxies high attenuations are due to the presence of active nuclear star formation events. Despite that, the average and scatter obtained for the $a_{IR}$ coefficient in these galaxies are not very different from those obtained at higher masses neither $a_{IR}$ reaches very low values.

\section{Summary and conclusions}
\label{Conclusions}

In this work, we present the analysis of the Star Formation Rate in a sample of 380 galaxies from the diameter-limited CALIFA survey. A total of 272 galaxies shows detected emission in both H$\beta$ and H$\alpha$ and are listed in Table \ref{table_fluxes} for reference. The availability of wide-field IFS for all the galaxies in the sample is a major advantage over other techniques. Using IFS data we can recover the flux in galaxies with low equivalent widths and separate H$\alpha$ and the [NII] without assuming a [NII]/H$\alpha$ ratio avoiding problems associated with narrow-band imaging or long-slit spectroscopy. It also ensures a proper determination of the underlying stellar continuum and, consequently, of the extinction-corrected H$\alpha$ luminosity. 

We have combined the aperture-corrected H$\alpha$ measurements from CALIFA with those measured in other bands that are also used to estimate the SFR, including luminosity measurements in the UV from GALEX (200 galaxies), 22\,$\mu$m from WISE (265 galaxies) and TIR luminosities from WISE+IRAS+AKARI SED fitting (221 galaxies).

We first compare the extinction-corrected H$\alpha$ SFR with measurements from single-band (FUV, 22\,$\mu$m and TIR) and hybrid-tracers (H$\alpha$+22\,$\mu$m, H$\alpha$+TIR, FUV+22\,$\mu$m, FUV+TIR). In this part of the paper, we use recent compilations of SFR recipes by \citet{Calzetti_2012}. The good correlation between the SFR surface density obtained with extinction-corrected H$\alpha$ tracer and FUV+22\,$\mu$m hybrid tracer guarantees that potential linear correlations between different SFR tracers (some of them not resolved, such as those relying on TIR measurements) are not driven by scaling effects and that global values of the SFR can be used reliably. Our results indicate that, overall, the extinction-corrected H$\alpha$ luminosity (once underlying stellar absorption and dust-attenuation effects are properly accounted for) matches the SFR obtained from hybrid tracers combining the observed FUV or H$\alpha$ and the IR (22\,$\mu$m and TIR) luminosities with dispersions around $\sim$0.20 dex being found. In the case of the comparison with single-band tracers we conclude (1) that the use of IR measurements clearly underestimates the SFR below $\sim$1\,M$_{\odot}$ yr$^{-1}$ and (2) the large uncertainty in the correction for attenuation when only FUV$-$NUV color (similar to the UV slope, $\beta$) information is available. This factor introduces a very large scatter, particularly at SFR$>$5\,M$_{\odot}$yr$^{-1}$, where the $\beta$-corrected FUV luminosity also tend to underestimate the SFR. This prevents the use of the UV luminosity alone as a SFR tracer.

\paragraph{}
We also provide a new set of single-band calibrators anchored to the extinction-corrected H$\alpha$ luminosities. The values for these coefficients appear in Table \ref{single_table}. 
In the case of the hybrid calibrators we determine the best (median) fit for the coefficient that weights the amount of IR luminosity reprocessed by dust, {\it a$_{IR}$}. We assume an energetic balance and calculate the {\it a$_{IR}$} coefficients for different combinations of observed (UV or H$\alpha$) and dust-reprocessed (22\,$\mu$m or TIR) SFR contributions anchored to the extinction-corrected H$\alpha$ luminosities. These values appear in Table \ref{hybrids_global_table} and are calculated with and without galaxies hosting type-2 AGN being considered. 
\begin{itemize}
\item[(1)] 
This analysis allows us to provide, for the first time, a set of hybrid calibrations for different morphological types and masses. These are particularly useful in case that the sample to be analyzed shows a different bias in terms of morphology or, more commonly, luminosity or stellar mass (see Tables \ref{hybrids_type_table} and \ref{hybrids_mass_table}). 
\item[(2)] 
We also study the dependence of this coefficient not only with morphological type and mass but also with color (SDSS $g-r$), axial ratio and ionized-gas attenuation.
\item[(3)] 
The distributions of {\it a$_{IR}$} values (for each of the hybrid tracers) are quite wide in all cases. While part of the spread can be attributed to changes in morphological type, stellar mass, color, and attenuation among the galaxies in the sample, there is no single physical property that can by itself explain the entire variation in {\it a$_{IR}$} from galaxy to galaxy. 
\item[(4)] 
The analysis of the dependence of {\it a$_{IR}$} with galaxy properties indicates that galaxies hosting type-2 AGN tend to reduce the median value of {\it a$_{IR}$}, likely due to the contribution of obscured AGN to the infrared emission. The fact that {\it a$_{IR}$} does not show a particularly low value at high ionized-gas attenuations nor low axial ratios, suggests that obscured star formation is, comparatively, playing a minor role. Part of the dependence of the median value of {\it a$_{IR}$} with the morphological type disappears once the AGN contribution is removed although early spirals still show a somewhat lower {\it a$_{IR}$} than intermediate- and late-type spirals. This behavior, also present when comparing massive with less massive systems, can be explained in part as due to the enhanced contribution of optical photons to the heating of the dust in both early-type spirals and massive systems. 
\end{itemize}

These conclusions will allow us to make use of the CALIFA IFS data to explore the distribution of the SFR with spatial resolution in a future work \cite[see][]{Catalan_Torrecilla_2014}. We emphasize that the impact of potential differences in the selection criteria should be addressed carefully when extrapolating these results to other samples of galaxies and, particularly, to other redshifts.

\begin{acknowledgements}

This study makes uses of the data provided by the Calar Alto Legacy Integral Field Area (CALIFA) survey (http://califa.caha.es). CALIFA is the first legacy survey being performed at Calar Alto. The CALIFA collaboration would like to thank the IAA-CSIC and MPIA-MPG as major partners of the observatory, and CAHA itself, for the unique access to telescope time and support in manpower and infrastructures. The CALIFA collaboration thanks also the CAHA staff for the dedication to this project.
C. C.-T. thanks the support of the Spanish {\it Ministerio de Educaci\'on, Cultura y Deporte} by means of the FPU fellowship program. The authors also thank the support from the {\it Plan Nacional de Investigaci\'on y Desarrollo} funding programs, AYA2012-30717 and AyA2013-46724P, of Spanish {\it Ministerio de Econom\'ia y Competitividad} (MINECO). P.G. P-G. acknowledges support from the AYA2012-30717 and AYA2012-31277. J.I.P. acknowledges financial support from the Spanish MINECO under grant AYA2010-21887-C04-01 and from Junta de Andaluc\'{\i}a Excellence Project PEX2011-FQM7058. R.A. M. is funded by the Spanish program of International Campus of Excellence Moncloa (CEI). MAPT acknowledges support from the Spanish MINECO through grant AYA2012-38491-C02-02. AdO acknowledge financial support from the Spanish grant AYA2013-42227-P. Support for LG is provided by the Ministry of Economy, Development, and Tourism's Millennium Science Initiative through grant IC120009, awarded to The Millennium Institute of Astrophysics, MAS. LG acknowledges support by CONICYT through FONDECYT grant 3140566. JMG acknowledges support from the Funda\c{c}\~{a}o para a Ci\^encia e a Tecnologia (FCT) through the Fellowship SFRH/BPD/66958/2009 from FCT (Portugal) and POPH/FSE (EC) by FEDER funding through the program Programa Operacional de Factores de Competitividade (COMPETE). JMG also acknowledges support by FCT under project FCOMP-01-0124-FEDER-029170 (Reference FCT PTDC/FIS-AST/3214/2012), funded by FCT-MEC (PIDDAC) and FEDER (COMPETE). This research has made use of the NASA/IPAC Extragalactic Database (NED) which is operated by the Jet Propulsion Laboratory, California Institute of Technology, under contract with the National Aeronautics and Space Administration. This publication makes use of data products from the Wide-field Infrared Survey Explorer, which is a joint project of the University of California, Los Angeles, and the Jet Propulsion Laboratory/California Institute of Technology, funded by the National Aeronautics and Space Administration. This research has made use of the NASA/ IPAC Infrared Science Archive, which is operated by the Jet Propulsion Laboratory, California Institute of Technology, under contract with the National Aeronautics and Space Administration. Funding for the SDSS and SDSS-II has been provided by the Alfred P. Sloan Foundation, the Participating Institutions, the National Science Foundation, the U.S. Department of Energy, the National Aeronautics and Space Administration, the Japanese Monbukagakusho, the Max Planck Society, and the Higher Education Funding Council for England. The SDSS Web Site is http://www.sdss.org/. The SDSS is managed by the Astrophysical Research Consortium for the Participating Institutions. The Participating Institutions are the American Museum of Natural History, Astrophysical Institute Potsdam, University of Basel, University of Cambridge, Case Western Reserve University, University of Chicago, Drexel University, Fermilab, the Institute for Advanced Study, the Japan Participation Group, Johns Hopkins University, the Joint Institute for Nuclear Astrophysics, the Kavli Institute for Particle Astrophysics and Cosmology, the Korean Scientist Group, the Chinese Academy of Sciences (LAMOST), Los Alamos National Laboratory, the Max-Planck-Institute for Astronomy (MPIA), the Max-Planck-Institute for Astrophysics (MPA), New Mexico State University, Ohio State University, University of Pittsburgh, University of Portsmouth, Princeton University, the United States Naval Observatory, and the University of Washington. GALEX (Galaxy Evolution Explorer) is a NASA Small Explorer, launched in April 2003. We gratefully acknowledge NASA's support for construction, operation, and science analysis for the GALEX mission, developed in cooperation with the Centre National d'Etudes Spatiales (CNES) of France and the Korean Ministry of Science and Technology.

\end{acknowledgements}

\bibliographystyle{aa} 
\bibliography{paper_v8} 

\twocolumn


\onecolumn

\begin{landscape}
\tiny
\begin{longtable}{lllrrrrrrrrrrr}
\caption{(1) ID CALIFA identifier; (2) Galaxy name; (3) Morphological type; (4) Distance in Mpc calculated from redshift; (5) Minor-to-major axis ratio as given by the RC3 catalog, i.e. measured in the D25 B-band isophote; (6) Observed-H$\alpha$ flux using an elliptical aperture with a semi-major axis of 36 arcsec; (7) Attenuation obtained using the Balmer decrement. As explained in Section \ref{Flux corrections}, we have assumed an intrinsic H$\alpha$/H$\beta$ ratio of 2.86 (although lower values are also physically possible in HII regions, depending on the electron density, effective temperature and therefore on the chemical abundance. This leads to a number of galaxies for which we assumed A(H$\alpha$) $=$ 0 that also appeared in this table). The errors listed here use the H$\alpha$ flux uncertainty only, an additional 7$\%$ uncertainty in the H$\beta$ flux have been taken into account for the errors in the H$\alpha$ luminosities, col. (8), and SFRs, col. (9) and in the attenuations used in the paper; (8) Observed H$\alpha$ luminosity corrected for aperture effects; (9) Balmer-corrected and aperture-corrected H$\alpha$ SFR values; (10) Observed FUV luminosity; (11) Observed NUV luminosity; (12) UV color in AB system; (13) TIR luminosity defined as the total infrared emission in the range 8$-$1000\,$\mu$m; (14) 22\,$\mu$m (aperture photometry) luminosity. As explained in Section \ref{WISE mid-infrared imaging}, it has been pointed out by several authors \citep{wright_2010,Jarrett_2013,Brown_2014} that SF galaxies measured with the WISE 22\,$\mu$m filter are systematically brighter by $\sim$ 10 $\%$ than what is inferred from Spitzer IRS and 24\,$\mu$m data. For that reason, the 22\,$\mu$m luminosity values given in column (14) have to be multiplied by 1/1.1. Note that galaxies UGC 00987 and UGC 03973 are classified as type-1 AGN within our sample. As a consequence, they have been excluded from the SFR analysis.} \\
\end{longtable}
\footnotesize{This table is available in its entirety in an online version. A portion is shown here for reference regarding its form and content.}

\end{landscape}

\setcounter{table}{0}

\onllongtab{
\begin{landscape}
\tiny
\begin{longtable}{lllrrrrrrrrrrr}
\caption{(1) ID CALIFA identifier; (2) Galaxy name; (3) Morphological type; (4) Distance in Mpc calculated from redshift; (5) Minor-to-major axis ratio as given by the RC3 catalog, i.e. measured in the D25 B-band isophote; (6) Observed-H$\alpha$ flux using an elliptical aperture with a semi-major axis of 36 arcsec; (7) Attenuation obtained using the Balmer decrement. As explained in Section \ref{Flux corrections}, we have assumed an intrinsic H$\alpha$/H$\beta$ ratio of 2.86 (although lower values are also physically possible in HII regions, depending on the electron density, effective temperature and therefore on the chemical abundance. This leads to a number of galaxies for which we assumed A(H$\alpha$) $=$ 0 that also appeared in this table). The errors listed here use the H$\alpha$ flux uncertainty only, an additional 7$\%$ uncertainty in the H$\beta$ flux have been taken into account for the errors in the H$\alpha$ luminosities, col. (8), and SFRs, col. (9) and in the attenuations used in the paper; (8) Observed H$\alpha$ luminosity corrected for aperture effects; (9) Balmer-corrected and aperture-corrected H$\alpha$ SFR values; (10) Observed FUV luminosity; (11) Observed NUV luminosity; (12) UV color in AB system; (13) TIR luminosity defined as the total infrared emission in the range 8$-$1000\,$\mu$m; (14) 22\,$\mu$m (aperture photometry) luminosity. As explained in Section \ref{WISE mid-infrared imaging}, it has been pointed out by several authors \citep{wright_2010,Jarrett_2013,Brown_2014} that SF galaxies measured with the WISE 22\,$\mu$m filter are systematically brighter by $\sim$ 10 $\%$ than what is inferred from Spitzer IRS and 24\,$\mu$m data. For that reason, the 22\,$\mu$m luminosity values given in column (14) have to be multiplied by 1/1.1. Note that galaxies UGC 00987 and UGC 03973 are classified as type-1 AGN within our sample. As a consequence, they have been excluded from the SFR analysis.} \\
 \hline\hline 
  \multicolumn{1}{c}{ID} &
  \multicolumn{1}{c}{name} &
  \multicolumn{1}{c}{type} &
  \multicolumn{1}{c}{dist} &
  \multicolumn{1}{c}{b/a} &
  \multicolumn{1}{c}{\centering {\it f}(H$\alpha$$_{obs}$)} &
  \multicolumn{1}{c}{A(H$\alpha$)} &
  \multicolumn{1}{c}{L(H$\alpha$$_{obs}$)} &
  \multicolumn{1}{c}{SFR[H$\alpha$$_{corr}$]} &
  \multicolumn{1}{c}{L(FUV$_{obs}$)} &
  \multicolumn{1}{c}{L(NUV$_{obs}$)} &
  \multicolumn{1}{c}{F$-$N} &
  \multicolumn{1}{c}{L(TIR)} &
  \multicolumn{1}{c}{L(22$\mu$m) } \\
 &  &  & \centering (Mpc) & & \multicolumn{1}{p{2.0cm}}{\centering($\times$ 10$^{-16}$ cgs)}  & \multicolumn{1}{p{1.2cm}}{\centering(mag)}  & ($\times$ 10$^{40}$ erg\,s$^{-1}$) & \multicolumn{1}{p{1.5cm}}{\centering(M$_{\odot}$ yr$^{-1}$)}  & ($\times$ 10$^{42}$ erg\,s$^{-1}$) & ($\times$ 10$^{42}$ erg\,s$^{-1}$)  & (AB mag) & ($\times$ 10$^{43}$ erg\,s$^{-1}$) & ($\times$ 10$^{42}$ erg\,s$^{-1}$)\\
(1) & \multicolumn{1}{p{2.1cm}}{\centering (2)} & (3) & \multicolumn{1}{p{0.8cm}}{\centering(4)} & \centering (5) & \multicolumn{1}{p{2.0cm}}{\centering(6)} & \multicolumn{1}{p{1.2cm}}{\centering(7)} & \multicolumn{1}{p{1.7cm}}{\centering(8)} & \multicolumn{1}{p{1.5cm}}{\centering(9)}& \multicolumn{1}{p{1.7cm}}{\centering(10)} & \multicolumn{1}{p{1.7cm}}{\centering(11)} & \multicolumn{1}{p{1.1cm}}{\centering(12)} & \multicolumn{1}{p{1.7cm}}{\centering(13)} & \multicolumn{1}{p{1.7cm}}{\centering(14)} \\ 
\hline\hline 
\endfirsthead
\hline\hline 
  \multicolumn{1}{c}{ID} &
  \multicolumn{1}{c}{name} &
  \multicolumn{1}{c}{type} &
  \multicolumn{1}{c}{dist} &
  \multicolumn{1}{c}{b/a} &
  \multicolumn{1}{c}{\centering {\it f}(H$\alpha$$_{obs}$)} &
  \multicolumn{1}{c}{A(H$\alpha$)} &
  \multicolumn{1}{c}{L(H$\alpha$$_{obs}$)} &
  \multicolumn{1}{c}{SFR[H$\alpha$$_{corr}$]} &
  \multicolumn{1}{c}{L(FUV$_{obs}$)} &
  \multicolumn{1}{c}{L(NUV$_{obs}$)} &
  \multicolumn{1}{c}{F$-$N} &
  \multicolumn{1}{c}{L(TIR)} &
  \multicolumn{1}{c}{L(22$\mu$m) } \\
   &  &  & \centering (Mpc) & & \multicolumn{1}{p{2.0cm}}{\centering($\times$ 10$^{-16}$ cgs)}  & \multicolumn{1}{p{1.2cm}}{\centering(mag)}  & ($\times$ 10$^{40}$ erg\,s$^{-1}$) & \multicolumn{1}{p{1.5cm}}{\centering(M$_{\odot}$ yr$^{-1}$)}  & ($\times$ 10$^{42}$ erg\,s$^{-1}$) & ($\times$ 10$^{42}$ erg\,s$^{-1}$)  & (AB mag) & ($\times$ 10$^{43}$ erg\,s$^{-1}$) & ($\times$ 10$^{42}$ erg\,s$^{-1}$)\\
(1) & \multicolumn{1}{p{2.1cm}}{\centering (2)} & (3) & \multicolumn{1}{p{0.8cm}}{\centering(4)} & \centering (5) & \multicolumn{1}{p{2.0cm}}{\centering(6)} & \multicolumn{1}{p{1.2cm}}{\centering(7)} & \multicolumn{1}{p{1.7cm}}{\centering(8)} & \multicolumn{1}{p{1.5cm}}{\centering(9)}& \multicolumn{1}{p{1.7cm}}{\centering(10)} & \multicolumn{1}{p{1.7cm}}{\centering(11)} & \multicolumn{1}{p{1.1cm}}{\centering(12)} & \multicolumn{1}{p{1.7cm}}{\centering(13)} & \multicolumn{1}{p{1.7cm}}{\centering(14)} \\ 
\hline\hline
\endhead
\\[0.5pt]
1 & IC 5376 & Sb & 71.97 & 0.17 & 719.75 $\pm$ 10.36 & 0.67 $\pm$ 0.07 & 11.47 $\pm$ 0.88 & 1.17 $\pm$ 0.22 & 3.20 $\pm$ 0.44 & 5.71 $\pm$ 0.18 & 0.63 &  \ldots & 1.88 $\pm$ 0.18 \\  
2 & UGC 00005 & Sbc & 104.31 & 0.51 & 2676.95 $\pm$ 22.90 & 0.65 $\pm$ 0.05 & 40.99 $\pm$ 1.87 & 4.10 $\pm$ 0.71 & 20.10 $\pm$ 0.19 & 31.57 $\pm$ 0.78 & 0.49 & 20.08 $\pm$ 0.66 & 20.55 $\pm$ 0.66 \\  
3 & NGC 7819 & Sc & 71.47 & 0.76 & 2617.25 $\pm$ 7.13 & 0.43 $\pm$ 0.04 & 25.42 $\pm$ 0.75 & 2.07 $\pm$ 0.35 &  \ldots &  \ldots & \ldots & 9.42 $\pm$ 1.80 & 9.33 $\pm$ 0.22 \\  
5 & IC 1528 & Sbc & 54.77 & 0.48 & 3146.26 $\pm$ 12.74 & 0.54 $\pm$ 0.03 & 19.78 $\pm$ 0.40 & 1.79 $\pm$ 0.30 & 17.37 $\pm$ 0.16 & 21.66 $\pm$ 0.13 & 0.24 & 5.22 $\pm$ 0.24 & 5.58 $\pm$ 0.17 \\   
7 & UGC 00036 & Sab & 89.82 & 0.46 & 839.79 $\pm$ 10.40 & 0.16 $\pm$ 0.11 & 9.57 $\pm$ 0.66 & 0.61 $\pm$ 0.11 & 3.74 $\pm$ 0.62 & 7.47 $\pm$ 0.28 & 0.75 & 5.31 $\pm$ 0.33 & 5.03 $\pm$ 0.29 \\
\end{longtable}
\end{landscape}
}

\begin{table*}
\tiny
\caption{Values of the a, a$'$ and b coefficients for the calibration of single-band tracers: Linear and non-linear fits, as explained in Section \ref{Single-band updated}. The numbers in brackets refer to the number of galaxies used in each case (same notation applied in Tables \ref{hybrids_global_table}, \ref{hybrids_type_table} and \ref{hybrids_mass_table}). The luminosities are expressed in erg s$^{-1}$ and the values of the SFR are in M$_{\odot}$yr$^{-1}$.}
\label{single_table}
\begin{center}
\begin{tabular}{lcccccccc}
\hline
\multirow{1}{4.0cm}{Single-band Tracers}  & \multicolumn{1}{p{4.0cm}}{\centering Without type-2 AGN} & \multicolumn{1}{p{4.0cm}}{\centering With type-2 AGN}  \\
\hline\hline
SFR $=$ a $\times$ L(22\,$\mu$m) & a $=$  \big(3.0$^{+2.1}_{-0.4}$ \big) $\times$ 10$^{-43}$ [164]  & a $=$ \big(2.8$^{+1.6}_{-0.6}$ \big) $\times$ 10$^{-43}$ [263]    \bigstrut \\
SFR $=$ a $\times$ L(TIR) & a $=$ \big(2.8$^{+1.7}_{-0.5}$ \big) $\times$ 10$^{-44}$ [135]   & a $=$ \big(2.3$^{+1.6}_{-0.5}$ \big) $\times$ 10$^{-44}$ [218]    \bigstrut \\
SFR $=$ a$'$ $\times$ [L(22\,$\mu$m)/10$^{43}$]$^{b}$ (log scale fit) & a$'$ $=$ 2.12 $\pm$ 0.18 $;$ b $=$ 0.733 $\pm$ 0.053  & a$'$ $=$ 1.86 $\pm$ 0.11 $;$ b $=$ 0.702 $\pm$ 0.039   \bigstrut \\
SFR $=$ a$'$ $\times$ [L(22\,$\mu$m)/10$^{43}$]$^{b}$ (curve\_fit) & a$'$ $=$ 2.60 $\pm$ 0.12 $;$ b $=$ 0.697 $\pm$ 0.051 & a$'$ $=$ 2.38 $\pm$ 0.12 $;$ b $=$ 0.564 $\pm$ 0.043  \bigstrut \\
SFR $=$ a$'$ $\times$ [L(TIR)/10$^{43}$]$^{b}$ (log scale fit) & a$'$ $=$ 0.403 $\pm$ 0.032 $;$ b $=$ 0.720 $\pm$ 0.042 & a$'$ $=$ 0.359 $\pm$ 0.028  $;$ b $=$ 0.719 $\pm$ 0.038 \bigstrut \\
SFR $=$ a$'$ $\times$ [L(TIR)/10$^{43}$]$^{b}$ (curve\_fit) & a$'$ $=$ 0.350 $\pm$ 0.063 $;$ b $=$ 0.845 $\pm$ 0.064 & a$'$ $=$ 0.285 $\pm$ 0.054 $;$ b $=$ 0.881 $\pm$ 0.061 \bigstrut \\
\hline
\end{tabular}
\end{center}
\end{table*}

\begin{table*}
\tiny
\caption{Values of the a$_{IR}$ coefficients for the calibration of hybrid tracers: GLOBAL VALUES (see Section \ref{Hybrid tracers} for a detailed explanation)\\
The errors quoted here are the 1 $\sigma$ dispersions measured as the interval that includes 68\% of the data points around the median and correspond with the spread of the histograms in Figure \ref{histograms_constant}. Note that the standard error of the median, computed from the asymptotic variance formula as 1.253$\times$$\sigma$$/$$\sqrt{N}$, where $\sigma$ is referred to the values listed here and N is the number of galaxies shown in brackets, decreases these errors considerably (black tick-marks shown at the top in Figure \ref{histograms_constant}). The luminosities in these expressions are in erg s$^{-1}$ and the values of the SFR are expressed in M$_{\odot}$yr$^{-1}$.}
\label{hybrids_global_table}
\begin{center}
\begin{tabular}{lcccccccc}
\hline
\multirow{1}{2.0cm}{Hybrid Tracers}  & \multicolumn{1}{p{4.0cm}}{\centering Without type-2 AGN} & \multicolumn{1}{p{4.0cm}}{\centering With type-2 AGN}  \\
\hline\hline
SFR $=$ 5.5 $\times$ 10$^{-42}$ [L(H$\alpha$)$_{obs}$  + a$_{IR}$ $\times$ L(22\,$\mu$m)] & 0.018$^{+0.018}_{-0.006}$  [164]  & 0.015$^{+0.018}_{-0.006}$ [263]   \bigstrut \\
SFR $=$ 5.5 $\times$ 10$^{-42}$ [L(H$\alpha$)$_{obs}$ + a$_{IR}$ $\times$ L(TIR)] & 0.0019$^{+0.0015}_{-0.0005}$ [135]   & 0.0015$^{+0.0016}_{-0.0006}$ [218]   \bigstrut \\
SFR $=$ 4.6 $\times$ 10$^{-44}$ [L(FUV)$_{obs}$  + a$_{IR}$ $\times$ L(22\,$\mu$m)] &  4.52$^{+3.55}_{-1.14}$ [113]  & 3.55$^{+3.38}_{-0.95}$ [187]  \bigstrut \\
SFR $=$ 4.6 $\times$ 10$^{-44}$ [L(FUV)$_{obs}$  + a$_{IR}$ $\times$ L(TIR)]  &  0.40$^{+0.33}_{-0.09}$ [94]  & 0.33$^{+0.29}_{-0.07}$ [156]  \bigstrut \\
\hline
\end{tabular}
\end{center}
\end{table*}

\begin{table*}
\tiny
\caption{Values of the a$_{IR}$ coefficients for the calibration of hybrid tracers: BY MORPHOLOGICAL TYPE (see Sections \ref{Hybrid tracers}  and \ref{morphological dependence} for a detailed explanation). The recipes to compute the SFR in M$_{\odot}$yr$^{-1}$ shown on the left column are the same as in Table \ref{hybrids_global_table}. The luminosities are in units of erg s$^{-1}$.}
\label{hybrids_type_table}
\begin{center}
\begin{tabular}{lcccccccc}
\hline
\multirow{2}{2.0cm}{Hybrid Tracers}  & \multicolumn{3}{p{6.0cm}}{\centering Without type-2 AGN} & \multicolumn{3}{p{6.0cm}}{\centering With type-2 AGN}  \\
\cline{2-7} & \multicolumn{1}{p{2.0cm}}{\centering S0/a-Sab } & \multicolumn{1}{p{2.0cm}}{\centering Sb-Sbc } & \multicolumn{1}{p{2.0cm}}{\centering Sc-Irr } & \multicolumn{1}{p{2.0 cm}}{\centering S0/a-Sab } & \multicolumn{1}{p{2.0cm}}{\centering Sb-Sbc } & \multicolumn{1}{p{2.0cm}}{\centering Sc-Irr }  \bigstrut \\
\hline\hline
L(H$\alpha$)$_{obs}$  + a$_{IR}$ $\times$ L(22\,$\mu$m) & 0.010$^{+0.037}_{-0.005}$ [21] & 0.022$^{+0.016}_{-0.007}$ [76] & 0.021$^{+0.016}_{-0.009}$ [67] & 0.006$^{+0.026}_{-0.006}$ [58] & 0.017$^{+0.017}_{-0.005}$ [131] & 0.019$^{+0.015}_{-0.008}$ [74] \bigstrut \\
L(H$\alpha$)$_{obs}$ + a$_{IR}$ $\times$ L(TIR) & 0.0014$^{+0.0029}_{-0.0006}$ [16] & 0.0020$^{+0.0013}_{-0.0006}$ [68] & 0.0022$^{+0.0012}_{-0.0006}$ [51] & 0.0008$^{+0.0022}_{-0.0004}$ [44] & 0.0016$^{+0.0012}_{-0.0004}$ [116] & 0.0018$^{+0.0013}_{-0.0005}$ [58] &  \bigstrut \\
L(FUV)$_{obs}$  + a$_{IR}$ $\times$ L(22\,$\mu$m) &  4.17$^{+2.76}_{-0.86}$ [14] & 4.67$^{+4.52}_{-0.88}$ [51] & 4.75$^{+3.97}_{-1.37}$ [48] & 2.55$^{+4.37}_{-0.46}$ [39] & 3.65$^{+2.90}_{-0.78}$ [93] & 4.04$^{+4.61}_{-1.04}$ [55] \bigstrut \\
L(FUV)$_{obs}$  + a$_{IR}$ $\times$ L(TIR)  &  0.37$^{+0.24}_{-0.15}$ [11] & 0.42$^{+0.25}_{-0.06}$ [46] & 0.43$^{+0.32}_{-0.12}$ [37] & 0.27$^{+0.31}_{-0.07}$ [30] & 0.34$^{+0.21}_{-0.07}$ [82] & 0.36$^{+0.39}_{-0.08}$ [44] \bigstrut \\
\hline
\end{tabular}
\end{center}
\end{table*}

\begin{table*}
\tiny
\caption{Values of the a$_{IR}$ coefficients for the calibration of hybrid tracers: BY STELLAR MASS (see Sections \ref{Hybrid tracers}  and \ref{mass dependence} for a detailed explanation). The stellar masses are in units of M$_{\star}$/M$_{\odot}$. Expressions to compute the SFR (M$_{\odot}$yr$^{-1}$) appeared on the left column. These recipes are the same as the ones in Table \ref{hybrids_global_table}. The luminosities are in units of erg s$^{-1}$.}
\label{hybrids_mass_table}
\begin{center}
\begin{tabular}{lcccccccc}
\hline
\multirow{2}{2.0cm}{Hybrid Tracers}  & \multicolumn{3}{p{6.0cm}}{\centering Without type-2 AGN} & \multicolumn{3}{p{6.0cm}}{\centering With type-2 AGN}  \\
\cline{2-7} & \multicolumn{1}{p{2.0cm}}{\centering log[M$_{\star}$]$>$10.5 } & \multicolumn{1}{p{2.0cm}}{\centering 10.0$<$log[M$_{\star}$]$<$10.5 } & \multicolumn{1}{p{2.0cm}}{\centering log[M$_{\star}$]$<$10.0 } & \multicolumn{1}{p{2.0 cm}}{\centering log[M$_{\star}$]$>$10.5 } & \multicolumn{1}{p{2.0cm}}{\centering 10.0$<$log[M$_{\star}$]$<$10.5 } & \multicolumn{1}{p{2.0cm}}{\centering log[M$_{\star}$]$<$10.0  }  \bigstrut \\
\hline\hline
L(H$\alpha$)$_{obs}$  + a$_{IR}$ $\times$ L(22\,$\mu$m) & 0.014$^{+0.024}_{-0.006}$ [44] & 0.021$^{+0.012}_{-0.004}$ [60] & 0.021$^{+0.018}_{-0.010}$ [60] & 0.009$^{+0.019}_{-0.007}$ [108] & 0.017$^{+0.016}_{-0.005}$ [88] & 0.019$^{+0.020}_{-0.007}$ [66] & \bigstrut \\
L(H$\alpha$)$_{obs}$ + a$_{IR}$ $\times$ L(TIR) & 0.0018$^{+0.0017}_{-0.0008}$ [37] & 0.0020$^{+0.0008}_{-0.0006}$ [54] & 0.0021$^{+0.0017}_{-0.0006}$ [44] & 0.0010$^{+0.0018}_{-0.0006}$ [92] & 0.0018$^{+0.0011}_{-0.0004}$ [77] & 0.0019$^{+0.0018}_{-0.0005}$ [48] & \bigstrut \\
L(FUV)$_{obs}$  + a$_{IR}$ $\times$ L(22\,$\mu$m) & 4.06$^{+1.93}_{-1.61}$ [30] & 4.75$^{+2.21}_{-0.63}$ [39] & 4.77$^{+5.07}_{-1.39}$ [44] & 2.93$^{+2.34}_{-0.84}$ [81] & 4.25$^{+2.17}_{-1.09}$ [57] & 4.55$^{+4.87}_{-1.56}$ [48] &  \bigstrut \\
L(FUV)$_{obs}$  + a$_{IR}$ $\times$ L(TIR)  &  0.36$^{+0.22}_{-0.11}$ [26] & 0.44$^{+0.16}_{-0.06}$ [36] & 0.41$^{+0.37}_{-0.08}$ [32] & 0.25$^{+0.25}_{-0.03}$ [70] & 0.41$^{+0.18}_{-0.10}$ [50] & 0.40$^{+0.38}_{-0.10}$ [35]  \bigstrut \\
\hline
\end{tabular}
\end{center}
\end{table*}

\end{document}